\documentclass[11pt,a4paper]{article}
\usepackage{jheppub}
\usepackage{amsmath}
\usepackage{graphicx}
\usepackage{indentfirst}
\usepackage{physics}
\usepackage{verbatim}

\usepackage{titlesec}
\titleformat*{\section}{\normalfont\Large\bfseries\boldmath}
\titleformat*{\subsection}{\normalfont\large\bfseries\boldmath}

\newcommand{\Tau}{\mathcal{T}}
\numberwithin{equation}{section}
\newcommand{\HirD}[3]{D_{#2}^{(#3)}\left(#1,#1\right)}
\newcommand{\uns}{{\bf u}}
\newcommand{\T}{{\bf T}}
\newcommand{\gII}{{\bf g_2}}
\newcommand{\gIII}{{\bf g_3}}

\title{
Surface observables in gauge theories,\\ 
modular Painlevé tau functions\\ and 
non-perturbative topological strings
}

\author{Giulio Bonelli, Pavlo Gavrylenko, Ideal Majtara and Alessandro Tanzini}

\affiliation{SISSA, Via Bonomea 265, 34136 Trieste, Italy}
\affiliation{INFN, Sezione di Trieste, Trieste, Italy}
\affiliation{Institute for Geometry and Physics, IGAP, via Beirut 2, 34136 Trieste, Italy}
    
    \emailAdd{bonelli@sissa.it}
    \emailAdd{pasha.145@gmail.com}
    \emailAdd{imajtara@sissa.it}
    \emailAdd{tanzini@sissa.it}

\abstract{We study BPS surface observables of ${\cal N}=2$ four dimensional $SU(2)$ gauge theory in gravitational $\Omega$-background at perturbative and at Argyres-Douglas superconformal fixed points. This is done by formulating the equivariant gauge theory on the blow-up of ${\mathbb C}^2$ and considering the decoupling Nekrasov-Shatashvili limit. We show that in this limit the blow-up equations are solved by corresponding Painlev\'e $\Tau$-functions and exploit operator/state correspondence to compute their expansion in an integer basis, given in terms of the moduli of the quantum Seiberg-Witten curve. We study the modular properties of these solutions and show that they do directly lead to BCOV holomorphic anomaly equations for the corresponding topological string partition function. The resulting $\Tau$-functions are holomorphic and modular and as such they provide a natural non-perturbative completition of topological strings partition functions.
}

\begin{document}
\maketitle

\section{Introduction}

The study of the behaviour of gauge and string theories under variations of the topology or the Riemann geometry of the 
manifold over which these are formulated
is a tool to explore their non-perturbative properties.

For example, the handle gluing operator in topological quantum field theory in two dimensions is a prototype of topology changing operator and it is relevant to understand the link between topological two dimensional gauge theories and  
quantum integrable systems \cite{Nekrasov:2014xaa}. As we will review later, the handle gluing operator formula can be obtained as a consequence of the state/operator correspondence. This is an example of the link between topology changing operators and local operators via state/operator correspondence.

In four dimensions, a minimal topology changing operator is the operation of blowing up a point.
Blowup equations are relations describing how the gauge theory on a manifold $X$ behaves under the replacement of a regular (or singular) point of $X$ with the local geometry of a two-sphere. In the case of a regular point, this 
corresponds to replacing the four manifold $X$ with
$\hat X= X \# \overline{\mathbb{CP}^2}$, the connected sum of $X$ with the projective complex plane.

For supersymmetric gauge theories and topological strings the behaviour under
blowup is controlled by strong integrability constraints which turn out to be very effective in determining the partition function and BPS observables of these theories. 
A new perspective on the relation between $\mathcal{N}=2$ supersymmetric gauge theories and integrable systems 
opened with the discovery that the general solution of Painlev\'e equations can be obtained in terms of supersymmetric partition functions \cite{Gamayun:2013auu,Litvinov:2013sxa}.
Since then there have been several further studies concerning late time expansion \cite{Its:2014lga}
and its relation to strongly coupled phases of gauge theory \cite{Bonelli:2016qwg}, extension to more general isomonodromic deformation problems
\cite{Gavrylenko:2016zlf,Bonelli:2019boe,Bonelli:2019yjd,Bonelli:2021rrg,
Bonelli:2022iob,DelMonte:2022nem,Gavrylenko_2019} and to q-difference
equations \cite{Gamayun:2013auu,Bershtein:2016aef,Bonelli:2017gdk,Bershtein:2017swf,Bonelli:2020dcp,jimbo2017cftapproachqpainlevevi}. Moreover, the
relation to Fredholm determinants \cite{McCoy:1976cd,1996CMaPh.179....1T,Zamolodchikov:1994uw} allowed to estabilish a link to 
the proposal of \cite{Grassi:2014zfa} for the non-perturbative completion of topological strings in terms of spectral theory \cite{Bonelli:2016idi,Bonelli:2017ptp,Bonelli:2017gdk,Gavrylenko:2023ewx,Francois:2023trm}. Other proposals for a non-perturbative formulation of topological strings in terms of isomonodromic $\Tau$-functions appeared in \cite{Coman:2020qgf,Coman:2022igv,Bridgeland:2023eka}.

For toric manifolds one can study the gauge theory in presence of the so called $\Omega$-background \cite{Nekrasov:2003rj}, corresponding to considering an equivariant extension of the supersymmetric charges with respect to the action of the algebraic torus $\left(\mathbb{C}^*\right)^2$ with parameters $(\epsilon_1,\epsilon_2)$.
This led Nakajima and Yoshioka (NY) to formulate an equivariant version of blowup equations \cite{Nakajima:2003pg,gottsche2010donaldson} on $\mathbb{C}^2$.
The appearence of Painlev\'e equations in supersymmetric gauge theories is strictly related to equivariant blowup
equations. Indeed, proofs of the original result of \cite{Gamayun:2013auu} were provided in \cite{Bershtein:2014yia} in terms of blowup equations
on $\mathbb{C}^2/\mathbb{Z}_2$ \cite{Bonelli:2011jx,Bonelli:2011kv}, and in \cite{Nekrasov:2020qcq,Jeong:2020uxz} in terms of blowup equations on
$\mathbb{C}\times \mathbb{C}/\mathbb{Z}_2$ where the orbifold in the second factor models a surface defect in the gauge theory.
Moreover, the relation between holomorphic anomaly equations and NY equivariant blowup equations in theories without surface observables was investigated in
\cite{Sun:2021lsq}.

In this paper we propose a new road to Painlev\'e equations from the blowup, based on taking the so-called Nekrasov-Shatashvili (NS) limit of the original Nakajima-Yoshioka blowup equations on $\mathbb{C}^2$. 

We consider the generating function of surface observables on the blow-up of the plane, called \emph{blow-up factor} \eqref{aut,4}, which, in the NS limit, is essentially the Painlev\'e $\Tau$-function itself.
We show that this produces a new expansion of the latter, in terms of {\it gauge invariant} quantities
describing the infrared (quantum) effective Seiberg-Witten geometry. This new expansion displays a number of remarkable features:
\begin{itemize}
    \item it is a convergent expansion around one of its zeros, being thus {\it analytic}
    \item it is a {\it Hurwitz integral}  series\footnote{The integrality of the coefficients is presently only a conjectural property that we 
    experimentally verified to very high orders in the expansion (see Appendix \ref{appE}). For earlier observations on this theme, see \cite{hone2013properties,hone2017hirota}.}
    \item the resulting $\Tau$-function has manifest {\it modular} properties
\end{itemize}
The third property allows to establish a clear and direct link to the holomorphic anomaly equations governing the topological string amplitudes which geometrically engineer the corresponding gauge theory.
In particular, we obtain Painlev\'e $\Tau$-functions which are holomorphic and modular in their arguments, and are such that when expanded around a saddle point in the $\Omega$-background
they reproduce the standard genus expansion of topological strings, satisfying holomorphic anomaly equations. Furthermore, the above properties do not rely on the genus expansion of topological strings being exact in the string coupling.
We thus propose these $\Tau$-functions as realising a non-perturbative (therefore, background independent) formulation of topological string theory for these geometries. Schematically we have
\begin{equation}
\partial_{E_2}\Tau=0 \Leftrightarrow \begin{gathered} \text{BCOV holomorphic}\\ \text{anomaly equation.}\end{gathered}\ ,
\end{equation}
where the latter directly appears in the 
master form \cite{Bershadsky:1993cx}.
The physical principle at the basis of our derivation is operator/state correspondence. Indeed, this correspondence implies that the surface observable insertion can be expressed through a dependence on the chiral ring generator. As a consequence of this, the modular properties of the equivariant gauge theory on the blow-up
are be dictated by its dependence on this generator only. See \eqref{holNS,1} and 
section \ref{section4}
for the specific discussion of this.

The fact that the expansion is in terms of gauge invariant quantities allows to easily study it around any point of the Coulomb moduli space. We display expansions around the Argyres-Douglas points of $SU(2)$ gauge theory with massive hypers. 

In the Seiberg-Witten (SW) limit the expansion we find has a universal structure. 
Indeed, in this case the $\Tau$-function is proportional to the Weierstrass $\sigma$-function (see \cite{Moore:1997pc})
\begin{equation}\label{intro,0}
\Tau(s) \to e^{-\frac{1}{2} T s^2}\sigma(s,g_2,g_3)\ ,
\end{equation}
and is completely fixed  by two data, the elliptic invariants $g_2,g_3$ of the Weierstrass parametrization of the SW curve
\begin{equation}
y^2=4x^3-g_2x-g_3 \ ,    
\end{equation}
and the contact term $T$ which arises because the algebra of surface observables gets changed under the renormalization group flow \cite{Moore:1997pc}. This happens because at the intersection of two surfaces $\Sigma_1$ and $\Sigma_2$ there can be singularities which are given by a local observable $\tilde T$
\begin{equation}
I(\Sigma_1)I(\Sigma_2) \rightarrow I_{IR}(\Sigma_1)I_{IR}(\Sigma_2) + \tilde T\Sigma_1 \cdot \Sigma_2 \ , 
\end{equation}
where $\Sigma_1 \cdot \Sigma_2$ is the intersection number and $\tilde T$ is related to $T$ by a suitable shift as in \eqref{aut,T}. As a consequence we have that the generating function of the surface observable $I(\Sigma)$ has extra contributions given by Wick contractions of $\tilde T(u)$
\begin{equation}
e^{I(\Sigma)}\rightarrow e^{I_{IR}(\Sigma)+\tilde T(u)\Sigma^2}\ . 
\end{equation}
The remarkable feature of the result \eqref{intro,0} is that the details of the theory are completely encoded in the invariants $g_2,g_3$ and in $T$, so that different theories differ just by the specific form of the parametrization of the geometrical data $g_2,g_3, T$ in terms of the physical parameters of theory. The universal structure of the SW result implies the remarkable features of the $\Tau$-functions. Indeed, these are realized in a universal common way because of the classical power expansion of the Weierstrass $\sigma$-function itself in terms of integer polynomials of the elliptic invariants $g_2$ and $g_3$.
Painlev\'e $\Tau$-functions preserve the above properties, but in a more general basis of polynomials.
We indeed find that the Painlev\'e $\Tau$-function can be universally written as a quantum   
Weierstrass $\sigma$-function, that is, it can be obtained from this latter by acting with a suitable differential operator whose explicit form depends on the corresponding Painlev\'e equation, see \eqref{adesso}.

The content of the paper is the following.
In the next subsection we list several problems
which stay open to us.
In Sect.\ref{section2} we review the geometry of topology changing operators in topological
quantum field theory, 
provide a brief reminder of blowup equations in supersymmetric gauge theory in four dimensions and
focus on the NS limit of the blowup factors.
In Sect.\ref{section3} we discuss the
chiral ring expansion of the NS blowup factor by explaining its
autonomous (i.e. Seiberg-Witten) limit with respect to the Weierstrass $\sigma$-function 
and the blowup equations. We also explain the expansion of the blowup factor according to state/operator correspondence.
In Sect.\ref{section4}
we discuss the modular properties of the Painlev\'e $\Tau$-function, which lead to the proposal that they provide
a non-perturbative description of topological string. 
We propose concrete expansion formulas as universal
$\epsilon$-deformations of the Weierstrass $\sigma$-function in the form of quantum versions of the latter
and show how the 
holomorphic anomaly equations follow from the modular properties of the blowup factor implied by state/operator correspondence.
In Sect.\ref{section5} we explicitly give the Hurwitz expansions of the Painlev\'e $\Tau$-functions 
around their generic zeros. We specifically discuss the cases of 
PVI, PIV, PIII$_2$, PI and PIII$_3$, which according to Painlev\'e/gauge correspondence describe respectively $SU(2)$ supersymmetric gauge theories with 
$N_f=4$ fundamental hypers, $H_2$ AD-point, $N_f=1$, $H_0$ AD theory and pure SYM.
In App.\ref{appA} we recap our conventions on Weierstrass elliptic functions, 
in App.\ref{appB} we collect some details of the computation of modular invariants appearing in the 
expansion of PVI and in App.\ref{appD} we report some technical points relevant for section \ref{section4}.
In the extra App.\ref{appE} (located after the bibliography) we collect numerical tables of integer polynomial coefficients corresponding to the Hurwitz
expansions of section \ref{section5}.

\subsection{Open problems}

Let us comment on some open problems and further directions worth to explore according to our opinion.

 The $\Tau$-function we propose as non-perturbative completition of topological strings share some features with earlier proposals, such as
 \cite{Aganagic:2003qj,
 Eynard_2011} and the trans-series solution of holomorphic anomaly equations discussed in
 \cite{Couso-Santamaria:2014iia,Gu:2022sqc}. It would be important to analyse in detail the relation among these various approaches.
 As we already noticed, Painlev\'e 
$\Tau$-functions can be regarded as a quantum version of the classical Weierstrass $\sigma$-function. The resulting expansion in the $\epsilon$ $\Omega$-background parameter
can be possibly compared with the results of 
\cite{Grassi_2016,Iwaki_2020,Eynard_2011,Iwaki:2023cek,LesDiablerets,KLIZ}.
We also notice that this expansion should interestingly simplify at the singular strongly coupled points where the  Weierstrass functions reduce to trigonometric ones.

The blowup factor plays a key r\^ole in mathematics to study 
 Donaldson invariants of four manifolds \cite{fintushel1994blowup}. These are integral-valued by their very definition, as the 
    singular part of the instanton moduli space does not contribute to their evaluation.  In the algebraic geometry description, even if the moduli space is singular, assuming that semistability implies stability, this should not create rational numbers as the virtual class is anyhow an integral class\footnote{M. Kool, private communication.}.
    As a consequence of this, the Donaldson invariants on the blowup are also integers and indeed 
    these are counted by the integer coefficients of the $\sigma$-function expansion. By extending these assumptions to the equivariant setting,
    this would explain the integrality of the coefficients of expansions of the Painlev\'e $\Tau$-function that we conjecturally propose as equivariant Donaldson invariants on the blowup. It would be important then to provide a proof of the integrality of such coefficients.
    This issue can be also addressed in the AGT dual picture by making use of representation theory of Virasoro algebras \cite{bershtein2013coupling}.

     An obvious generalization is to study the higher rank case. For pure super Yang-Mills theory with general gauge group this corresponds to the $\Tau$-functions of non-autonomous Toda system \cite{Bonelli:2021rrg,Bonelli:2022iob}. The non-equivariant blowup equations for the $SU(N)$ case were studied in \cite{Edelstein:2000aj} were a relation to KdV hierarchy was highlighted and the blow-up factors computed to be the analog of $\sigma$ functions on higher genus Riemann surfaces. We remark that also higher genus $\sigma$ functions are known to enjoy Hurwitz integrality (see for example \cite{ayano2024hurwitzintegralitypowerseries}), so one can expect this to be the case also for the $\Tau$-functions of the non-autonomous Toda system.
 The Hurwitz integrality of the generating functions of higher rank equivariant Donaldson invariants seems indeed a non trivial prediction and a conjecture to further verify, e.g. against \cite{Gottsche:2021dye,Gottsche:2021ihz}.

    Another natural extension is to linear $SU(N)^n $ quiver gauge theories whose blowup equations are related to the $\Tau$-function of isomonodromic deformations on the sphere with 
    $n+3$-punctures \cite{Gavrylenko:2016zlf}. Also circular quivers can be considered by formulating the isomonodromic deformation problem on an elliptic curve \cite{Bonelli:2019boe,Bonelli:2019yjd}. 
    For more general class S theories \cite{Gaiotto:2009we} the corresponding $\Tau$-functions are the ones of the isomonodromic deformation problem of linear systems with rational coefficients on higher genus Riemann surfaces. 

 The analysis of the zeroes of the $\Tau$-functions we calculate in this paper contains a direct combinatorial relation between the gauge theory prepotential in the self-dual $(\epsilon,-\epsilon)$ and NS $(\epsilon,0)$ $\Omega$-backgrounds, see \eqref{bl,1} at $s=0$. The geometrical interpretation of the latter is given by a relation between the ordinary and relative Gromov-Witten invariants of local del Pezzo geometries \cite{Bousseau:2020ckw} in the geometric engineering limit. It would be interesting to uplift our analysis to five-dimensional gauge theories to fully establish this correspondence for the above local geometries and to extend this analysis to higher rank gauge theories. Indeed, for higher rank Super-Yang Mills theories in five dimensions these relations were studied in \cite{Grassi:2016nnt}.

 The five-dimensional uplift of the blowup equations is related to cluster integrable systems \cite{Bershtein:2017swf,Bershtein:2018srt,Bonelli:2020dcp}. It would be interesting to study the expansion of the corresponding $\Tau$-functions around their zeroes and the corresponding K-theoretic equivariant Donaldson invariants.
 
 In this paper the blowup equations have been studied in the NS limit connecting them to Painlev\'e equations. It would be interesting to study them in the full $\Omega$-background, which should correspond to a suitable quantum version of the Painlev\'e equations. In this context it would be very interesting to compare our results for PI with two-dimensional quantum gravity and possibly find an interpretation of the general $\Omega$-background from this perspective.

In the Nekrasov-Shatashvili (NS) limit of vanishing of one of the two equivariant parameters,
it is known that the gauge theory prepotential $\mathcal{F}_{NS} =\log Z_{NS}$ can be computed in terms of the quantum periods \cite{Mironov:2009uv,Maruyoshi:2010iu,Bonelli:2011na,Aganagic:2011mi}. This is defined by a Schrödinger operator obtained with the replacement $y \to -i\epsilon\partial_x$. It is then natural to ask if the expansion of the $\Tau$ function can be fixed by the invariants of the quantum curve which intuitively will be operators $\hat g_2,\hat g_3$ acting on the NS wavefunction. This would also require to understand how the contact term gets modified in this case. 

Related to the above
question, we remark that 
in the foundational paper \cite{Moore:1997pc},
    Donaldson invariants are obtained from an integral over the Coulomb moduli space of the gauge theory, namely 
    the $u$-plane integral, see \cite{Korpas:2019cwg,Manschot:2021qqe,Aspman:2022sfj,Aspman:2023ate} for the extension to massive theories. This is proposed as an effective reduction of the gauge theory path integral
    over the low energy modes and computed by exploiting the implication of the topological nature of the twisted supersymmetric gauge theory. It sounds natural to expect that an equivariant version of this construction exists and that it might reproduce our results.
     
 More recently, the
 computation of partition functions of AD theories on four manifolds has been considered 
    in \cite{Moore:2017cmm} along with the question if these define new differential invariants. 
    The results presented in the above paper are given in the form of a $u$-plane integral, where the relevant contributions to the integrand are computed by assuming that their form follows a universal expression independent on the specific theory, but it is dictated by the underlying SW integrable system. This is consistent with our results as the structure of the SW blow-up factor is universally given by the specific evaluation of the modular invariants and the contact term coefficient. It would be interesting to lift also the results of \cite{Moore:2017cmm} to an equivariant set-up for toric manifolds.

It would be very interesting to find a derivation of the Hirota 
bilinear form of the Painlev\'e equations for the blowup factor
 by making use of fusion rules for the surface observable. These should express 
the consequences of the relation between surface observables and magnetic $U(1)$ 1-form symmetry.

{\bf Acknowledgments:} 
We would like to thank 
M.~Bershtein,
F.~Bousseau,
A.~Grassi,
N.~Iorgov,
M.~Kool,
O.~Lisovyy,
M.~Mariño,
A.~Marshakov,
G.~Moore,
N.~Nekrasov,
A.~Shchechkin,
C.~Vafa, 
E.~Witten,
Yu.~Zhuravlev,
for useful discussions and comments.

We also thank G. Cotti as local organiser of the CaLISTA 
School on Integrable System ``Contemporary trends in integrable systems'' in Lisbon, during which part of this paper was written, for having provided crucial technical tools for the implementation of the figures.

A preliminary communication on the results contained in this paper was given by Ideal Majtara at the gong show of Stringmath24 (ICTP-IGAP).

The research of G.B. and I.M. is partly supported by the INFN Iniziativa Specifica ST\&FI and by the PRIN project “Non-perturbative Aspects Of Gauge Theories And Strings”. The research of  P.G. and A.T. is partly supported by the INFN Iniziativa Specifica GAST and Indam GNFM. 
The research is partly supported by the MIUR PRIN Grant 2020KR4KN2 ``String Theory as a bridge between Gauge Theories and Quantum Gravity''.  
All the authors acknowledge funding from the EU project Caligola HORIZON-MSCA-2021-SE-01), Project ID: 101086123, and CA21109 - COST Action CaLISTA.

\section{Topology changing operations in TQFT}\label{section2}

\subsection{Handle gluing operators in two dimensions}

Topology changing operators are naturally studied in connection to operator/state correspondence.
Let us discuss this with an example in two dimensional topological field theory (TFT2). 
A well-known topology changing operator is the handle-gluing operator on the Riemann surface 
$\Sigma_g$
where the TFT2 is formulated.
This is given by the operator changing the Riemann surface by the addition of a handle.
This operation can be realised in two equivalent ways related by operator/state correspondence. 

One is the creation of an handle on $\Sigma_g$ by replacing it with $\Sigma_g\# T^2$, the connected sum of $\Sigma_g$ and the two torus $T^2$.
This is obtained by excising a disk on $\Sigma_g$ and on $T^2$ and gluing the two components
along the boundary circles. 
From this view point the string partition function at genus $g+1$ is expressed as
the pairing of the two  
component TFT2 wave functions along the boundaries
$${\cal Z}_{g+1}(\Sigma_g\# T^2)
=<T^2|\Sigma_g>.$$ 
Elongating the cylinder containing the gluing circle to conformal infinity defines the TFT2
handle operator ${\cal H}$ via operator/state correspondence
$$<T^2|\Sigma_g>=<{\cal H}>_{\Sigma_g}.$$
This depends on three complex parameters, namely
the modulus of the elliptic curve, the center of the excised disk in $\Sigma_g$ and the
length/twist parameter of the gluing. 
These are the three extra Teichmuller parameters needed to 
correctly account for the complex structure of the genus $g+1$ Riemann surface $\Sigma_g\# T^2$.

The second way is the creation of a handle on the string world sheet $\Sigma_g$ by
excising two non overlapping disks on the Riemann surface and by gluing their boundary circles 
generating the new Riemann surface
${\cal H}\Sigma_g$. 
The corresponding TFT2 amplitude depends on the three extra Teichmuller parameters needed to 
correctly account for the complex structure of the genus $g+1$ Riemann surface 
${\cal H}\Sigma_g$, namely the position of the centers of the two excised disks in $\Sigma_g$
and the length/twist parameter of their gluing.
The state/operator correspondence relates then the resulting TFT2 
partition function to the normalized trace of two point functions 
$${\cal Z}_{g+1}({\cal H}\Sigma_g)= 
C^{AB}<{\cal O}_A{\cal O}_B>_{\Sigma_g}\ .
$$

\begin{figure}[htb]
\centerline{\includegraphics[scale=.3]{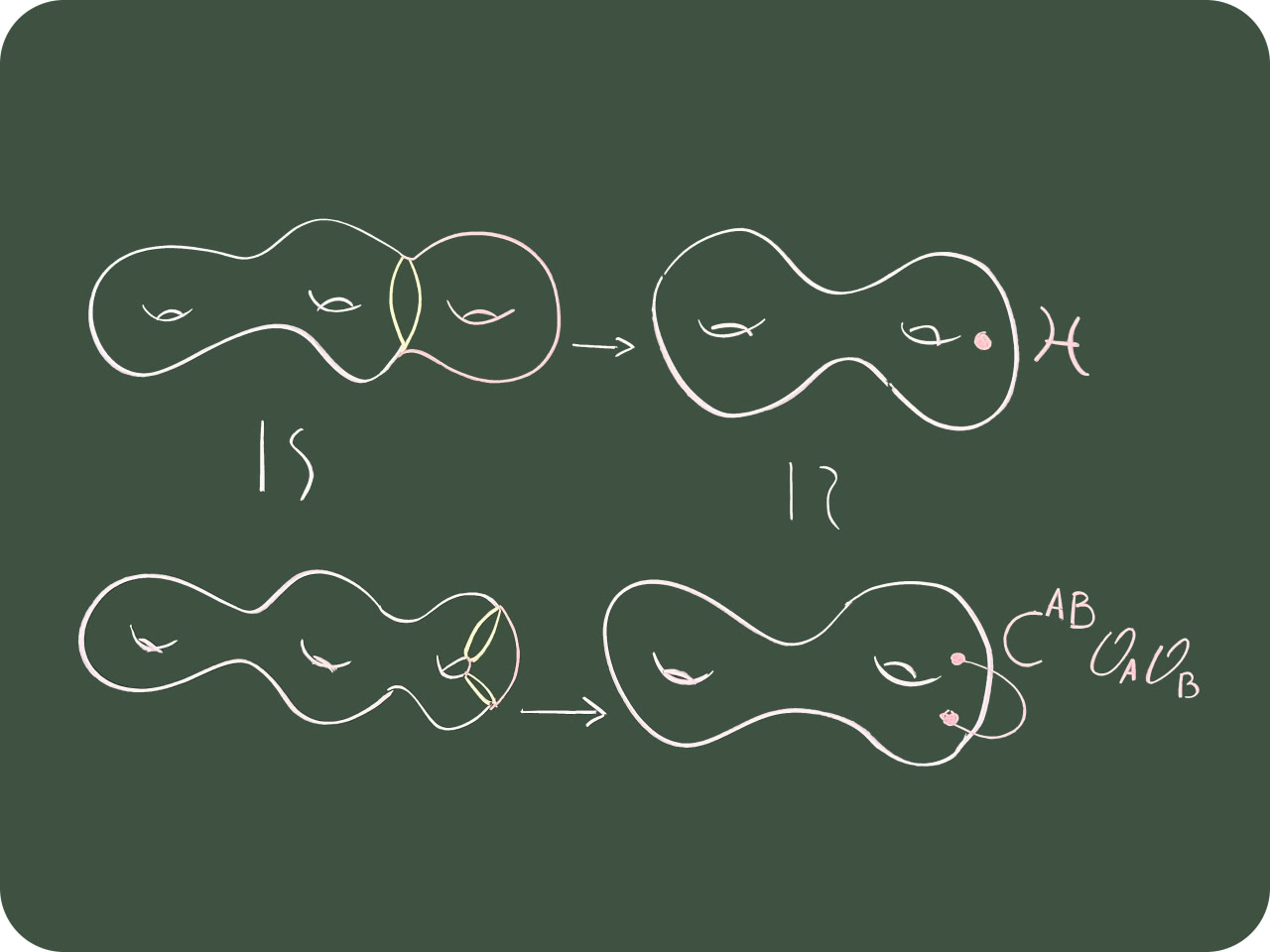}}
\caption[]{\label{fig:glue} Handle gluing operator and state/operator correspondence.}
\end{figure}

These two ways are indeed equivalent by operator/state correspondence and define the handle gluing operator in terms of local ones. Indeed, the circle in the first construction can be deformed from the base of the handle to a pair of circles linking the handle itself, but with opposite orientations.
While elongating at conformal infinity the first was defining the handle operator ${\cal H}$
as a topology changing operator, elongating to conformal infinity the pair of resulting circles
expresses the handle operator in terms of local ones. Indeed one obtains the ``handle gluing formula'' 
$${\cal H}=C^{AB}{\cal O}_A{\cal O}_B$$
and one gets
$${\cal Z}_{g+1}({\cal H}\Sigma_g)={\cal Z}_{g+1}(\Sigma_g\# T^2)$$
as depicted in Fig.\ref{fig:glue}.
The above construction was stated in more general terms in 
\cite{Nekrasov:2014xaa}. 

\subsection{Blow-up operator in four dimensions}

The blow up of a smooth point of a four dimensional Riemannian manifold $X$ is obtained by 
removing a four-dimensional ball around that point and gluing back the component at 
infinity of $\overline{\mathbb{CP}}^2$. This is the connected sum $\hat X=X\# \overline{\mathbb{CP}}^2$.
The local description of 
${\mathbb{CP}}^2$ around its irreducible divisor
is the total space of ${\mathcal O}(-1)_{\mathbb{CP}^1}$, which as also a toric manifold and can be described as the GIT quotient $\left({\mathbb{C}^3\setminus {0}}\right)/\mathbb{C}^*$
with weights $(1,1,-1)$ on the three coordinates. 
Equivalently, it can be described by the K\"ahler quotient of $\mathbb{C}^3$ w.r.t. a $U(1)$ action whose moment map is set to
$|z_1|^2+|z_2|^2-|z_0|^2=\zeta>0$.
This space is the blow up of the complex plane, usually denoted as $\hat{\mathbb{C}}^2$.
The geometry of $\hat{\mathbb{C}}^2$ can be analysed in terms of its K\"ahler potential.
This is obtained by extremising the ${\mathbb C}^*$ invariant potential
$$
U[v, z_0,z_1,z_2] =  e^{-2 v} (|z_1|^2 + |z_2|^2) + e^{2 v} |z_0|^2 + 2 v \zeta\ ,
$$
with respect to $v$. At the stable critical point
$$
v_*=\frac{1}{2} {\rm ln}\left[((-\zeta + \sqrt{\zeta^2 + 4 |z_0|^2 (|z_1|^2 +|z_2|^2)})/(2 |z_0|^2))\right]\ ,
$$
and setting $K[z_0,z_1,z_2] =U[v_*, z_0, z_1, z_2]$.
One has 
$$
K[z_0,z_1,z_2] =\sqrt{\zeta^2 + 4 |z_0|^2 (|z_1|^2 + |z_2|^2)} + 
 \zeta \,{\rm ln}\left[(-\zeta + \sqrt{\zeta^2 + 4 |z_0|^2 (|z_1|^2 + |z_2|^2)})/(2 |z_0|^2)\right]\ ,
$$
In the blow down limit 
$
\zeta\sim 0
$
one can reduce to the region
$
 z_0\sim\sqrt{|z_1|^2+|z_2|^2}
$
and obtain
$$
K \sim 2(|z_1|^2+|z_2|^2) -\frac{\zeta^2}{4(|z_1|^2+|z_2|^2)} + O(\zeta^4)\ ,
$$
which is the flat K\"ahler potential on ${\mathbb C}^2$ with corrections due to the infinitesimal blow up.
In the blown up limit 
$
\zeta\sim +\infty
$
one can instead rescale to 
$
z_0\sim 1
$
and expand
$$
K \sim 
\zeta(1-\ln(\zeta))+
\zeta\, \ln(|z_1|^2+|z_2|^2) + \zeta^{-1} (|z_1|^2+|z_2|^2) + O(\zeta^{-3})\ ,
$$
which is the Kahler potential of the Burns metric (up to a constant).

\begin{figure}[htbp]
\centerline{\includegraphics[scale=.3]{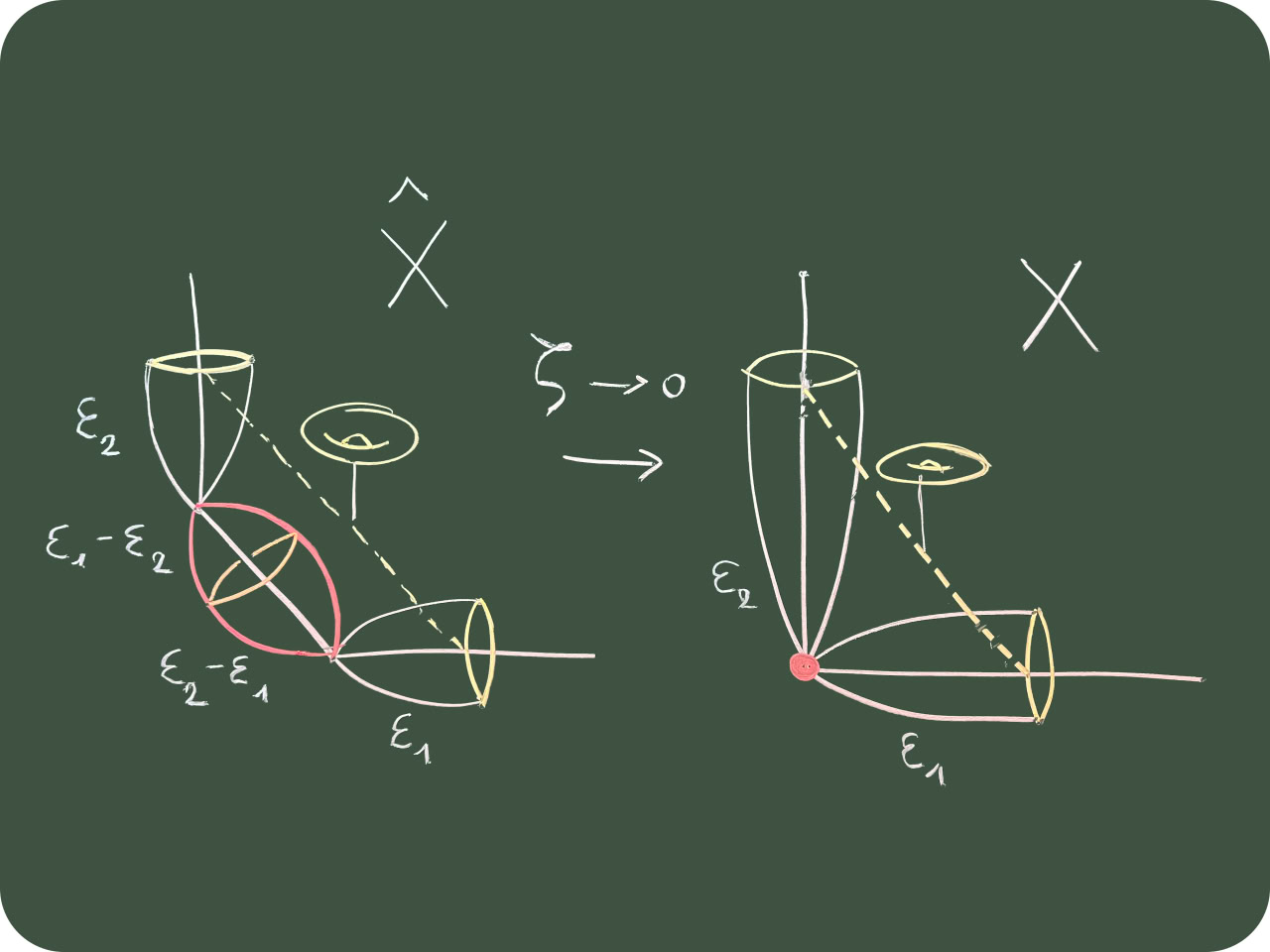}}
\caption[]{\label{fig:blowup} Toric diagram of the blow down of the exceptional divisor $E$ (in red) and state/operator correspondence.}
\end{figure}

The topology of the blow up $\hat X$ differs by that of $X$ by the addition of a non contractible 
two-sphere $E$, which is called exceptional divisor.
The BPS partition function of the ${\cal N}=2$ gauge theory can be formulated after the topological twist
\cite{Witten:1988ze}. 
As the stress-energy tensor of the twisted theory is exact under the scalar supersymmetry defining the BPS partition function, the latter does not depend on the K\"ahler parameter $\zeta$.
Therefore the partition functions on $\hat X$ and $X$ coincide\footnote{This should be carefully normalized and holds in the appropriate topological sector. See later for a more precise statement.}.  
Moreover, on $\hat X$ one can study BPS surface observables located on the exceptional divisor $E$.
Since the theory is topological, 
the  multiple insertions of the BPS surface observable in the theory on $\hat X$ can be traded for the insertions of the corresponding local operators on $X$ 
\cite{Moore:1997pc}. 
This realizes the topological version of state/operator correspondence. 
Indeed, the limit 
$\zeta\to0$ corresponds to elongating the location of the surface observable on the exceptional divisor
along an infinite cylinder and its insertion is therefore equivalent to that of a local operator located at the point at infinity. See figure \ref{fig:blowup}.
The equivalence of the two pictures implies the blow up equations of Fintushel and Stern 
\cite{fintushel1994blowup}.

When $X$ is a toric manifold and one considers a toric blowup, the set of toric divisors of $\hat X$ gets correspondingly augmented by the element corresponding to the exceptional divisor\footnote{See \cite{Hori:2003ic} for details.}.
${\cal N}=2$ gauge theories on $X$ and on $\hat X$ can be formulated equivariantly under the toric isometry. The exceptional divisor lifts to an equivariant one and therefore defines an extra set of equivariant 
surface observables for the gauge theory on $\hat X$.
Indeed state/operator correspondence lifts to the equivariant set up and the corresponding blow-up equations were studied by Nakajima and Yoshioka on $\hat{\mathbb{C}}^2$ in \cite{nakajima2003lectures}.

For the 
generating function of the $SU(2)$ surface observable $I(E)$ 
\begin{equation}\label{bl,0}
\hat Z(a,\vec{m},\Lambda,\epsilon_1,\epsilon_2,s) = \ev{e^{s I(E)}}_{\hat{\mathbb{C}}^2}\ ,
\end{equation}
they read\footnote{Here we consider 
the gauge theory on the blowup with
first Chern class $c_1=1$. This fixes the odd lattice in \eqref{bl,1}. The other sector $c_1=0$ produces the non-vanishing blowup equations \cite{nakajima2003lectures}.} 
\begin{equation}\label{bl,1}
\hat Z(a,\vec{m},\Lambda,\epsilon_1,\epsilon_2,s) = \sum_{n\in \mathbb{Z}+\frac{1}{2}} Z(a+n\epsilon_1,\vec{m},\Lambda_{\epsilon_1s},\epsilon_1,\epsilon_2-\epsilon_1)Z(a+n\epsilon_2,\vec{m},\Lambda_{\epsilon_2s},\epsilon_1-\epsilon_2,\epsilon_2)\ ,
\end{equation}
where $\hat Z|_{s=0}=0$ and we defined the shifted coupling
\begin{equation} 
\label{lambdaepsiloni}
\Lambda_{\epsilon_is} = \Lambda \exp(\epsilon_i s)\ .
\end{equation}
In \eqref{bl,1} $a$ is the $SU(2)$ Cartan parameter, 
$\vec{m}=\{m_i\}$ are the masses of the hypermultiplets, $\Lambda$ is the gauge coupling, and $\epsilon_1,\epsilon_2$ are the $\Omega$-background parameters.
The structure of the formula \eqref{bl,1} is simple: the two factors correspond to the contributions of the two patches of the  exceptional divisor $E$, and are convoluted through the sum over the magnetic fluxes supported on $E$. The effect of the exponentiated surface observable is to shift the couplings as in
\eqref{lambdaepsiloni}.
The non-equivariant limit of the above blowup equations was studied in detail in \cite{nakajima2003lectures}, showing that they indeed reduce to the equation of Fintushel and Stern. The blowup factor in the latter is given by the Weierstrass $\sigma$-function. This is naturally arising in the Seiberg-Witten 
description of the low-energy effective field theory in the Coulomb phase
and was indeed used in \cite{Nakajima:2003pg,nakajima2003lectures} to provide a proof of Nekrasov's conjecture about the recontruction of the Seiberg-Witten prepotential from supersymmetric localisation.

\subsection{The blowup factor in the NS limit}

In the following we will study the NS limit of the blowup factor in the $\Omega$-background parameters, namely
$(\epsilon_1,\epsilon_2)\to (\epsilon,0)$ and show that it is related to Painlev\'e $\Tau$-function expanded around its zero.
This will allow us to study the new expansions of Painlev\'e $\Tau$-functions which correspond to study the gauge theory around weakly coupled, monopole and Argyres-Douglas points, and their integrality and modularity properties.

The $SU(2)$ theory can be naturally associated to an integrable system, the Hitchin system, if we further compactify on a circle. In this way we obtain a $3d\ \mathcal{N}=4$ sigma model whose target space $\mathcal{M}$ is the moduli space of the Hitchin system \cite{Gaiotto:2009hg}. It turns out that this Hitchin system can be related to a Painlevé Lax pair and in this way we get a direct relation between gauge theory and Painlevé equations, the so called ``Painlevé-gauge theory correspondence'' \cite{Bonelli:2016qwg}.
This will serve us as a guide to identify the relevant Painlev\'e equation for a given gauge theory.

Let us define the blowup factor
\begin{equation}\label{aut,4}
\mathcal{B}(a,\vec{m},\Lambda,\epsilon_1,\epsilon_2,s) = \frac{ \hat Z(a,\vec{m},\Lambda,\epsilon_1,\epsilon_2,s)}{ Z(a,\vec{m},\Lambda,\epsilon_1,\epsilon_2)} \ ,
\end{equation}
as the normalised generating functional of the correlation functions of the surface observable $I(E)$.
The partition function $Z$ of gauge theory in the $\Omega$-background can be expanded as a refined topological string theory
\begin{equation}\label{aut,3}
\log Z(a,\vec{m},\Lambda,\epsilon_1,\epsilon_2) \sim \sum_{g=0}^{+\infty}\sum_{k\in \{\frac{1}{2}\}\cup\mathbb{Z}_{\geq0}} (-\epsilon_1\epsilon_2)^{g-1} (\epsilon_1+\epsilon_2)^{2k} \mathcal{F}_{g,k}(a,\vec{m},\Lambda) \ .
\end{equation}
The tree level term $\mathcal{F}_0 \equiv \mathcal{F}_{0,0}$ corresponds to the Seiberg-Witten prepotential of the gauge theory.
The only term with half-integer \(n\) is given explicitly by\footnote{This is true for $N_f\leq 3$. In the case of $N_f=4$ we have an extra contribution coming from the instanton corrections which is independent on $a$.}
\begin{equation}
\mathcal{F}_{0,\frac12}(a,\vec{m},\Lambda)=-i\pi a\ .
\end{equation}
Substituting \eqref{aut,3} in \eqref{aut,4} and taking the NS limit $\epsilon_2 \to 0$ with fixed $\epsilon_1\equiv \epsilon$ we obtain
\begin{align}\label{aut,5}
&\mathcal{B}_{NS}(a,\vec{m},\Lambda,\epsilon,s) \equiv 
\lim_{\epsilon_2 \to 0} \mathcal{B}(a,\vec{m},\Lambda,\epsilon,\epsilon_2,s)= \nonumber\\ 
&=  e^{\alpha-\frac{\uns s}{\epsilon}} \sum_{n\in \frac{1}{2}+\mathbb{Z}} e^{-\frac{n\rho}{\epsilon}} Z_{SD}(a+n \epsilon,\vec{m},\Lambda_{\epsilon s},\epsilon)= \nonumber\\
&= e^{\alpha-\frac{\uns s}{\epsilon}} Z_D(a,\vec{m},\rho,\Lambda_{\epsilon s},\epsilon) \ ,
\end{align}
 where $Z_{SD}$ is the Nekrasov partition function in the self-dual background $\epsilon_1=-\epsilon_2=\epsilon$ and we have defined
\begin{equation}\label{aut,5.1}
\rho = \epsilon\pdv{a}W(a,\Lambda,\epsilon)\ ,\quad  \alpha = \pdv{\epsilon}W(a,\Lambda,\epsilon) \ , \quad \uns= \epsilon\Lambda\pdv{\Lambda}W(a,\Lambda,\epsilon) \ ,
\end{equation}
and $W$ is the twisted superpotential
\begin{equation}\label{aut,5.2}
W(a,\Lambda,\epsilon)=\sum_{k\in \{\frac{1}{2}\}\cup\mathbb{Z}_{\geq0}} \mathcal{F}_{0,k}(a,\Lambda) \epsilon^{2k-1} \ .
\end{equation}
We recall that the identification between the dual partition function in \eqref{aut,5} and the Painlev\'e $\Tau$-function was found in \cite{Gamayun:2013auu} - known as {\it Kyiv formula}. Their $\Tau$-function is expanded around a generic initial datum for the Painlev\'e equation.
The blowup formulae we are considering here instead lead to an expansion around a {\it zero} of the $\Tau$-function. This imposes a reparametrization of the time variable, that is of the gauge coupling, as the one determined by the insertion of the surface observable \eqref{lambdaepsiloni},
$s$ being the Painlev\'e time.

From formula \eqref{aut,5} we see that in the limit $\epsilon_2 \to 0 $ the blowup factor reduces simply to the dual partition function, up to an exponential prefactor $\exp(\alpha-\uns s/\epsilon )$. The factor $\exp(\alpha)$ fixes the normalization of the Painlev\'e $\Tau$-function. We finally obtain
\begin{equation}\label{aut,7}
\Tau= e^{\frac{\uns s}{\epsilon}}\mathcal{B}_{NS}\ .
\end{equation}

\section{Chiral ring expansion of the blowup factor}\label{section3}
\subsection{Autonomous limit and Weierstrass $\sigma$-function}

In this subsection we preliminarly discuss the limit $\epsilon \to 0$ of the the NS blowup factor $\mathcal{B}_{NS}$. 
This is the SW limit, already discussed in \cite{nakajima2003lectures}, and corresponds 
to the autonomous limit of the Painlev\'e dynamics. 
It turns out that in this limit the Painlev\'e $\Tau$-function universally 
reduces to the Weierstrass $\sigma$-function
and correspondingly the blow up factor reduces to the one of Fintushel and Stern, that is the sigma function itself - up to a Gaussian prefactor - 
computed in the modular invariants parametrizing the SW curve of the specific gauge theory. 

Let us underline that here one does not have to assume the theory to be Lagrangian 
and therefore the analysis holds also at the Argyres-Douglas points in the moduli space.

In the following, for simplicity, we omit the explicit dependence on the masses $\vec{m}$
$ \mathcal{F}_0(a,\Lambda) \equiv \mathcal{F}_0(a,\vec{m},\Lambda)$, as it 
plays no role in the derivation.

We compute now the SW limit $\epsilon \to 0$. We start from the expression of the NS blowup factor in \eqref{aut,5} 
\begin{align}
\mathcal{B}_{NS}(a,\Lambda,\epsilon,s)=e^{\alpha-\frac{\uns s}{\epsilon}} \sum_{n\in \mathbb{Z}+\frac{1}{2}} e^{-\frac{n\rho}{\epsilon}} Z_{SD}(a+n \epsilon,\Lambda_{\epsilon s},\epsilon) \ .
\end{align}
with $\rho=\rho(a,\Lambda,\epsilon)$ given by \eqref{aut,5.1}. Using the genus expansion \eqref{aut,3} we obtain
\begin{equation}
\mathcal{B}_{NS}(a,\Lambda,\epsilon,s)=e^{\alpha-\frac{\uns s}{\epsilon}}\sum_{n\in\mathbb{Z}+\frac{1}{2}}  \exp\left(-\frac{n}{\epsilon}\rho+\frac{1}{\epsilon^2}\mathcal{F}_0(a+n\epsilon,\Lambda_{\epsilon s})+\mathcal{F}_{1,0}(a+n\epsilon, \Lambda_{\epsilon s})+O(\epsilon)\right) \ .
\end{equation}
Expanding in $\epsilon$ the $a$-dependence we get
\begin{align}
&\mathcal{B}_{NS}(a,\Lambda,\epsilon,s) =  \\
&=e^{\alpha-\frac{\uns s}{\epsilon}}\sum_{n\in\mathbb{Z}+\frac{1}{2}}  \exp\left[\frac{1}{\epsilon^2}\mathcal{F}_0(a,\Lambda_{\epsilon s})-\frac{n}{\epsilon}\left(\rho-\pdv{\mathcal{F}_0}{a}\left(a,\Lambda_{\epsilon s}\right)\right)+\frac{1}{2}n^2\pdv[2]{\mathcal{F}_0}{a}\left(a,\Lambda_{\epsilon s}\right)+\mathcal{F}_{1,0}\left(a, \Lambda_{\epsilon s}\right)+O(\epsilon)\right] \ .
\end{align}
Expanding further in the instanton counting scale $\Lambda_{\epsilon s} = \Lambda+\epsilon\Lambda s +O(\epsilon^2)$ we obtain
\begin{align}
&\mathcal{B}_{NS}(a,\Lambda,\epsilon,s) =  \nonumber\\
&=  \exp\left[\alpha+\frac{1}{\epsilon^2}\mathcal{F}_0(a,\Lambda)+\mathcal{F}_{1,0}(a, \Lambda)\right]\times \\
&\times \exp\left[\frac{1}{\epsilon}\left(u\left(a,\Lambda\right)-\uns \right)s+\frac{1}{2}u\left(a,\Lambda\right)s^2+\frac{1}{2}\Lambda^2\pdv[2]{\mathcal{F}_0}{\Lambda}\left(a,\Lambda\right)s^2\right]\times \nonumber\\ 
&\times\sum_{n\in\mathbb{Z}+\frac{1}{2}} \exp\left[-\frac{n}{\epsilon}\left(\rho-\pdv{\mathcal{F}_0}{a}\left(a,\Lambda\right)\right)+n\pdv{u}{a}\left(a,\Lambda\right)s+\frac{1}{2}n^2\pdv[2]{\mathcal{F}_0}{a}\left(a,\Lambda\right)+O(\epsilon)\right] \ , \label{4.3}
\end{align}
where we defined $u(a,\Lambda) = \Lambda \partial_\Lambda\mathcal{F}_0(a,\Lambda)$. Finally, we expand in $\epsilon$ all the NS quantities
\begin{equation}
\rho=\pdv{\mathcal{F}_0}{a}+\epsilon \partial_a\mathcal{F}_{0,\frac{1}{2}}+O(\epsilon^2) \ ,\ \alpha=-\frac{1}{\epsilon^2}\mathcal{F}_0+\mathcal{F}_{0,1}+O(\epsilon) \ ,\ \uns=u+\epsilon \Lambda \partial_\Lambda\mathcal{F}_{0,\frac{1}{2}}+O(\epsilon^2) \ ,
\end{equation}
and using the following relations\footnote{We follow the convention of \cite{nakajima2003lectures} where a factor $\pm\frac{1}{2\pi i}$ is included in the definition of the observables of the gauge theory to get the correct normalization for the sigma function.}
\begin{align}
&\tau\equiv\tau(a,\Lambda) =\frac{1}{2\pi i}\pdv[2]{\mathcal{F}_0}{a}\left(a,\Lambda\right) \ , \\ 
& \frac{1}{2\omega_1} = \frac{1}{2\pi i}\pdv{u}{a}\left(a,\Lambda\right) =\frac{1}{2\pi i}\Lambda \pdv[2]{\mathcal{F}_0}{\Lambda}{a}\left(a,\Lambda\right) \ , \\
&\partial_a\mathcal{F}_{0,\frac{1}{2}}\left(a,\Lambda\right)=-i\pi \ , \quad \Lambda \partial_\Lambda\mathcal{F}_{0,\frac{1}{2}}=0 \ , \quad \exp(\mathcal{F}_{1,0}+\mathcal{F}_{0,1})=\frac{2\omega_1\Lambda }{\theta_1'(0)},
\end{align}
where $\tau$ is the complexified gauge coupling of the IR theory and $\omega_1\equiv\omega_1(a,\Lambda)$ is the lattice half-period associated to $a$ of the Seiberg-Witten curve, we obtain
\begin{align}
\mathcal{B}_{NS}(a,\Lambda,\epsilon,s)=&\frac{2\omega_1\Lambda}{\theta_1'(0)}\exp\left[\frac{1}{2}us^2+\frac{1}{2}\Lambda^2\pdv[2]{\mathcal{F}_0}{\Lambda}\left(a,\Lambda\right)s^2\right]\times\nonumber\\
\times&
\sum_{n\in\mathbb{Z}+\frac{1}{2}} (-1)^n\exp\left[2in\left(\frac{\pi s}{2\omega_1}\right) +\pi i\tau n^2+O(\epsilon)\right]  \ .
\end{align}
Using the relations \eqref{2.1} and \eqref{2.2} we finally get
\begin{equation}\label{aut,8}
\mathcal{B}_{SW}(a,\Lambda,s)=
 \lim_{\epsilon \to 0}\mathcal{B}_{NS}(a,\Lambda,\epsilon,s)=\Lambda e^{-\frac{1}{2}T s^2} \sigma(s;\omega_ 1,\omega_2)\ .
\end{equation}
Here we have defined $\omega_2 = \omega_1\tau$ and the contact term
\begin{equation}
T=-\Lambda \pdv{u}{\Lambda}+\frac{\pi^2}{12}\frac{E_2(\tau)}{\omega_1^2}\ ,
\end{equation}
which can be computed from SW theory and depends on the specific theory, see appendix~\ref{appB}. Notice that in \cite{Moore:1997pc} the contact term is given by
\begin{equation}\label{aut,T}
\tilde T=-\Lambda \partial_\Lambda u=T-\frac{\pi^2}{12}\frac{E_2(\tau)}{\omega_1^2}\ .    
\end{equation}
We isolated the contribution coming from $E_2(\tau)$ to express the result in terms of the Weierstrass $\sigma$ function and $T$ is then modular invariant. 
The previous result shows that the blowup factor is ``universal'' in the sense that, up to contact terms, it always reduces to the Weierstrass $\sigma$-function in the SW limit. The information about the gauge theory is all encoded in the parametrization of the half-periods $\omega_j(a, \vec{m},\Lambda), j=1,2$, or equivalently of the elliptic invariants $g_2(u,\vec{m},\Lambda), g_3(u, \vec{m},\Lambda)$ of the SW curve, where we have reintroduced the dependence on the masses $\vec{m}$ of the hypermultiplets.

\subsection{Blowup equations and expansion in the chiral ring}
In the previous section we verified that in the autonomous limit the $\Tau$-function given by the Kyiv formula reduces to the Weierstrass $\sigma$-function up to Gaussian prefactors.
It is interesting to observe that $\sigma(s,g_2,g_3)$ admits the following 
Hurwitz integral 
expansion in terms of $s$, see for example \cite{onishi2010universalellipticfunctions}
\begin{equation}\label{blowup,-1}
\sigma(s;g_2,g_3) = \sum_{m,n=0}^{\infty} a_{mn} \left(\frac{g_2}{2}\right)^m\left(2g_3\right)^n \frac{s^{4m+6n+1}}{(4m+6n+1)!} \ ,
\end{equation}
where $a_{mn}$ are integer coefficients (see appendix \ref{appA}). 

Since $\sigma$ gives the SW limit of $\mathcal{B}_{NS}$, it is natural to ask if a similar expansion holds also in the presence of $\Omega$-background. 

This is indeed the case: in this section we propose that the chiral ring has the natural basis to be exploited, and in the next sections we provide evidence of the existence of a Hurwitz integral expansion for the NS
blow-up factor $\mathcal{B}_{NS}$.
An expansion similar to \eqref{blowup,-1} can  be 
derived from the NY blowup equations. To this end we rewrite the partition function on the blowup as 
\begin{equation}\label{NY,0}
\hat Z(a,\Lambda,\vec{m},\epsilon_1,\epsilon_2,s) = D_{NY}(s) Z(a,\Lambda,\vec{m},\epsilon_1,\epsilon_2)
\end{equation} 
which allows to compute it by acting on a single copy of the Nekrasov partition function on $\mathbb{C}^2$ with a suitable linear differential operator $D_{NY}(s)$. The structure of $D_{NY}(s)$ has been derived from the AGT correspondence using a CFT analysis in \cite{bershtein2013coupling}\footnote{It would be interesting to derive it also from a pure gauge theory treatment, for example by analysing an equivariant version of the $u$-plane integral. The expression \eqref{NY,0} is valid for $SU(2)$ gauge theories. By operator/state correspondence, an higher rank extension of this should exist. A possible mathematical setup for this generalization could be the wall crossing approach to the blow-up of
\cite{Nakajima_2011}.}.
\begin{equation}
D_{NY}(s)=\sum_{n=0}^{\infty}D_n(-\epsilon_1\epsilon_2\Lambda\pdv{\Lambda} ,\Lambda,\vec{m},\epsilon_1,\epsilon_2)\frac{s^{n+1}}{(n+1)!} \ ,
\end{equation}
and the operator vanishes for $s=0$. The $c_n$ are normal-ordered polynomials in which all the derivatives $\Lambda\partial_{\Lambda}$ are moved to the right. In the NS limit $\epsilon_2\to 0, \epsilon_1=\epsilon$ we observe that using \eqref{aut,3} and \eqref{aut,5.1} we get
\begin{equation}
\lim_{\epsilon_2\to 0} \frac{1}{Z} \left(-\epsilon\epsilon_2\Lambda\pdv{\Lambda}\right)^k Z= 
\uns^k
\ ,
 \end{equation}
correspondingly, the blowup factor in the NS limit has the following structure
\begin{equation}\label{NY,1}
\mathcal{B}_{NS}(s)=\lim_{\epsilon_2\to 0}\frac{1}{Z}D_{NY}(s)Z=b_0\sum_{n=0}^{\infty}c_n(\uns,\Lambda,\vec{m},\epsilon)\frac{s^{n+1}}{(n+1)!} \ ,
\end{equation}
where $b_0=\langle I(E)\rangle$ is the one-point function of the surface observable and $c_0=1$.
Such limit of the blowup equations was studied before in cases with surface defect in~\cite{Nekrasov:2020qcq,Jeong:2020uxz}, and also in cases without surface defect, which we are studying here, in~\cite{Grassi:2016nnt,Gavrylenko:2020gjb,Bershtein:2021uts,Lisovyy:2021bkm,Lisovyy:2022flm}.

The physical interpretation of the above expansion is given by the operator/state correspondence. The blowup factor is an observable in the topologically twisted gauge theory. The blowup is a local change of the topology of $X$, therefore in the limit of very high distances, that is in the low-energy theory, it must be possible to reproduce the effect of the blowup $\hat X$ as a local operator of the theory on $X$. As the local observables of the equivariant chiral ring are generated by $\uns$, we get correspondingly a series in this variable
resulting from the OPE of the local equivariant observables. 
The coefficients of this OPE are precisely the polynomials $c_n$.
Let us underline that the above analysis is valid independently whether the theory is Lagrangian or not. 

From the previous analysis we obtain the following dictionary
\begin{center}
\begin{tabular}{|c|c|}
\hline
\bf{Gauge theory}                                          &  \bf{Painlevé equation} \\
\hline\hline
blowup factor $\mathcal{B}_{NS}$                           & $\Tau$-function \\
\hline
gauge coupling $\Lambda$                                   & position of the zero of $\Tau$    \\
\hline
surface observable source $s$                              & Painlev\'e time \\
\hline
chiral ring expansion of $\mathcal{B}_{NS}$                & expansion around a zero of $\Tau$ \\
\hline
SW theory                                                  & autonomous limit\\
\hline
\end{tabular}
\end{center}

In section \ref{section5} we will apply this map to compute the coefficients of  the expansion \eqref{NY,1} from Painlevé equations.

Before ending this section we observe that in the SW limit $\epsilon\to 0$, the coefficients $c_n$ are given precisely by the expansion \eqref{blowup,-1} of the Weierstrass $\sigma$-function. Therefore they are modular polynomials  with integer coefficients in the elliptic invariants $g_2,g_3$ of the SW curve.   Modularity is a consequence of the electromagnetic duality of the low-energy theory.
A thorough analysis of the modular properties of the Painlev\'e $\Tau$-function and its relation with the holomorphic anomaly equations will be performed in the next sections.

\section{Modular properties of the $\Tau$-function} \label{section4}

The Nekrasov partition function $Z_{SD}(a,\Lambda,\epsilon)$ in the self-dual $\Omega$-background corresponds to the holomorphic limit $\bar a \to \infty$ of the partition function of topological strings $\mathcal{Z}_X(a,\bar a,g_s,\Lambda)$ in the local Calabi-Yau (CY) $X$ which geometrically engineers the corresponding gauge theory. Namely, denoting by $a,\bar a$ the moduli of $X$ and by $g_s=\epsilon$, we have  
\begin{equation}
Z_{SD}(a,\Lambda,\epsilon)=\lim_{\bar a \to \infty} \mathcal{Z}_X(a,\bar a, \Lambda,g_s)\ .
\end{equation}
The $\bar a$ dependence of $\mathcal{Z}_X$ is controlled by the BCOV holomorphic anomaly equation \cite{Bershadsky:1993cx}. This takes a particularly simple form for $SU(2)$ gauge theories where all the non-holomorphic dependence enters only through the non-holomorphic extension of the second Eisenstein series \cite{Grimm_2007}
\begin{equation}
\hat E_2(\tau,\bar \tau)=E_2(\tau)-\frac{3}{\pi\Im \tau}\ ,
\end{equation}
where $\tau$ is the IR gauge coupling. This is modular thanks to the non-holomorphic contribution coming from $\Im \tau$. In the holomorphic limit  $\Im \tau \to \infty$ $\hat E_2(\tau,\bar \tau)$ reduces to $E_2(\tau)$ which is indeed holomorphic but not modular. Therefore, the holomorphic limit 
of $\mathcal{Z}_X$ is not modular, the lack of modularity being encoded in its $E_2$ dependence.

Let us recall the relation \eqref{aut,7} between the NS blowup factor $\mathcal{B}_{NS}$ and the Painlevé $\Tau$-function
\begin{equation}
\mathcal{B}_{NS}(a,\Lambda,\epsilon,s)=e^{-\frac{\uns s}{\epsilon}}\Tau(a,\Lambda,\epsilon,s)\ ,
\end{equation}
where
\begin{equation}\label{holNS,0}
\Tau(a,\Lambda,\epsilon,s)= e^{\pdv{W}{\epsilon}}\sum_{n\in \mathbb{Z}+\frac{1}{2}} e^{-\frac{n\rho}{\epsilon}} Z_{SD}(a+n \epsilon,\Lambda_{\epsilon s},\epsilon)\ ,
\end{equation}
$W$ is the twisted superpotential \eqref{aut,5.2}, and $\rho=\epsilon\partial_a W$ is the dual quantum period.

In the previous section we showed that, thanks to the topological operator/state correspondence, the NS blowup factor $\mathcal{B}_{NS}$ can be interpreted as a local observable of the equivariant chiral ring, and therefore is generated by $\uns $. This is a direct consequence of the blowup equations in presence of the surface observable $I(E)$ \eqref{NY,0} on the exceptional divisor $E$. Since $\mathcal{B}_{NS}\leftrightarrow \Tau$ is a function of $\uns $ only, all the $E_2$ dependence is carried by $\uns $ and there is no further explicit $E_2$ dependence. Therefore the $E_2$ derivative of the $\Tau$ function is
\begin{equation}
D_{E_2}\Tau=\pdv{\Tau}{\uns }D_{E_2} \uns =\partial_a\Tau\frac{D_{E_2} \uns }{\partial_a \uns }\ ,
\end{equation}
that is
\begin{equation}\label{holNS,1}
D_{E_2}\Tau\partial_a \uns -\partial_a\Tau D_{E_2} \uns =0\ .
\end{equation}

We will show that the condition \eqref{holNS,1} implies the holomorphic anomaly equations both for the SD and NS partition functions. The derivation holds for general $SU(2)$ gauge theories, but we omit the explicit dependence on the masses because it is not relevant for the analysis.

The reason why we can obtain both the equations, starting from the single condition \eqref{holNS,1}, is due to the insertion of the surface observable $I(E)$ which shifts the coupling $\Lambda \to \Lambda_{\epsilon s}$ of the SD contribution in \eqref{holNS,0}. This effectively decouples the scales of the SD and NS contributions and allows to study separately their $E_2$ dependence. 

For this purpose it is convenient to write the $\Tau$-function as a series in the Weierstrass $\sigma$-function and its derivatives. To do this we define
\begin{equation}\label{hol,1}
\mathcal{F}_{\mathrm{st}}^{x,s}(a, \Lambda,\epsilon) = \mathcal{F}(a+\epsilon x, \Lambda_{\epsilon s},\epsilon)-
\frac1{\epsilon^2} \mathcal{F}_0(a, \Lambda)
-\frac{x}{\epsilon} \partial_a\mathcal{F}_0(a, \Lambda)
-\frac{x^2}{2} \partial_a^2\mathcal{F}_0(a, \Lambda)\ ,
\end{equation}
where the subscript ``st'' stands for ``stable'' and
\begin{equation}
\mathcal{F}(a+\epsilon x, \Lambda_{\epsilon s},\epsilon)=\log Z_{SD}(a+\epsilon x, \Lambda_{\epsilon s},\epsilon)\ . 
\end{equation}
The prepotential $\mathcal{F}_{\mathrm{st}}^{x,s}$ can be interpreted as the BCOV generating functional for the correlation functions of the chiral ring of $X$. The correlation functions are obtained taking derivatives with respect to $x,s$ and setting $x=0,s=0$ at the end. Therefore $\mathcal{F}_{\mathrm{st}}^{x,s}$ encodes the full physical information about the perturbative topological string theory.

\subsection{Algebra of operators on the modular ring}

In order to study \(E_2\)-dependence of the tau function we first study the action of \(\partial_{\tau}\) derivative on the space of modular functions with respect to some subgroup $\Gamma$ of \(SL(2,\mathbb{Z})\) (e.g. the duality group of $SU(2)$ Seiberg-Witten theory).
It is easy to see~\cite{Grimm_2007} that if \(f_k\) is a modular form of weight \(k\), then
\begin{equation}
D_{\tau}f_k = \frac{\partial_{\tau}}{2\pi i} f_k-\frac{k}{12}E_2 f_k\ ,
\end{equation}
is again modular. We also have the relation
\begin{equation}
\frac{\partial_{\tau}}{2\pi i} E_2 = \frac1{12}E_2^2 - \frac1{12}E_4\ .
\end{equation}
We now introduce the weight operator \(\widehat{d}\)
\begin{equation}
\widehat{d} f_k=k f_k\ ,\qquad \widehat{d} E_2 = 2 E_2\ , \qquad \widehat{d}\tau = 0\ , \qquad \widehat{d}(f g) = \widehat{d}(f) g + f \widehat{d}(g)\ ,
\end{equation}
and the derivation with respect to \(E_2\)
\begin{equation}
D_{E_2}F(E_2,\tau,g_1,\ldots g_n) = \left.\frac{\partial F(E_2,\tau,g_1,\ldots g_n)}{\partial E_2}\right|_{\tau, g_1,\ldots, g_n}\ ,
\end{equation}
where \(g_i\) are the generators of the ring of holomorphic modular forms with respect to $\Gamma$.
One can check by explicit computation that
\begin{equation}
[D_{E_2},\frac{\partial_{\tau}}{2\pi i}] = \frac1{12}\widehat{d}\ .
\end{equation}
We emphasize that \(D_{E_2}\) and $\hat d$ are defined only on the ring generated by \(\tau\), \(E_2\), and modular forms, but not on the space of arbitrary functions of \(\tau\).
We also introduce explicit \(\tau\) derivative \(\partial'_{\tau}\) by
\begin{equation}
\partial_{\tau}'F(E_2,\tau,g_1,\ldots g_n) = \left.\frac{\partial F(E_2,\tau,g_1,\ldots g_n)}{\partial \tau}\right|_{E_2, g_1,\ldots, g_n}\ .
\end{equation}
As
\begin{equation}
\tau = \frac1{2\pi i} \frac{\partial^2 }{\partial a^2}\mathcal{F}(a,\Lambda,\vec{m})\ ,
\end{equation}
we can express the \(a\), \(\Lambda\), and \(m_i\) derivatives by
\begin{eqnarray}
\label{eq:OpDef}
\partial_a = \frac{\partial^3 \mathcal{F}_0}{\partial a^3}\frac{\partial_{\tau}}{2\pi i}\ ,
\quad \Lambda\partial_\Lambda = \Lambda\frac{\partial^3 \mathcal{F}_0}{\partial\Lambda\partial a^2}\frac{\partial_{\tau}}{2\pi i} +\Lambda\partial_{\Lambda}' \ , \quad
\partial_{m_i} = \frac{\partial^3 \mathcal{F}_0}{\partial m_i\partial a^2}\frac{\partial_{\tau}}{2\pi i} +\partial_{m_i}'\ .
\end{eqnarray}
In general, \(\partial'\) means the derivative with all other variables kept fixed. We notice that \(\partial_a^3\mathcal{F}_0\) is modular of weight \(-3\), since
\begin{equation}
\partial_{a_D}\frac{-1}{\tau}=\frac1{\tau^2} \frac{\partial a}{\partial a_D} \partial_a\tau = \frac1{\tau^3}\partial_a \tau\ .
\end{equation}
Analogously, since $u=\Lambda\partial_\Lambda {\mathcal F}_0$ is modular invariant, by using \eqref{eq:OpDef} one gets that \(\pi i \omega_1^{-1}=\Lambda \partial_{\Lambda}\partial_a \mathcal{F}_0\) is modular of weight \(-1\). The canonically normalized holomorphic differential is given by 
\begin{equation}
d\omega = \partial_a dS^{SW}=\Lambda\partial_{\Lambda}\partial_a \mathcal{F}_0 \frac{\partial }{\partial u}dS^{SW}\ ,
\end{equation}
where $dS^{SW}$ is the SW differential. By using the above and \eqref{aut,T} one gets
\begin{equation}
D_{E_2}\Lambda \partial_{\Lambda}\partial_a^2 \mathcal{F}_0 = -\frac1{12} \Lambda \partial_{\Lambda}\partial_a\mathcal{F}_0\cdot \partial_a^3 \mathcal{F}_0\ .
\end{equation}
Using this one can check that the above operators satisfy the following algebra
\begin{equation}\begin{gathered}
\label{eq:OpAlgebra}
[\partial_a, \Lambda \partial_{\Lambda}] = 0\ ,\\
[D_{E_2}, \partial_a] =  \frac1{12} \partial_a^3\mathcal{F}_{0}(a,\Lambda) \widehat{d}\ ,\\
[D_{E_2}, \Lambda\partial_\Lambda] =
\frac1{12} \Lambda \partial_\Lambda\partial_a^2\mathcal{F}_{0}(a,\Lambda) \widehat{d}
- \frac1{12} \Lambda \partial_\Lambda\partial_a\mathcal{F}_{0}(a,\Lambda) \partial_a\ ,\\
[\widehat{d}, \Lambda \partial_{\Lambda}] = -\Lambda\partial_{\Lambda}\partial_a^2\mathcal{F}_0(a,\Lambda) \frac{\partial'_{\tau}}{\pi i}\ ,\qquad
[\widehat{d}, \partial_a] = -\partial_a - \partial_a^3\mathcal{F}_0(a,\Lambda) \frac{\partial'_{\tau}}{\pi i}\ ,\qquad
[\widehat{d}, D_{E_2}] = -2 D_{E_2}\ .
\end{gathered}\end{equation}
In order to be able to compute the action of this algebra on the \(\mathcal{F}_{g,n}\)'s  we
display the modular properties of the first terms of the prepotential
\begin{equation}\begin{gathered}
\label{eq:3main}
D_{E_2}\Lambda\partial_\Lambda\mathcal{F}_0(a,\Lambda) = 0\ ,\quad
D_{E_2}\partial_a^3\mathcal{F}_0(a,\Lambda)= 0\ ,\\
\widehat{d}\Lambda\partial_{\Lambda}\mathcal{F}_0=0\ , \quad 
\widehat{d}\partial_a^3\mathcal{F}_0=-3 \partial_a^3\mathcal{F}_0\ , \quad
D_{E_2}\mathcal{F}_{1,0}(a,\Lambda)=0\ ,\quad
D_{E_2}\mathcal{F}_{0,1}(a,\Lambda)=0\ ,
\end{gathered}\end{equation}
while the modular weights of the $\mathcal{F}_{g,n}$ are fixed by
\begin{equation}
\widehat{d} \mathcal{F}_{g,n}(a,\Lambda) = -\frac12\delta_{g,1}\delta_{n,0}\ ,\ g+n\geq 1\ .    
\end{equation}
Now we study the genus expansion of the prepotential in more detail.
For 
$g+n>1$ we can write \cite{Grimm_2007}
\begin{equation}\label{fgn}
\mathcal{F}_{g,n}=\left(\Delta(\tau)\right)^{2-2g-2n}\sum_{\substack{l+k\le 3g+3n-3\\k\le 3g+2n-3}}P_{l,k}(\Lambda,\vec{m})u^lX^k\ ,
\end{equation}
where $\Delta(\tau)=g_2^3-27g_3^2$ is the discriminant of the SW curve, $P_{l,k}$ are polynomials of the appropriate scaling dimensions, and
\begin{equation}
X= \frac{\pi^2}{64}\frac{E_2(\tau)}{\Lambda^2\omega_1^2}\ .
\end{equation}
For a gauge theory $\tt T$ we will express our solution \eqref{fgn} in the following ring $R_{\tt T}$ of polynomials
\begin{equation}\label{ring}
R_{\tt T}=\mathbb{C}[X, \{SW\}_{\tt T}, \Delta_{\tt T}^{-1},\omega_1^{-1}][\omega_1,\tau]\ ,
\end{equation}
where $\{SW\}_{\tt T}$ is the set of coefficients of the SW curve and the generator $\omega_1^{-1}$ guarantees that $R_{\tt T}$ is closed under the action of derivatives.
$\mathcal{F}_{g,n}$'s with $g+n>1$ belong to the first part of \eqref{ring}, the only terms that involve the extension by \(\omega_1, \tau\) are $\mathcal{F}_0$ and $\mathcal{F}_{0,\frac12}$.
The terms $\mathcal{F}_{1,0}$ and $\mathcal{F}_{0,1}$ do not lie in the ring, but $e^{12\mathcal{F}_{1,0}}$ and $e^{24\mathcal{F}_{0,1}}$ do (see the example below).

For the pure theory \(N_f=0\) this ring can be described by\footnote{See \eqref{tc} for definiteness.}
\begin{equation}
\label{eq:UXbasis}
R_{N_f=0}=\mathbb{C}[X,U,\Lambda,\Delta^{-1},(\theta_2\theta_3)^{-1}][\theta_2\theta_3,\tau]\ ,
\end{equation}
with
\begin{equation}
\qquad U=-\frac12 \left( \frac{\theta_4^2}{\theta_2^2}+\frac{\theta_2^2}{\theta_4^2}\right)=\frac{u}{8\Lambda^2}\ , 
\quad X= \frac{E_2}{\theta_2^2\theta_3^2}\ , \quad \Delta=\Lambda^{12}(U^2-1)\ ,\quad\omega_1=\frac{\pi}{8\Lambda}\theta_2\theta_3\ .
\end{equation}
Explicit formulas for the first \(\mathcal{F}_{g,n}\)'s in terms of these functions for the pure $SU(2)$ gauge theory are:
\begin{eqnarray}
&\mathcal{F}_0&=\frac{4}{3}\Lambda^2(5U-2X)-\frac 49 \pi i \tau \theta_2^2\theta_3^2\Lambda^2(X-U)^2\ ,\\
&\mathcal{F}_{0,\frac12}&=-i\pi a=-\frac 23 \pi (U-X)\theta_2\theta_3 \Lambda\ ,\\
&\mathcal{F}_{1,0}&=-\frac1{12} \log(U^2-1) - \frac12 \log \left( \theta_2\theta_3 \right) + \frac1{6} \log \Lambda\ ,\\
&\mathcal{F}_{0,1}&=-\frac1{24} \log(U^2-1) - \frac1{6} \log \Lambda\ ,
\end{eqnarray}
while, for the elements of the ring \eqref{ring} we have
\begin{eqnarray}
&\mathcal{F}_{0,2}&=\frac{U (75 + 4 U^2 + 5 U X)}{17280 (U^2-1)^2 \Lambda^2}\ ,\\
&\ldots &  \ldots \nonumber \\
&\mathcal{F}_{g,n}&=\left(\Lambda(U^2-1)\right)^{2-2g-2n}\sum_{\substack{m+k\le 3g+3n-3\\k\le 3g+2n-3\\2\nmid g+n+k+m}}c_{k,m}X^kU^m\ ,
\end{eqnarray}
and according to \eqref{fgn} all other \(\mathcal{F}_{g,n}\)'s are rational functions of \(U\) and  \(X\) multiplied by the appropriate power of \(\Lambda\).
The explicit action of the derivatives in \(a\) and \(\Lambda\)  \eqref{eq:OpDef} on the ring \(R_{N_f=0}\) is given by
\begin{eqnarray}
\partial_a &=& \frac{\partial^3 \mathcal{F}_0}{\partial a^3}\frac{\partial_{\tau}}{2\pi i} =
\frac{2i}{\Lambda \theta_2^3\theta_3^3(U^2-1)}\frac{\partial_{\tau}}{2\pi i}\ ,\\
\Lambda\partial_\Lambda &=& \Lambda\frac{\partial^3 \mathcal{F}_0}{\partial\Lambda\partial a^2}\frac{\partial_{\tau}}{2\pi i} +\Lambda\partial_{\Lambda}' = \frac{4 (X-U)}{3 (U^2-1) \theta_2^2\theta_3^2} \frac{\partial_{\tau}}{2\pi i} + \Lambda \partial_{\Lambda}'\ ,
\end{eqnarray}
where \(\partial_{\Lambda}\) is the derivative in \(\Lambda\) with fixed \(a\), while \(\partial_{\Lambda}'\) is the derivative in $\Lambda$ with fixed \(\theta_2\), \(\theta_3\), \(\tau\), \(U\), \(X\).
The modular weight operator is given explicitly by
\begin{equation}
\widehat{d}=\frac12 \left( \theta_3\partial_{\theta_3}'+\theta_2\partial_{\theta_2}' \right)\ ,
\end{equation}
and the 
derivative with respect to \(E_2\) has the following expression
\begin{equation}
D_{E_2}=\frac{1}{\theta_2^2\theta_3^2}\partial_X' \ .
\end{equation}
The explicit expression of the \(\partial_{\tau}\) operator acting on \(R_{N_f=0}\) \eqref{eq:UXbasis} is given by 
\begin{multline}
\frac1{\theta_2^2 \theta_3^2}\frac{\partial_{\tau}}{2\pi i}=
\frac1{12}\left(3 - 4 U^2 + 2 U X - X^2\right)\partial_X' +
\frac12 (U^2-1)\partial_U' \\+
\frac1{24}(X - U)(\theta_2\partial_{\theta_2}'+\theta_3\partial_{\theta_3}') +
\frac1{8}\sqrt{U^2-1}(\theta_3\partial_{\theta_3}'-\theta_2\partial_{\theta_2}')+
\frac1{\theta_2^2\theta_3^2}\frac{\partial'_{\tau}}{2\pi i}\ .
\end{multline}
One can check that the operators \(\partial_a\), \(\Lambda \partial_{\Lambda}\), \(D_{E_2}\), and \(\widehat{d}\) acting as derivations on \(R_{N_f=0}\) actually satisfy the algebra~\eqref{eq:OpAlgebra}.
Let us also display the explicit expression of the first few derivatives of the prepotential:
\begin{eqnarray}
\partial_a\mathcal{F}_0 &= & a_D = \frac{8i\Lambda}{\theta_2\theta_3} +\frac{4}{3}\pi \tau \Lambda \theta_2\theta_3(U-X)\ ,\\
\partial_a^2\mathcal{F}_0 & = & 2\pi i \tau\ ,\\
\partial_a^3 \mathcal{F}_0 & = &\frac{2i}{\Lambda \theta_2^3\theta_3^3(U^2-1)}\ ,\\
\Lambda \partial_{\Lambda} \mathcal{F}_0&=& 8U\Lambda^2\ ,\\
\Lambda\partial_{\Lambda}\partial_a \mathcal{F}_0&=&\frac{8i\Lambda}{\theta_2\theta_3}\ ,\\
\left(\Lambda\partial_{\Lambda}\right)^2\mathcal{F}_0& =& \frac{16}{3} (2U+X) \Lambda^2\ .
\end{eqnarray}
All the other derivatives are in the ring $R_{N_f=0}$.

\subsection{$\Tau$-function as a quantum Weierstrass $\sigma$-function}

Using \eqref{hol,1} the $\Tau$-function \eqref{holNS,0} can be written as 
\begin{equation}\label{pair,0}
\Tau(a,\Lambda,\epsilon,s)= e^{\pdv{W_{\mathrm{st}}}{\epsilon}}\sum_{n\in \mathbb{Z}+\frac{1}{2}} e^{i\pi \tau n^2}e^{\mathcal{F}_{\mathrm{st}}^{n,s}(a, \Lambda_{\epsilon s},\epsilon)-\frac{\rho_{\mathrm{st}}n}{\epsilon}}\ ,
\end{equation}
where we defined
\begin{equation}
W_{\mathrm{st}}=W-\frac{1}{\epsilon}\mathcal{F}_0\ , \quad \rho_{\mathrm{st}}=\rho-a_D\ .  
\end{equation}
We can now rewrite \eqref{pair,0} in terms of the $\sigma$-function using \eqref{2.1}, \eqref{2.2}
\begin{align}
&\sum_{n\in \mathbb{Z}+\frac{1}{2}} e^{i\pi \tau n^2+inx}=\theta_1\left(\frac{x}{2};\tau\right)=-\pi\eta(\tau)^3\sigma\left(\frac{x}{\pi};1,\tau\right)e^{-\frac{1}{24}E_2 x^2} \ ,\\
&\sum_{n\in \mathbb{Z}+\frac{1}{2}} e^{i\pi \tau n^2}f(n) = f(-i\partial_x)\theta_1\left(\frac{x}{2};\tau\right)\eval_{x=0}=\theta_1\left(-\frac{i}{2}\partial_x;\tau\right)f(x)\eval_{x=0}\ ,
\end{align}
then we get
\begin{equation}\label{holNS,2}
\Tau=e^{\partial_\epsilon W_{\mathrm{st}}}\eta(\tau)^3\left(\sigma(ip)e^{\frac{E_2}{24}p^2},
e^{\mathcal{F}_{\mathrm{st}}^{x,s}(a, \Lambda,\epsilon)-\frac{\rho_{\mathrm{st}}x}{\epsilon}}\right),
\end{equation}
where $\eta(\tau)$ is the Dedekind eta function, $\sigma(ip)\equiv \sigma(ip;\pi,\pi \tau)$ is the Weierstrass $\sigma$-function, and we defined the pairing
\begin{equation}\label{hol,3}
(f(p), g(x)) = \left.f(\partial_x)g(x)\right|_{x=0}\ .
\end{equation}

The formula \eqref{holNS,2} can be used to compute the \(\epsilon\)-expansion of the $\Tau$-function and present it as a quantum Weierstrass $\sigma$-function.
In order to do this we separate all functions in the exponentials into singular, constant, and higher order corrections in $\epsilon$ as
\begin{equation}
\partial_{\epsilon}W_{\mathrm{st}}+\mathcal{F}^{x,s}_{\mathrm{st}}-\frac1{\epsilon}\rho_{\mathrm{st}}x=
\frac{s}{\epsilon} \Lambda \partial_{\Lambda} \mathcal{F}_0 +
xs \Lambda\partial_{\Lambda}\partial_a \mathcal{F}_0+
\frac{s^2}{2} \left(\Lambda\partial_{\Lambda}\right)^2 \mathcal{F}_0
+ \mathcal{F}_{0,1}+\mathcal{F}_{1,0}+\widetilde{\mathcal{F}}_{\mathrm{st}}(x,s)\ ,
\end{equation}
where
\begin{multline}
\label{eq:pndef}
\widetilde{\mathcal{F}}_{\mathrm{st}}(x,s)=\sum_{n=2}^{\infty}\epsilon^{2n-2}(2n-1)\mathcal{F}_{0,n}-
\sum_{n=1}^{\infty}\epsilon^{2n-1}x\partial_a \mathcal{F}_{0,n}\\+
\sum_{k+l\ge3}^{\infty}\frac{\epsilon^{k+l-2}s^kx^l}{k!\ l!}\left(\Lambda\partial_{\Lambda}\right)^k\partial_a^l\mathcal{F}_0+
\sum_{k,l,g}\frac{\epsilon^{k+l+2g-2}s^kx^l}{k!\ l!}\left(\Lambda\partial_{\Lambda}\right)^k\partial_a^l\mathcal{F}_{g,0}=\\=
\sum_{n=1}^{\infty}\epsilon^n p_{n+2}(i\pi x/\omega_1,s)\ ,
\end{multline}
where \(p_{n+2}(x,s)\) are some polynomials of degree at most \(n+2\) in \(x,s\).
\begin{equation}
\mathcal{T}=\eta(\tau)^3e^{\frac{s}{\epsilon}\Lambda\partial_{\Lambda}\partial_a\mathcal{F}_0+
\mathcal{F}_{0,1}+\mathcal{F}_{1,0}+
\frac{s^2}{2} \left(\Lambda\partial_{\Lambda}\right)^2 \mathcal{F}_0}
\left(\sigma(ip)e^{\frac{E_2}{24}p^2}, e^{\pi i \frac{xs}{\omega_1}}e^{\widetilde{\mathcal{F}}_{\mathrm{st}}(x,s)}\right)\ .
\end{equation}
Now we introduce
\begin{equation}
\sigma(-\pi s/\omega_1) = -\frac{\pi}{\omega_1} \widetilde{\sigma}(s)\ ,\quad 
\mathcal{T}_0 =
-\frac{\pi}{\omega_1}\eta(\tau)^3
e^{\frac{s}{\epsilon}\Lambda\partial_{\Lambda}\partial_a\mathcal{F}_0+
\mathcal{F}_{0,1}+\mathcal{F}_{1,0}+
\frac{s^2}{2} \left(\Lambda\partial_{\Lambda}\right)^2 \mathcal{F}_0}\ ,
\end{equation}
and we move the term \(e^{\pi i \frac{xs}{\omega_1}}\) to the left
\begin{equation}
\mathcal{T}=\mathcal{T}_0\cdot
\left(\widetilde{\sigma}(s-ip \omega_1/\pi)e^{-\frac{\pi^2E_2}{24\omega_1^2}\left(s-i p \omega_1/\pi\right)^2}, e^{\widetilde{\mathcal{F}}_{\mathrm{st}}(x,s)}\right)\ .
\end{equation}
Rescaling the \(x\) and \(p\) variables simultaneously
\begin{equation}
\mathcal{T}=\mathcal{T}_0\cdot
\left(\widetilde{\sigma}(s+p)e^{-\frac{\pi^2E_2}{24\omega_1^2}(s+p)^2}, e^{\widetilde{\mathcal{F}}_{\mathrm{st}}(-ix \omega_1/\pi,s)}\right)\ ,
\end{equation}
exchanging the two arguments of the pairing, and substituting the expansion \eqref{eq:pndef}, we get
\begin{equation}
\mathcal{T}=\mathcal{T}_0\cdot
\left(e^{\sum\limits_{n=1}^{\infty}\epsilon^n p_{n+2}(p,s)}, \widetilde{\sigma}(s+x)e^{-\frac{\pi^2E_2}{24\omega_1^2}(s+x)^2}\right) \ .
\end{equation}
This expression can be rewritten equivalently as
\begin{equation}\label{adesso}
\mathcal{T}=\mathcal{T}_0\ \colon e^{\sum\limits_{n=1}^{\infty}\epsilon^n p_{n+2}\left(\partial_s,s\right)}\colon e^{-\frac{\pi^2 E_2}{24\omega_1^2}s^2}\widetilde{\sigma}(s)\ ,
\end{equation}
where the normal ordering \(\colon\ \colon\) moves all the \(\partial_s\) to the right.
In this way we obtain a representation of the $\Tau$-function as a differential operator acting on the $\sigma$-function
\begin{equation}
\label{eq:sigmaDerivatives}
\mathcal{T} =
-\frac{\pi}{\omega_1}\eta(\tau)^3
e^{\frac{s}{\epsilon}\Lambda\partial_{\Lambda}\partial_a\mathcal{F}_0-
T\frac{s^2}{2}+
\mathcal{F}_{0,1}+\mathcal{F}_{1,0}}
\exp\left(\sum\limits_{n=1}^{\infty}\sum\limits_{k,l=0}^{n+2}\epsilon^n\Delta^{-n}c_{n,kl}s^k\partial_s^l\right)\widetilde{\sigma}(s) \ ,
\end{equation}
where \(\Delta\) is the discriminant of the elliptic curve.
For the \(N_f=0\) case it is
\begin{equation}
\Delta=64\Lambda^8(u^2-\Lambda^4) \ .
\end{equation}
The explicit form of the first few matrices of coefficients for the $N_f=0$ case is\footnote{In these formulas \(s\) actually denotes \(s'=\frac{s}{4}\). Otherwise we have huge powers of 2.}
\begin{equation}
c^{N_f=0}_1=\left(
\begin{array}{cccc}
 0 & 0 & 0 & -\frac{8 \Lambda ^8}{3} \\
 \frac{16}{3} \Lambda ^8 \left(u^2-\Lambda ^4\right) & 0 & \frac{8 \Lambda ^8 u}{3} & 0 \\
 0 & \frac{8}{9} \Lambda ^8 \left(8 u^2-9 \Lambda ^4\right) & 0 & 0 \\
 \frac{8}{81} \Lambda ^8 u \left(28 u^2-27 \Lambda ^4\right) & 0 & 0 & 0 \\
\end{array}
\right)\ ,
\end{equation}
\begin{multline}
\hspace{-1cm}c^{N_f=0}_2=\\
\hspace{-1cm}\footnotesize\left(
\begin{array}{ccccc}
 \frac{256}{9} \Lambda ^{16} u \left(2 \Lambda ^4+u^2\right) & 0 & -\frac{128}{3} \Lambda ^{16} \left(\Lambda ^4-3 u^2\right) & 0 & \frac{128 \Lambda ^{16} u}{3} \\
 0 & \frac{128}{9} \Lambda ^{16} u \left(3 u^2-7 \Lambda ^4\right) & 0 & \frac{64}{9} \Lambda ^{16} \left(7 u^2-15 \Lambda ^4\right) & 0 \\
 -\frac{256}{27} \Lambda ^{16} u^2 \left(3 u^2-4 \Lambda ^4\right) & 0 & -\frac{128}{9} \Lambda ^{16} u \left(u^2-3 \Lambda ^4\right) & 0 & 0 \\
 0 & -\frac{64}{81} \Lambda ^{16} \left(27 \Lambda ^8+44 u^4-63 \Lambda ^4 u^2\right) & 0 & 0 & 0 \\
 -\frac{128}{243} \Lambda ^{16} u \left(27 \Lambda ^8+20 u^4-48 \Lambda ^4 u^2\right) & 0 & 0 & 0 & 0 \\
\end{array}
\right)\ .
\end{multline}
We can compute such expansions for other theories by using the ansatz \eqref{eq:sigmaDerivatives} for the corresponding Painlevé equation and finding the coefficients of the expansion.
For example, such expansion for the case of $H_0$ Argyres Douglas theory (corresponding to the Painlev\'e I case), is
\begin{equation}
\Tau=\exp\left(\sum\limits_{n=1}^{\infty}\sum\limits_{k,l=0}^{n+2}\epsilon^n\Delta^{-n}c^{H_0}_{n,kl}s^k\partial_s^l\right)\sigma(s;g_2,g_3)\ ,
\end{equation}
where
\begin{equation}
\Delta=g_2^3-27 g_3^2\ ,
\end{equation}
and
\begin{equation}
c^{H_0}_1=\left(
\begin{array}{cccc}
 0 & 0 & 0 & -g_2 \\
 0 & 0 & \frac{9 g_3}{2} & 0 \\
 0 & -\frac{g_2^2}{4} & 0 & 0 \\
 \frac{g_2 g_3}{8} & 0 & 0 & 0 \\
\end{array}
\right)\ ,
\end{equation}

\begin{equation}
c_2^{H_0}=\left(
\begin{array}{ccccc}
 \frac{81 g_2^2 g_3}{2} & 0 & \frac{21 g_2^3}{4}+\frac{405 g_3^2}{4} & 0 & 27 g_2 g_3 \\
 0 & -27 g_2^2 g_3 & 0 & -\frac{13 g_2^3}{4}-\frac{297 g_3^2}{4} & 0 \\
 \frac{5 g_2^4}{16}+\frac{189 g_2 g_3^2}{16} & 0 & \frac{27 g_2^2 g_3}{2} & 0 & 0 \\
 0 & -\frac{3 g_2^4}{16}-\frac{135 g_2 g_3^2}{16} & 0 & 0 & 0 \\
 \frac{g_2^3 g_3}{8}+\frac{27 g_3^3}{16} & 0 & 0 & 0 & 0 \\
\end{array}
\right)\ ,
\end{equation}

\begin{multline}
\hspace{-2cm}c_3^{H_0}=\\
\hspace{-2cm}\tiny
\left(
\begin{array}{cccccc}
 0 & -\frac{467 g_2^5}{80}-\frac{34047 g_2^2 g_3^2}{80} & 0 & -\frac{4149 g_2^3 g_3}{8}-\frac{27945 g_3^3}{8} & 0 & -\frac{507 g_2^4}{40}-\frac{32967
   g_2 g_3^2}{40} \\
 \frac{3531 g_2^4 g_3}{160}+\frac{44631 g_2 g_3^3}{160} & 0 & \frac{451 g_2^5}{16}+\frac{34479 g_2^2 g_3^2}{16} & 0 & \frac{4095 g_2^3
   g_3}{16}+\frac{29403 g_3^3}{16} & 0 \\
 0 & -\frac{3525 g_2^4 g_3}{32}-\frac{44793 g_2 g_3^3}{32} & 0 & -\frac{141 g_2^5}{16}-\frac{11745 g_2^2 g_3^2}{16} & 0 & 0 \\
 \frac{85 g_2^6}{192}+\frac{3951 g_2^3 g_3^2}{64}+\frac{3159 g_3^4}{16} & 0 & \frac{1137 g_2^4 g_3}{32}+\frac{15957 g_2 g_3^3}{32} & 0 & 0 & 0 \\
 0 & -\frac{27 g_2^6}{128}-\frac{3879 g_2^3 g_3^2}{128}-\frac{243 g_3^4}{2} & 0 & 0 & 0 & 0 \\
 \frac{207 g_2^5 g_3}{1280}+\frac{9963 g_2^2 g_3^3}{1280} & 0 & 0 & 0 & 0 & 0 \\
\end{array}
\right)\ .
\end{multline}

\subsection{Derivation of the holomorphic anomaly equations}
We can now proceed with the derivation of the holomorphic anomaly equations. In view of the condition \eqref{holNS,1}
we start computing the derivatives of the $\Tau$-function.
Taking the $E_2$ derivative we get 
\begin{align}\label{holNS,3.5}
&D_{E_2}\Tau=e^{\partial_\epsilon W_{\mathrm{st}}}\eta(\tau)^3\left(\sigma(ip)e^{\frac{E_2}{24}p^2},
\left(D_{E_2}+D_{E_2}\partial_\epsilon W_{\mathrm{st}}+\frac{\partial_x^2}{24}\right)e^{\mathcal{F}_{\mathrm{st}}^{x,s}-\frac{\rho_{\mathrm{st}}x}{\epsilon}}\right)=\\ 
&=e^{\partial_\epsilon W_{\mathrm{st}}}\eta(\tau)^3\left(\sigma(ip)e^{\frac{E_2}{24}p^2}\ , 
e^{-\frac{\rho_{\mathrm{st}}x}{\epsilon}}\left(D_{E_2}-\frac{x}{\epsilon}D_{E_2}\rho_{\mathrm{st}}+D_{E_2}\partial_\epsilon W_{\mathrm{st}}+\frac{\partial_x^2}{24}-\frac{\rho_{\mathrm{st}}}{12\epsilon}\partial_x+\frac{(\rho_{\mathrm{st}})^2}{24 \epsilon^2}\right)e^{\mathcal{F}_{\mathrm{st}}^{x,s}}\right)\ , \nonumber\label{holNS,3}
\end{align}
while the $a$ derivative reads
\begin{equation}\label{holNS,4}
\partial_a\Tau=e^{\partial_\epsilon W_{\mathrm{st}}}\eta(\tau)^3\left(\sigma(ip)e^{\frac{E_2}{24}p^2},
\left(\partial_a+\partial_a\partial_\epsilon W_{\mathrm{st}}+i\pi\pdv{\tau}{a}x^2\right)e^{\mathcal{F}_{\mathrm{st}}^{x,s}(a, \Lambda,\epsilon)-\frac{\rho_{\mathrm{st}}x}{\epsilon}}\right)   \ .
\end{equation}
Here we just used the relations
\begin{align}
&\partial_a \left(\eta(\tau)^3\sigma(s)e^{-\frac{1}{24}E_2s^2}\right)=-i\pi\pdv{\tau}{a} \pdv[2]{s}\left(\eta(\tau)^3\sigma(s)e^{-\frac{1}{24}E_2s^2}\right)\ ,\\
&(f''(ip),g(x))=(f(ip),-x^2g(x))\ .
\end{align}
The previous expression can be further simplified using the following equation
\begin{equation}
\partial_a e^{\mathcal{F}_{\mathrm{st}}^{x,s}(a, \Lambda,\epsilon)} =\left(\frac{1}{\epsilon}\partial_x -i \pi \pdv{\tau}{a}x^2\right)e^{\mathcal{F}_{\mathrm{st}}^{x,s}(a, \Lambda,\epsilon)}\ , 
\end{equation}
which gives the relation between $a$ and $x$ derivatives and corresponds to the equation which imposes $\mathcal{F}_{\mathrm{st}}^{x,s}$ to be the generating function of the correlation functions of the chiral ring\footnote{See \cite{Bershadsky:1993cx} appendix B.}. Substituting in \eqref{holNS,4} we obtain
\begin{equation}\label{holNS,5}
\partial_a\Tau=e^{\partial_\epsilon W_{\mathrm{st}}}\eta(\tau)^3\left(\sigma(ip)e^{\frac{E_2}{24}p^2},
e^{-\frac{\rho_{\mathrm{st}}x}{\epsilon}}\left(\frac{1}{\epsilon}\partial_x+\partial_a\partial_\epsilon W_{\mathrm{st}}-\frac{x}{\epsilon}\partial_a\rho_{\mathrm{st}}\right)e^{\mathcal{F}_{\mathrm{st}}^{x,s}(a, \Lambda,\epsilon)}\right) \ .
\end{equation}
Inserting \eqref{holNS,3.5}, \eqref{holNS,5} in \eqref{holNS,1} we obtain the following equation
\begin{align}\label{holNS,6}
&\left(D_{E_2}-\frac{x}{\epsilon}D_{E_2}\rho_{\mathrm{st}}+D_{E_2}\partial_\epsilon W_{\mathrm{st}}+\frac{\partial_x^2}{24}-\frac{\rho_{\mathrm{st}}}{12\epsilon}\partial_x+\frac{(\rho_{\mathrm{st}})^2}{24 \epsilon^2}\right)e^{\mathcal{F}_{\mathrm{st}}^{x,s}(a, \Lambda,\epsilon)}\\
&-\frac{D_{E_2}\uns }{\partial_a \uns }\left(\frac{1}{\epsilon}\partial_x+\partial_a\partial_\epsilon W_{\mathrm{st}}-\frac{x}{\epsilon}\partial_a\rho_{\mathrm{st}}\right)e^{\mathcal{F}_{\mathrm{st}}^{x,s}(a, \Lambda,\epsilon)}=0\ ,
\nonumber
\end{align}
where, thanks to the dependence on the source $s$, we removed the pairing\footnote{The reason for this is that we can move the linear term $x s \Lambda \partial_\Lambda\partial_a\mathcal{F}_0$ coming from the $\epsilon$ expansion of $\mathcal{F}_{\mathrm{st}}^{x,s}$, to the lhs of the pairing. This corresponds to the substitution $p\to p+\Lambda \partial_\Lambda\partial_a\mathcal{F}_0 s, $. Expanding in $p$ we have that the pairing must vanish for arbitrary linear combinations of $\sigma(s)$ and its derivatives, which are linearly independent functions of $s$.}.
Dividing \eqref{holNS,6} by $\exp(\mathcal{F}_{\mathrm{st}}^{x,s}(a, \Lambda,\epsilon))$ and rearranging some terms we can rewrite it as 
\begin{align}
&\left(D_{E_2}\mathcal{F}_{\mathrm{st}}^{x,s}(a, \Lambda,\epsilon)+\frac{1}{24}\partial_x^2\mathcal{F}_{\mathrm{st}}^{x,s}(a, \Lambda,\epsilon)+\frac{1}{24}\left(\partial_x\mathcal{F}_{\mathrm{st}}^{x,s}(a, \Lambda,\epsilon)\right)^2\right)\nonumber\\
&-\left(\frac{\rho_{\mathrm{st}}}{12}+\frac{D_{E_2}\uns }{\partial_a \uns }\right)\frac{1}{\epsilon}\partial_x\mathcal{F}_{\mathrm{st}}^{x,s}(a, \Lambda,\epsilon)\label{holNS,7.2}\\
&-\frac{x}{\epsilon}\left(D_{E_2}\rho_{\mathrm{st}}-\frac{D_{E_2}\uns }{\partial_a \uns }\partial_a\rho_{\mathrm{st}}\right)+\left(D_{E_2}\partial_\epsilon W_{\mathrm{st}}+\frac{(\rho_{\mathrm{st}})^2}{24 \epsilon^2}-\frac{D_{E_2}\uns }{\partial_a \uns }\partial_a\partial_\epsilon W_{\mathrm{st}}\right)=0 \nonumber \ .
\end{align}
Finally, by conjugating the first line of the above equation with the shift operators arising from the algebra of the vector fields \eqref{eq:OpAlgebra} and expressing the ratio $D_{E_2}\uns /\partial_a \uns$ in terms of $W_{\mathrm{st}},\rho_{\mathrm{st}}$, we can rewrite the above equation as (see appendix \ref{appD,1} and \ref{appD,2} for details)
\begin{align}
&\left(D_{E_2}\widehat{\mathcal{F}}(y,\Lambda',\epsilon)+\frac{\epsilon^2}{24}\partial_y^2\widehat{\mathcal{F}}(y,\Lambda',\epsilon)+\frac{\epsilon^2}{24}\left(\partial_y\widehat{\mathcal{F}}(y,\Lambda',\epsilon)\right)^2\right)\nonumber\\
&-\left(\left(\frac{y-a}{\epsilon}\right)\partial_a-\partial_\epsilon-\frac{\epsilon\partial_y\mathcal{F}(y,\Lambda',\epsilon)-(y-a)\partial_a\rho/\epsilon+\epsilon\partial_a\partial_\epsilon W}{\partial_a \uns }\Lambda\partial_\Lambda\right)\times \nonumber \\
&\times\left(D_{E_2}W_{\mathrm{st}}+\frac{\epsilon}{24}(\partial_a W_{\mathrm{st}})^2\right)=0\ , \label{holNS,16}
\end{align}
where we defined the shifted modulus $y=\epsilon x+a$,
the scale 
$\Lambda'=\Lambda_{\epsilon s}=\Lambda\exp(\epsilon s)$
and\footnote{The notation $\widehat{\mathcal{F}}(a,\Lambda,\epsilon)$ denotes here the stable part of the SD free energy. This is to distinguish it from the BCOV generating function $\mathcal{F}_{st}^{x,s}$ and is not related in any way to the blowup.}
\begin{equation}
\widehat{\mathcal{F}}(a,\Lambda,\epsilon)=\mathcal{F}(a,\Lambda,\epsilon)-\frac{1}{\epsilon^2}\mathcal{F}_0(a,\Lambda)\ . 
\end{equation}
By using the separation of scales one can show (see appendix \ref{appD,2} for details) that \eqref{holNS,16}
is equivalent to the holomorphic anomaly equations obeyed by the SD and NS free energies
\begin{align}
&D_{E_2}W_{\mathrm{st}}+\frac{\epsilon}{24}(\partial_a W_{\mathrm{st}})^2=0\ , \label{holNS,10.1} \\
&D_{E_2}\widehat{\mathcal{F}}(y,\Lambda',\epsilon)+\frac{\epsilon^2}{24}\partial_y^2\widehat{\mathcal{F}}(y,\Lambda',\epsilon)+\frac{\epsilon^2}{24}\left(\partial_y\widehat{\mathcal{F}}(y,\Lambda',\epsilon)\right)^2=0 \ . \label{holNS,10.2} 
\end{align}

We emphasize here that the formula \eqref{holNS,0} defining the \(\mathcal{T}\)-function can be obtained using the purely mathematical arguments~\cite{Gavrylenko:2016zlf,Lisovyy:2021bkm,Lisovyy:2022flm,Gavrylenko:2020gjb}, so that \(Z_{SD}\) is given by explicit combinatorial expression~\cite{Gavrylenko:2016zlf}, and \(W\) can be identified with the classical action on the trajectory~\cite{Litvinov:2013sxa}.
Therefore, our derivation of the holomorphic anomaly equations~\eqref{holNS,10.1}, \eqref{holNS,10.2} is actually independent from any physical arguments and uses only the fact that the \(\mathcal{T}\)-functions of the Painlev\'e equation can be parameterized by the position of its zero \(\Lambda\) and the value of the Hamiltonian \(\mathbf{u}\) at this point, and the fact that the formula~\eqref{holNS,0} exists.

\subsection{Painlev\'e $\Tau$-function as a non-perturbative completion of topological strings partition function}

As reviewed at the beginning of Sect. \ref{section4}, the perturbative topological strings partition function $\mathcal{Z}_X$ displays a non-holomorphic dependence on the moduli. This can be interpreted as a background dependence of this formulation \cite{witten1993quantumbackgroundindependencestring}. On the other hand, the partition function $Z_{SD}$
is obviously holomorphic but it is not modular and as such frame dependent, thus suffering the same problem of background dependence. In this subsection we show that the very structure of the Painlev\'e $\Tau$-function
contains non-perturbative corrections in the topological string coupling $\epsilon=g_s$
that are suited to cancel the modular anomaly.  
In particular, we show that there exists a suitable expansion point \eqref{expansion} for $\Tau$ which provides a natural non-perturbative completition of topological strings, whose modularity is explicitly implied by the BCOV holomorphic anomaly equation. 

Let us recall eq.\eqref{holNS,0}
\begin{equation}\label{holNSI,0}
\Tau(a,\Lambda,\epsilon,s)= e^{\partial_\epsilon W(a,\Lambda,\epsilon)}\sum_{n\in \mathbb{Z}+\frac{1}{2}} e^{-\frac{n}{\epsilon}\rho(a,\Lambda,\epsilon)} Z_{SD}(a+n \epsilon,\Lambda_{\epsilon s},\epsilon)\ .
\end{equation}
As we saw, the $E_2$ dependence of \eqref{holNSI,0} enters only through the $\uns$ modulus. Let us now show that this $E_2$ dependence can be reabsorbed by fixing the surface observable source $s=s_0$ such that
\begin{equation}\label{holint,2}
\rho(a,\Lambda,\epsilon)=\epsilon\partial_a W(a,\Lambda,\epsilon)=a_D(a,\Lambda_{SD})= a_D(a,\Lambda e^{\epsilon s_0(a,\Lambda,\epsilon)})\ ,  
\end{equation}
which relates the dual quantum period $\rho$ computed at the NS scale $\Lambda$ to the SW period $a_D$ at the $SD$ scale
\begin{equation}\label{expansion}
\Lambda_{SD}=\Lambda e^{\epsilon s_0(a,\Lambda,\epsilon)}\ .
\end{equation}
Thus, by expanding the $\Tau$-function \eqref{holNSI,0} around $s_0$, we get 
\begin{equation}
\Tau(a,\Lambda,\epsilon,s'+s_0)= e^{\partial_\epsilon W(a,\Lambda,\epsilon)}\sum_{n\in \mathbb{Z}+\frac{1}{2}} e^{-\frac{n}{\epsilon} a_D(a,\Lambda_{SD})} Z_{SD}(a+n \epsilon,\Lambda_{SD}e^{\epsilon s'},\epsilon)\ .
\end{equation}
Up to a change of the overall normalization of the above $\Tau$-function, we then define
\begin{equation}
\Tau_{SD}(a,\Lambda_{SD},\epsilon,s')= e^{-\frac{1}{\epsilon^2}\mathcal{F}_0(a,\Lambda_{SD})}\sum_{n\in \mathbb{Z}+\frac{1}{2}} e^{-\frac{n}{\epsilon} a_D(a,\Lambda_{SD})} Z_{SD}(a+n \epsilon,\Lambda_{SD}e^{\epsilon s'},\epsilon)\ .
\end{equation}
We will now show that this $\Tau$-function is $E_2$ independent and as such it is holomorphic and modular, being thus globally defined on the moduli space of $X$. It then provides a non-perturbative extension of the topological string partition function.

The dependence on $\tau$ is now measured at the appropriate scale $\Lambda_{SD}$, $\tau=\tau(a,\Lambda_{SD})$ . Correspondingly also the algebra of derivations is referred to this scale.
Applying the same manipulation as in \eqref{pair,0}-\eqref{holNS,3.5} we can write
\begin{equation}\label{hol,2}
\Tau_{SD}(a,\Lambda_{SD},\epsilon,s')=\eta(\tau)^3\left(\sigma(ip)e^{\frac{E_2}{24}p^2},
e^{\mathcal{F}_{\mathrm{st}}^{x,s'}(a, \Lambda_{SD},\epsilon)}\right),
\end{equation}
Computing the $E_2$ derivative of $\Tau_{SD}$ and insisting on modular invariance we get
\begin{equation}\label{hol,4}
0=D_{E_2}\Tau_{SD}= \eta(\tau)^3\left(\sigma(p)e^{\frac{E_2}{24}p^2},\left(D_{E_2}+\frac{\partial_x^2}{24}\right)
e^{\mathcal{F}_{\mathrm{st}}^{x,s'}(a, \Lambda_{SD},\epsilon)}\right).
\end{equation}
As we evaluate this quantity at arbitrary $s'$, this implies
\begin{equation}\label{hol,5}
\left(D_{E_2}+\frac{\partial_x^2}{24} \right)e^{\mathcal{F}_{\mathrm{st}}^{x,s'}(a, \Lambda_{SD},\epsilon)} = 0\ ,
\end{equation}
which is nothing but the holomorphic anomaly equation\footnote{In \cite{Bershadsky:1993cx} in the equation for the generating function an extra term appears due to the contributions of two insertions of colliding chiral fields. However, this term explicitly depends on the K\"ahler potential and decouples in the holomorphic limit.} for the generating function. In particular, setting to zero the sources $x,s'=0$ we obtain also the usual holomorphic anomaly equation
\begin{equation}\label{hol,6}
\left(D_{E_2} + \frac{\epsilon^2}{24}\partial_a^2\right) e^{\widehat{\mathcal{F}}(a,\Lambda_{SD},\epsilon)}=0\ ,
\end{equation}
while, computing derivatives with respect to $x$ of \eqref{hol,5} and then setting $x,s=0$ one obtains also the holomorphic anomaly equations for the correlation functions of the chiral ring. 

Let us remark that \eqref{hol,5} also appear in \cite{Billo:2014bja}, were a reference point for the expansion in the SW moduli is taken to be the massless point. The relation between \eqref{hol,5} and \eqref{hol,6} can be explained in terms of shift operators based on the algebra \eqref{eq:OpAlgebra}.
Namely, it is the non-trivial relation
\begin{equation}
\left(D_{E_2}+\frac{\partial_x^2}{24} \right)e^{\mathcal{F}_{0,\mathrm{st}}^{x,s}(a, \Lambda,\epsilon)}e^{\epsilon x\partial_a+\epsilon s\Lambda\partial_\Lambda}e^{\widehat{\mathcal{F}}(a,\Lambda,\epsilon)}=e^{\mathcal{F}_{0,\mathrm{st}}^{x,s}(a, \Lambda,\epsilon)}e^{\epsilon x\partial_a+\epsilon s\Lambda\partial_\Lambda}\left(D_{E_2} + \frac{\epsilon^2}{24}\partial_a^2\right)e^{\widehat{\mathcal{F}}(a,\Lambda,\epsilon)}
\end{equation}
proved in the appendix~\ref{appD}.
We also notice that \eqref{hol,5} is the usual heat equation, while \eqref{hol,6} is not, because the derivative operators involved do not commute, see the second line of~\eqref{eq:OpAlgebra}.

So we proved that $E_2$-independence of the $\Tau$-function implies the holomorphic anomaly equations.
Clearly also the converse is true, the holomorphic anomaly equations \eqref{hol,5} imply that $\Tau_{SD}$ is $E_2$ independent.
Thus, the invariant $\Tau$-function $\Tau_{SD}$ is a natural candidate for a non-perturbative partition function of the corresponding topological string
\begin{equation}
    \Tau_{SD}={\mathcal Z}_X^{NP}\ .
\end{equation}

\section{Hurwitz expansions of Painlev\'e $\Tau$-functions}
\label{section5}

In \cite{hone2013properties} and \cite{hone2017hirota} Hone, Ragnisco and Zullo (HRZ) found numerical evidence that the Painlevé $\Tau$-functions of PI, PII and PIV 
enjoy Hurwitz integrality when expanded 
around one of their zeros. Indeed, this can be seen as a natural deformation of the Hurwitz expansion \eqref{sigmaHRZ} of the Weierstrass $\sigma$-function, while the recurrence relations for the corresponding coefficients of the deformed expansions generalizes the one of the coefficients of $\sigma$. 

As we already observed in section \eqref{section3}, from the gauge theory point of view the expansion of the $\Tau$-function around a zero
corresponds to the expansion of the blowup factor in terms of the source $s$ of the surface observable $I(E)$ in the NS limit. It is then possible to determine all the coefficients of the blowup formula recursively from the Painlev\'e equations.

In this section we explicitly compute these recursion relations, generalizing the results of HRZ. As previously said, the relation \eqref{aut,7} between the Painlevé $\Tau$-function and the blowup factor $\mathcal{B}_{NS}$, being based on operator/state correspondence in the topological theory, holds in general, also in the Argyres-Douglas cases\footnote{A consistency check is to show that the solution of the AD case corresponds exactly to the limit of the Lagrangian theory solution around the AD point. We will show explicitly this for the limit $(N_f=1)\to H_0$.}.
We reproduce the Hurwitz integral expansions for the AD theories $H_0$ and $H_2$, which correspond to PI and PIV equations respectively, in a form which is more suitable to a gauge theory analysis and find new expansions for the Lagrangian theories with $N_f=0,1,4$ corresponding to PIII$_3$, PIII$_2$, PVI respectively. In the case of PIII$_2$ we will show explicitly how one can recover the PI Hurwitz integral expansion from the one of PIII$_2$. 

Before starting the detailed analysis we explain here the general strategy.
The $\Tau$-function of Painlevé equations obeys an Hirota bilinear differential equation of the form
\begin{equation}\label{genpanfor}
D(\Tau,\Tau)= 0 \ ,
\end{equation}
where $D(f,g)$ is generically a quartic order bilinear differential operator in the Painlevé time $t$ (whose specific form depends on the specific equation) which is a combination of ordinary derivatives and Hirota derivatives $D_x^{(k)}(f,g)$ defined as
\begin{equation}\label{HRZ,-1}
f(x+h)g(x-h)=\sum_{k=0}^{\infty} D_x^{(k)}(f,g)\frac{h^k}{k!}\ \Rightarrow 
\ D_x^{(k)}(f,g) = \dv[k]{h}f(x+h)g(x-h)\eval_{h=0}\ . 
\end{equation}

Following the gauge theory interpretation, we consider then a zero $t_0$ of the $\Tau$-function and change the time variable $t=t(\epsilon,s)$ in such a way that $s=0$ corresponds to the position of the zero $t_0$, and in the limit $\epsilon \to 0$ the equation becomes an autonomous equation in $s$. From the gauge theory point of view $t_0$ corresponds to the value of the gauge coupling $\Lambda$ and we interpret the map $t=t(\epsilon,s)$ as the shifted\footnote{The precise form of the shift depends on the theory. In the Lagrangian theories the shift is multiplicative $t=t_0\exp(\epsilon s)$. In the AD theories the shift is additive $t=t_0+\epsilon^k s$, with $[s]=-k$. } gauge coupling $\Lambda_{\epsilon s}$ due to the insertion of the surface observable $I(E)$.

From \eqref{aut,7} we observe that $\Tau$ has a singular prefactor in the limit $\epsilon \to 0$. Therefore, to get a sensible limit, it is convenient to rewrite the equation in terms of $\mathcal{B}_{NS}$ which is regular in the limit $\epsilon \to 0$
\begin{equation}
\mathcal{B}_{NS}(\epsilon,s) = e^{-\frac{\uns s}{\epsilon}}\Tau(\epsilon,s)\ .
\end{equation}
In all cases, in the autonomous limit $\epsilon \to 0$ the Hirota equation reduces to 
\begin{equation}\label{HRZ,0}
\HirD{\mathcal{B}_{SW}}{s}{4}+12T\HirD{\mathcal{B}_{SW}}{s}{2}+(12T^2-g_2) \mathcal{B}_{SW}^2= 0 \ ,
\end{equation}
where $g_2,T$ depend on the parameters of the theory. 
In this limit the solution $\mathcal{B}_{NS}(\epsilon,s)$ reduces, up to a Gaussian prefactor which gives the contact term $T$, to the Weierstrass $\sigma$-function as
\begin{equation}\label{HRZ,0''}
\mathcal{B}_{SW}(s)=
\lim_{\epsilon \to 0}\mathcal{B}_{NS}(\epsilon,s)= \Lambda e^{-\frac{1}{2}T s^2}\sigma(s;g_2,g_3) \ ,
\end{equation}
in agreement with the gauge theory result in the SW limit \eqref{aut,8}.
The parameter $g_3$ can be determined from the sigma-form of the Painlevé equation, which is the equation for the Painlevé hamiltonian $\eta=\partial_s \log \mathcal{B}_{NS}$. In the autonomous limit $\epsilon \to 0$ the hamiltonian $\eta$ always reduces to the form $\eta_{SW} = \zeta_W(s;g_2,g_3)-sT$ where $\zeta_W(s;g_2,g_3)$ is the Weierstrass zeta function (see appendix \ref{appA}) and the equation has the form\footnote{This equation follows simply from the equation of the SW curve in Weierstrass parametrization written in terms of $\eta$.}
\begin{equation}\label{HRZ,0'}
\ddot \eta_{SW}^2+4\dot \eta_{SW}^3+12T\dot \eta_{SW}^2+(12T^2-g_2)\dot \eta_{SW} +4T^3-g_2T+g_3=0 \ , \quad \eta_{SW} =\pdv{s} \log \mathcal{B}_{SW} \ ,
\end{equation}
from which we can easily read the value of the parameter $g_3$ once $T$ and $g_2$ are known. The parameters $g_2,g_3$ are exactly the elliptic invariants given by the Weierstrass parametrization of the SW curve.

In the Painlev\'e equation in Hirota bilinear form \eqref{genpanfor} we then substitute the 
power series expression
\begin{equation}\label{HRZ,1}
\mathcal{B}_{NS}(\epsilon,s) =b_0 \sum_{n=0}^{\infty} c_n \frac{s^{n+1}}{(n+1)!} \ ,
\end{equation}
which means that $\mathcal{B}_{NS}(\epsilon,s)$ is expanded (in $s$) around its zero $t_0$. This is motivated in the gauge theory by the possibility of expressing $\mathcal{B}_{NS}$ as a sum of local operators given by a series in $s$ as defined by the differential operator $D_{NY}$ in the NY blowup equations in presence of the surface observable $I(E)$ \eqref{NY,1}.

To construct the recursion relation it is useful to write the action of the Hirota derivative \eqref{HRZ,-1} in the power basis $s^n$. One has
\begin{equation}
\frac{1}{n!m!}D_s^{(k)}(s^n,s^m) = P_{nm}^{k}s^{n+m-k} \ ,
\end{equation}
where the coefficients $P_{nm}^k$ are given by\footnote{We use the convention that $\binom{n}{k}=0$ if $n<k$.}
\begin{align}
&P_{nm}^k = \frac{k!}{n!m!}\sum_{l=0}^k (-1)^l \binom{n}{l}\binom{m}{k-l} \ , \quad P_{mn}^k = (-1)^k P_{nm}^k \ , \\
&P_{nm}^0 = \frac{1}{n!m!} \ , P_{nm}^k=0 \text{ if }n+m < k \ .
\end{align}
Therefore, using for $\mathcal{B}_{NS}$\footnote{In the following we set $b_0=1$. It can be restored simply rescaling $\mathcal{B}_{NS}\to b_0\mathcal{B}_{NS}$ thanks to bilinearity of the Hirota equation.} 
the ansatz \eqref{HRZ,1}, we can write
\begin{equation}
\HirD{\mathcal{B}_{NS}}{s}{k}= \sum_{n,m=0}^{+\infty} P_{n+1,m+1}^k c_n c_m = s^{4-k}\sum_{n=0}^{+\infty} s^{n-2} F_n^k \ , \quad F_n^k = \sum_{l=0}^n P_{l+1,n-l+1}^k c_l c_{n-l} \ .
\end{equation}
The quantities $F_n^k$ are the basic building blocks and the recursive relation can be written as linear combinations of the $F_n^k$'s. To obtain more compact expressions we follow the convention $F_n^k = 0 \text{ if }n<0$.  
For all Painlevé equations the bilinear form \eqref{genpanfor} contains a quartic Hirota derivative $\HirD{\mathcal{B}_{NS}}{s}{4}$ and this is the term of maximal degree. We can write
\begin{equation}
F_n^4 ={F'}_n^4+2c_n P_{n+1,1}^4 \ , \quad {F'}_n^4 = \sum_{l=1}^{n-1} P_{l+1,n-l+1}^4c_l c_{n-l}\ , \quad P_{n+1,1}^4 = \frac{n(n^2-1)(n-6)}{(n+1)!} \ ,
\end{equation}
where we isolated the factor containing the coefficient $c_n$ form the others and
used $c_0=1$. The operators $s$ and $\partial_s$ act as shift operators in the basis\footnote{The reason for the shift by $-2$ is that the leading order in the recursion relation is given by $\HirD{\mathcal{B}_{NS}}{s}{4}$.} $s^{n-2}$
\begin{align}\label{HRZ,1.5}
&s \sum_{n=0}^{+\infty} s^{n-2} \psi_n = \sum_{n=0}^{+\infty} s^{n-2} \psi_{n-1} \equiv \sum_{n=0}^{+\infty} s^{n-2} \hat S \psi_n\ , \quad \hat S \psi_n = \psi_{n-1}\ ,\\
&\pdv{s}\sum_{n=0}^{+\infty} s^{n-2} \psi_n =\sum_{n=0}^{+\infty} s^{n-2} (n-1)\psi_{n+1} \equiv \sum_{n=0}^{+\infty} s^{n-2} \hat S^\dagger\psi_n\ , \quad \hat S^\dagger \psi_n =(n-1)\psi_{n+1}\ .
\end{align}
Therefore we obtain the following mapping
\begin{equation}\label{HRZ,1.75}
\HirD{\mathcal{B}_{NS}}{s}{k}\rightarrow\hat S^{4-k} F_n^k\ , \quad s \rightarrow \hat S, \quad \pdv{s} \rightarrow \hat S^\dagger \ .
\end{equation}
Using this representation we obtain a recursion relation of the form
\begin{equation}\label{HRZ,2}
\frac{2 n(n^2-1)(n-6)}{(n+1)!}c_n = -{F'}_n^4+\sum_{k<4}\sum_{j<n} q_{jk} F_j^k \ ,    
\end{equation}
where the coefficients $q_{jk}$ depends on $t_0, \epsilon$ and the parameters of the specific Painlevè equation at hand and will be analyzed in full detail in the following subsections. It turns out that for $n=0,1,6$ both sides of \eqref{HRZ,2} vanish identically and the corresponding coefficients $c_0,c_1,c_6$ are undetermined. These coefficients correspond to \emph{resonances} that give a parametric family of solutions and are related to the assignment of initial conditions for the equation. 
These free coefficients can be fixed from the gauge theory. The resonance at $c_0$ corresponds to the freedom of changing the normalization of the $\Tau$-function. With our choice of normalization it corresponds to the first coefficient of the blowup factor $\mathcal{B}_{NS}(s) =b_0 s+\dots$ therefore we have always $c_0=1$. The coefficient $c_1$ corresponds to the $s^2$ term and it is related to the pole in the prefactor $\exp(\uns s/\epsilon)$. We have $c_1=0$ because in the blowup factor $\mathcal{B}_{NS}$ this pole is removed. Finally, we consider the coefficient $c_6$. 

The origin of this resonance is that the Hirota equation for $\mathcal{B}_{NS}$ is obtained by one derivative of the corresponding sigma-form equation for the 
 Painlevé Hamiltonian $\eta$. This  produces a further integration constant.
  This constant, which is related to the elliptic invariant $g_3$, determines the value of the coefficient $c_6$ which must be then fixed imposing that the solution $\mathcal{B}_{NS}$ satisfies the original sigma-form equation.

It is clear from the recursion relation \eqref{HRZ,2} that $c_n$ are polynomials in the parameters of the theory and for dimensional reasons they are homogeneous polynomials. From the limit \eqref{HRZ,0''} we see that the coefficients $c_n$ are deformations in $\epsilon$ of the Weierstrass sigma coefficients $c_n^\sigma$. 

The non-trivial result, which we checked numerically\footnote{We checked this to very high order, $n\sim 100$.}, but for which we don't have a proof, is that the polynomials $c_n$ turn out to have integer coefficients. This property was already noticed in \cite{hone2013properties,hone2017hirota} for the cases of PI, PII and PIV and we checked that it remains true also in the other cases.

In the following we will apply the analysis outlined above to Painlevé equations in Hirota form to derive the explicit structure of the recursion relation and of the coefficients $c_n$ of the \eqref{HRZ,1} expansion showing that it is an integral Hurwitz series in the ring of polynomials with integer coefficients of the corresponding invariants and we will comment on their specific structure. 

Finally, sometimes it will be convenient to organize the expansion in terms of the IR parameters $g_2,g_3,T$ of the SW curve also when $\epsilon\neq0$. In this case, however, they will be evaluated in the NS modulus $\uns$ and we will denote them by $\gII,\gIII,\T$. We remark that in this section the normalization for the $u$ modulus is different from the previous ones and is fixed by the explicit parametrization of the SW curve we use.

\subsection{PVI alias $N_f=4$}
We start the analysis from $SU(2)$ gauge theory with $N_f=4$ which corresponds to the PVI equation. The theory depends on the four masses $m_i$ of the hypermultiplets and one dimensionless UV coupling $q=\exp(i\pi \tau_0)$. In the massless case $m_i=0$ the UV coupling has no perturbative corrections, because the perturbative beta function vanishes, but is renormalized non-perturbatively by instanton corrections\footnote{See e.g. \cite{Marshakov_2009}.} as 
\begin{equation}\label{PVI,-0.5}
q=\exp(i\pi \tau_0)=\frac1{16}\frac{\theta_2^4(\tau)}{\theta_3^4(\tau)} 
\ ,
\end{equation}
where $\tau$ is the IR effective coupling. The SW curve for this theory is
\begin{equation} 
y^2 = x(x-u)(x-qu)-x^2(1-q)^2 e_2-4x(1-q)q(2(1+q)p_4+(1-q)e_4)+16(1-q)q^2(up_4-(1-q)e_6),
\end{equation}
the parameters $e_k, p_4$ are symmetric polynomials in the $m_j$'s
\begin{align}
&e_2 = m_1^2+m_2^2+m_3^2+m_4^2\ , \quad e_4 = m_1^2 m_2^2+m_1^2 m_3^2+ m_1^2 m_4^2+m_2^2 m_3^2+ m_2^2 m_4^2+ m_3^2 m_4^2\ , \\
&e_6= m_1^2 m_2^2 m_3^2+m_1^2 m_2^2 m_4^2+m_1^2m_3^2m_4^2+m_2^2 m_3^2 m_4^2\ , \quad p_4 =  m_1 m_2 m_3 m_4\ ,   
\end{align}
and they make manifest the $SO(8)$ flavour symmetry of the theory. The PVI equation is the most general Painlevé equation, all the others can be obtained as coalescence limits of PVI scaling the parameters in a suitable way. In the class $S$ construction of the theory in terms of the Gaiotto curve this operation corresponds to colliding some punctures to form irregular ones. This is the geometrical realization of the renormalization group flow for the corresponding gauge theories. Later we will work in detail how to take the coalescence limit to obtain the simplest AD theory $H_0$ from the $N_f=1$ Lagrangian theory.

We start now the analysis of the PVI equation. The Hirota bilinear equation is \cite{Bershtein_2015}
\begin{align}\label{PVI,0}
& (t-1)^3\HirD{\Tau}{\log t}{4}+2(t-1)^2(1+t)\left(t\dv{t}\right)\HirD{\Tau}{\log t}{2}
\nonumber \\
&+(1-t)(e'_2t^2-t^2+t-1)\HirD{\Tau}{\log t}{2}+ t(t-1)\left(t\dv{t}\right)^2\Tau^2 \nonumber \\
&-e'_2 t^2\left(t\dv{t}\right)\Tau^2+t(t e'_4+2(t-2)p'_4)\Tau^2 = 0 \ ,
\end{align}
while the Hamiltonian $\zeta(t) =t(t-1)\partial_t \log \tau (t)$ satisfies the sigma-form equation
\begin{align}
&t^2 (1-t)^2\ddot\zeta^2-4t(1-t)\dot\zeta^3-(e'_2 t^2 + 4 (2t-1) \zeta)\dot\zeta^2 -(4 p'_4 - (e'_4 + 2 p'_4) t - 2 e'_2 t \zeta - 4 \zeta^2) \dot\zeta \nonumber \\
&- (e'_4+2 p'_4) \zeta - e'_2\zeta^2 -e'_6=0\ ,
\end{align}
where $e'_k, p'_4$ are the symmetric polynomials in the normalized mass parameters $\mu_i=m_i/\epsilon$. To get the autonomous limit we use the mass parameters $m_j$
\begin{equation}\label{PVI,1}
\quad \mu_j = \frac{m_j}{\epsilon}\ \Rightarrow \quad e'_k = \frac{e_k}{\epsilon^k} \ , \quad p'_4 =\frac{p_4}{\epsilon^4}\ .
\end{equation}
For the Painlevé time we use the following parametrization 
\begin{equation}\label{PVI,11}
t=q e^{(q-1)\epsilon s} \ ,     
\end{equation}
where we notice that in the formula an extra factor $(q-1)$ appears. From the point of view of the Painlevé VI equation this factor is required to cancel poles at $q=1$ which is a non-movable singularity of PVI. In the gauge theory the map \eqref{PVI,11} gives the shift of the coupling $q$ due to the insertion of the surface observable $I(E)$. This suggests that in the conformal case the surface observable contribution contains a non-trivial renormalization factor $(q-1)$, which is non-perturbative, similarly to the coupling renormalization $\eqref{PVI,-0.5}$. The expression for the $\Tau$-function \eqref{aut,7} and for the hamiltonian $\zeta$ becomes
\begin{equation}
\Tau(s)= e^{\frac{\uns(q-1)s}{\epsilon}}\mathcal{B}_{NS}(s)=e^{\frac{\uns_r s}{\epsilon}}\mathcal{B}_{NS}(s) \ , \quad \zeta(s)=\epsilon^{-1}r(t)\eta(s)+r(t)\epsilon^{-2}\uns_r,
\end{equation}
where we defined $\uns_r =\uns(q-1)$ and the ratio
\begin{equation}
r(t)=\frac{t-1}{q-1} \ , \quad r(t)\underset{\epsilon \to 0}{\rightarrow} 1 \ .
\end{equation}
Using \eqref{PVI,1}, \eqref{PVI,11} the Hirota equation reads
\begin{align}\label{PVI,2}
& r(t)^3\HirD{\mathcal{B}_{NS}}{s}{4}+2\epsilon r(t)^2(t+1)\pdv{s}\HirD{\mathcal{B}_{NS}}{s}{2}\nonumber\\
+&r(t)(4\uns_r r(t)(1+t)-e_2 t^2+\epsilon^2(t^2-t+1))\HirD{\mathcal{B}_{NS}}{s}{2} \nonumber \\
+&\epsilon^2 r(t) t\pdv[2]{s}\mathcal{B}_{NS}^2 +\epsilon t\left(4 r(t)\uns_r -e_2 t\right)\pdv{s}\mathcal{B}_{NS}^2\nonumber\\
+&\left(4 r(t)t \uns_r^2-2t^2 \uns_r e_2+t(q-1)(e_4 t+2(t-2)p_4)\right)\mathcal{B}_{NS}^2=0 \ ,
\end{align}
and the sigma-form equation becomes 
\begin{align}\label{PVI,2.5}
&r(t)^2\left(r(t)\ddot \eta +\epsilon (t+1)\dot\eta\right)^2 +4r(t)^4\dot\eta^3+r(t)^2\left(4r(t)(1+t)(\uns_r+\epsilon \eta)-e_2 t^2\right)\dot\eta^2 \nonumber\\
+&(4r(t)^2t(\uns_r+\epsilon \eta)^2-2r(t)t^2e_2(\uns_r+\epsilon \eta)+t(t-1)(2p_4(t-2)+e_4 t))\dot\eta \nonumber \\ 
-&t^2\left(e_6(q-1)^2+(2p_4-e_4)(\uns_r+\epsilon \eta)(q-1)+e_2(\uns_r+\epsilon \eta)^2\right)=0 \ .
\end{align}
In the limit $\epsilon \to 0$ we have $\uns\to u$ and the equations \eqref{PVI,2}, \eqref{PVI,2.5} reduce to the autonomous form \eqref{HRZ,0}, \eqref{HRZ,0'} respectively, with the following values of the parameters
\begin{align*}
&T_r^{PVI}=\frac{u_r}{3}(1+q)-\frac{1}{12}e_2 q^2 \ , \\
&g_{2,r}^{PVI} = -q(q-1)(2p_4 (q-2)+ e_4 q)+\frac{1}{12}e_2^2 q^4 - \frac{2}{3} e_2 q^2(q-2) u_r + \frac{4}{3}(1 - q + q^2) u_r^2 \ , \\
&g_{3,r}^{PVI}=-q^2(q-1)^2e_6+\frac{1}{12} e_2 (q-1) q^3 (2p_4 (q-2) + e_4 q)-\frac{1}{216}e_2^3 q^6\\
&\qquad        +\frac{1}{18} q \left( (q-2) q (-6e_4(q-1) + e_2^2 q^2) -12 p_4 (2 -4q + q^2 + q^3)\right)u_r\\
&\qquad        -\frac{1}{9} e_2 q^2 (5 -5q + 2 q^2) u_r^2 + \frac{4}{27} (2-3q-3q^2+2q^3)u_r^3\ .
\end{align*}
In the massless case $m_i=0$ they reduce to
\begin{equation}
T_r^{PVI}=\frac{u_r}{3}(1+q) \ , \quad g_{2,r}^{PVI} =\frac{4}{3}(1 - q + q^2) u_r^2 \ , \quad g_{3,r}^{PVI}=\frac{4}{27} (2-3q-3q^2+2q^3)u_r^3\ .
\end{equation}
The variables are not directly the physical parameters of the theory, because of the renormalization of the shifted coupling \eqref{PVI,11}, which corresponds to the replacement $s\to (q-1)s$ in the SW result. This means that in the SW limit we have
\begin{align}
&\lim_{\epsilon \to 0}\mathcal{B}^{PVI}_{NS}(\epsilon,s)\propto e^{-\frac{1}{2}T^{PVI}(q-1)^2 s^2}\sigma(s(q-1);g_2,g_3)= \\
&=(q-1)e^{-\frac{1}{2}T^{PVI}(q-1)^2 s^2}\sigma(s;(q-1)^4g_2,(q-1)^6 g_3)=(q-1)e^{-\frac{1}{2}T_r^{PVI}s^2}\sigma(s; g_{2,r}, g_{3,r})\ , \nonumber
\end{align}
where we used the homogeneity of $\sigma$. Therefore, the physical variables are
\begin{equation}\label{PVI,3}
T^{PVI}= \frac{1}{(q-1)^2}T_r^{PVI}= -\frac{u}{3}\frac{1+q}{1-q}-\frac{1}{12}e'_2 \left(\frac{q}{1-q}\right)^2 \ , \quad g_2^{PVI}=\frac{1}{(q-1)^4} g_{2,r}^{PVI} \ , \quad g_3^{PVI}=\frac{1}{(q-1)^6} g_{3,r}^{PVI} \ .
\end{equation}
in particular, in the massless limit we have (substituting back $u_r=(q-1)u)$
\begin{equation}\label{PVI,4}
T^{PVI}=-\frac{u}{3}\frac{1+q}{1-q}\ , \quad g_2^{PVI} =\frac{4}{3}\frac{(1 - q + q^2)}{(q-1)^2}u^2 \ , \quad g_3^{PVI}=\frac{4}{27}\frac{(2-3q-3q^2+2q^3)}{(q-1)^3}u^3\ ,
\end{equation}
which is consistent with the SW result\footnote{See appendix \ref{appB}.}. 

As previously discussed, the recurrence relation for the coefficients of the Hurwitz expansion can be obtained expressing all the operators in the power basis $s^n$. Applying the map \eqref{HRZ,1.75} to \eqref{PVI,2} we obtain the following recurrence relation
\begin{align}\label{PVI,5}
\frac{2 n(n^2-1)(n-6)}{(n+1)!}c_n &=-{F'}_n^4+(r(\hat t)^3-1)F_n^4+2\epsilon r(\hat t)^2(\hat t+1)(n-1)F_{n-1}^2\nonumber\\
+&r(\hat t)(4\uns_r r(\hat t)(1+\hat t)-e_2 {\hat t}^2+\epsilon^2({\hat t}^2-\hat t+1))F_{n-2}^2 \nonumber \\
+&\epsilon^2 r(\hat t) \hat t (n-1)nF_{n-2}^0 +\epsilon \hat t\left(4 r(\hat t)\uns_r -e_2 \hat t\right)(n-1)F_{n-3}^0\nonumber\\
+&\left(4 r(\hat t)\hat t \uns_r^2-2{\hat t}^2 \uns_r e_2+\hat t(q-1)(e_4+2(\hat t-2)p_4)\right)F_{n-4}^0 \ ,
\end{align}
where we defined the operator $\hat t =q\exp(\epsilon (q-1) \hat S)$. Solving the recurrence relation we obtain that the coefficients $c_n$ have the following structure.
\begin{equation}
\label{eq:cnPVI}
c_n^{PVI} = \sum_{2j+k+2i_1+4i_2+6i_3+h+l=n}a_{jkli_1i_2i_3h} \uns^j q^k {\left(\frac{e_2}{4}\right)}^{i_1} {\left(\frac{e_4}{2}\right)}^{i_2} {e_6}^{i_3}{p_4}^h \left(\frac{\epsilon}{2}\right)^l\ ,
\end{equation}
i.e. they are homogeneous polynomials of all the parameters of the theory $\uns,q,\vec{m}$, where the coefficients $a_{jkli_1i_2i_3h}$ are computed recursively using \eqref{PVI,5} and we verified that they are integers to very high order in $n$.
The values of the first \(c_n^{PVI}\)'s are written in the appendix~\ref{app:PVI}.

It is interesting to notice that this expansion actually reorganizes in a non-trivial way in the SW limit where the coefficients $c_n^{PVI}$ reduce just to the Weierstrass sigma coefficients $c_n^\sigma$,
after we removed the contact term Gaussian prefactor. This implies that in the SW limit the coefficients $c_n$ are completely fixed by the invariants $g_2,g_3$ of the SW curve and by the contact term $T$.

\subsection{PIV alias Argyres-Douglas $H_2$}
As a first example of AD theory we consider $H_2$ which is obtained from the $N_f=3$ theory (PV equation) taking the coalescence limit, i.e. namely running the renormalization group flow from $N_f=3$ to the AD $H_2$ singularity. The resulting theory has two relevant massive deformations $\tilde m$ and $m_-$. The operator $u$ has dimension $[u]=3/2$ and the corresponding coupling $\Lambda$ has dimension $[\Lambda]=1/2$. The SW curve is
\begin{equation}
y^2 = x^2 + 2\Lambda x+(2\tilde m+\Lambda^2)+\frac{u+2\Lambda m_-}{x}+\frac{m_-^2}{x^2}\ .   
\end{equation}
The correspondent Painlevé equation is PIV. Its Hirota bilinear form reads\footnote{The equation we use is a reparametrization of the one in \cite{hone2017hirota}. Our parametrization is convenient to take the autonomous limit of PIV and recover the SW theory for $H_2$.}
\begin{equation}\label{PIV,0}
\HirD{\Tau}{t}{4} - (t^2 - 8(\theta_\infty+3\theta_0)) \HirD{\Tau}{t}{2} + t \pdv{t}\Tau^2 + 
 32\theta_0 (\theta_0+\theta_\infty) \Tau^2 \ ,
\end{equation}   
where $\theta_0, \theta_{\infty}$ are the monodromy parameters. The sigma-form is
\begin{equation}
\ddot\zeta^2 - (t\dot\zeta - \zeta )^2 + 4\dot\zeta (\dot\zeta + 4\theta_0 ) (\dot\zeta+ 2\theta_0 + 2\theta_{\infty})=0 \ ,
\end{equation}
and the autonomous limit is obtained with the following change of variables
\begin{align}
&t=t_0+\epsilon^{\frac{1}{2}}s\ , \quad \theta_0 = \frac{m_-}{\epsilon} \ , \quad \theta_\infty=-\frac{\tilde m}{\epsilon}\ , \quad \Lambda = \epsilon^{\frac{1}{2}}t_0 \ , \\
&\Tau(s)=e^{\frac{\uns s}{\epsilon}}\mathcal{B}_{NS}(s) \ , \quad \zeta = \epsilon^{-\frac{1}{2}}\eta+\epsilon^{-\frac{3}{2}}\uns \ .   
\end{align}
In these variables the equation \eqref{PIV,0} reads
\begin{align}
&\HirD{\mathcal{B}_{NS}}{s}{4}+(12T+2\Lambda\epsilon s+\epsilon^2 s^2)\HirD{\mathcal{B}_{NS}}{s}{2}\nonumber\\
&+\epsilon(\Lambda+\epsilon s)\pdv{s} \mathcal{B}_{NS}^2+2\uns\epsilon s\mathcal{B}_{NS}^2+(12T^2-\gII){\mathcal{B}_{NS}}^2=0 \ ,
\end{align}
the sigma-form becomes
\begin{equation}
\ddot\eta^2 - ((\Lambda+\epsilon s)\dot\eta - \epsilon \eta-\uns )^2 +4\dot\eta (\dot\eta + 4m_- ) (\dot\eta+ 2m_- - 2\tilde m)=0 \ ,
\end{equation}
and the values of the parameters of the autonomous equation \eqref{HRZ,0} are
\begin{align}
&T^{PIV} = -\frac{\Lambda^2}{12}+2m_- -\frac{2}{3}\tilde m \ , \quad g_2^{PIV} = 16 m_-^2 - 2 u \Lambda - 4 m_- \Lambda^2 + 
 \frac{1}{12}(8 \tilde m + \Lambda^2)^2 \ , \\
 &g_3^{PIV}=-u^2-2 T^{PIV} u \Lambda+16 m_-^2 T^{PIV} - 4 (T^{PIV})^3  - 
 4 m_- T^{PIV} \Lambda^2 + \frac{1}{12} T^{PIV} (8 \tilde m + \Lambda^2)^2\ .
\end{align}
Let us observe that all the dependence on the masses is reabsorbed in the parameters $T,g_2,g_3$.
The recursion relation for the Hurwitz expansion \eqref{HRZ,1} is
\begin{align*}
&\frac{2 n(n^2-1)(n-6)}{(n+1)!}c_n = \\
&=-{F'}_n^4-12TF_{n-2}^2+\Lambda\epsilon(2F_{n-3}^2-(n-1)F_{n-3}^0)+\epsilon^2(F_{n-4}^2-(n-2)F_{n-4}^0)\\
& \quad +(\gII-12T^2)F_{n-4}^0-2\uns\epsilon F_{n-5}^0 \ ,    
\end{align*}
and the coefficients of the Hurwitz expansion have the following structure
\begin{equation}
\label{eq:cnPIV}
c_n^{PIV} = \sum_{4j+6k+2p+q+3r+2l=n}a_{jkpqrl} \left(\frac{\gII}{2}\right)^j\left(2\gIII\right)^k T^p \uns^r \Lambda^q \epsilon^l\ .
\end{equation}
As in the previous cases, the coefficients $a_{jkpql}$ are integers.
Values of these coefficients are written in the appendix~\ref{app:PIV}.
In the autonomous limit we recover again the Weierstrass $\sigma$-function coefficients $c_n^\sigma$ (up to contact term).

A similar analysis can be done for the AD theory $H_1$, decoupling one of the mass parameters. The corresponding Painlevé equation is PII and the structure of the expansion remains the same and we have only to change the parametrizations of $g_2,g_3,T$.

\subsection{PIII$_2$ alias $N_f=1$}
As an example of how we can get the AD theory from the coalescence limit of Painlevé equations, we consider now $N_f=1$ theory which is the simplest $SU(2)$ Lagrangian theory with matter. The corresponding Painlevé equation is PIII$_2$. 

The theory has an AD point where extra massless mutually non-local d.o.f. appear, and for this reason the theory at this point corresponds to an \emph{interacting} Coulomb phase. This is the AD theory $H_0$, and the corresponding Painlevé equation is PI. In this subsection we will show how we can obtain it from a limit of PIII$_2$ equation. In the following subsection we will study then in detail the $H_0$ theory using the PI equation.

We start now the analysis of PIII$_2$. The SW curve for $N_f=1$ is
\begin{equation}
y^2=(x^2-u)^2-4\Lambda^3(x+m)  \ ,
\end{equation}
 where $m$ is the mass of the hypermultiplet. The Hirota bilinear equation for PIII$_2$ is \cite{Bershtein_2015}
 \begin{equation}\label{PIII_2,0}
 \HirD{\Tau}{\log t}{4}-2t\frac{\partial }{\partial t}\HirD{\Tau}{\log t}{2}+\HirD{\Tau}{\log t}{2}+4t\Tau ^2=0 \ ,
\end{equation} 
and the sigma form equation for the hamiltonian $\zeta =t\partial_t \log \Tau$ is
\begin{equation}\label{PIII_2,1}
{(t\ddot\zeta)}^2 - 4{\dot\zeta}^2 (\zeta- t\dot\zeta) + 4\dot\zeta=\frac{1}{\theta^2} \ .
\end{equation}
We observe that the mass parameter $\theta=m/\epsilon$ does not appear directly in the Hirota form, but it appears in the parametrization of the elliptic invariants $g_2,g_3$ and in the sigma-form equation.

We now proceed with the derivation of the Hurwitz expansion. As usual, to obtain the autonomous limit we consider a time $t_0$ and we change variable as follows
\begin{equation}
    t= t_0e^{\epsilon s}\ , \quad \theta=\frac{m}{\epsilon}\ ,  \quad m\Lambda^3=\epsilon^4 t_0\ ,\quad \Tau(s) =e^{\frac{\uns s}{\epsilon}}\mathcal{B}_{NS}(s)\ , \quad \zeta(s)=\epsilon^{-1}\eta(s)+\epsilon^{-2}\uns\ .
\end{equation}
The Hirota equation becomes
\begin{equation}\label{PIII_2,2}
    \HirD{\mathcal{B}_{NS}}{s}{4}-2\epsilon\frac{\partial }{\partial s}\HirD{\mathcal{B}_{NS}}{s}{2}+(\epsilon^2-4\uns)\HirD{\mathcal{B}_{NS}}{s}{2}+4m\Lambda^3 e^{\epsilon s}{\mathcal{B}_{NS}}^2=0 \ ,
\end{equation}
the sigma-form in autonomous variables is
\begin{equation}
{(\ddot\eta-\epsilon \dot\eta )}^2+ 4{\dot\eta}^3 - 4\epsilon\dot\eta^2\eta -4\uns\dot\eta^2+ 4m\Lambda^3 e^{\epsilon s}\dot\eta=\Lambda^6 e^{2\epsilon s}\ ,
\end{equation}
and taking the limit $\epsilon \to 0$ we can find the contact term $T$ and the elliptic invariants $g_2,g_3$
\begin{equation}
T^{PIII_2}=-\frac{u}{3} \ , \quad g_2^{PIII_2}=\frac{4}{3}u^2-4 m\Lambda^3\ , \quad g_3^{PIII_2}=-\frac{8}{27}u^3+\frac{4}{3}u m\Lambda^3 -\Lambda^6 \ .
\end{equation}

Inserting the ansatz \eqref{HRZ,1} in \eqref{PIII_2,2} we obtain the recursion for the coefficients $c_n$
\begin{equation}\label{PIII_2,22}
\frac{2 n(n^2-1)(n-6)}{(n+1)!}c_n = -{F'}_n^4+2\epsilon(n-1)F_{n-1}^2-(\epsilon^2-4\uns)F_{n-2}^2-4m\Lambda^3e^{\epsilon \hat S}F_{n-4}^0 \ ,    
\end{equation}
where $\hat S$ is the shift operator\footnote{Notice that $F^k_n=0$ for $n<0$ therefore for fixed $n$ the action of $e^{\epsilon \hat S}$ is truncated.} defined in \eqref{HRZ,1.5}. Therefore the polynomials $c_n$ have the following\footnote{We observe that using the expression for $g_2$ the dependence on $m,\Lambda$ can be eliminated. Therefore, the expansion depends only on the parameters $g_2,g_3,T$ and $\epsilon$.} structure
\begin{equation}
\label{eq:cnPIII}
c_n^{PIII_2} = \sum_{4j+6k+2p+l=n}a_{jkpl} \left(\frac{\gII}{2}\right)^j\left(2\gIII\right)^k \T^p\left(\frac{\epsilon}{2}\right)^l\ ,
\end{equation}
and again numerically the coefficients $a_{jklp}$ turn out to be integers, see appendix~\ref{app:PIII}.

As anticipated, an interesting feature of the Hurwitz expansion of the blowup factor $\mathcal{B}_{NS}$ is that it is defined in terms of the IR parameters $\uns,\Lambda$ and therefore we can easily study the theory around strongly coupled points. In particular, we are interested in the AD superconformal point of $N_f=1$ which corresponds to PI equation.

The PI equation can be obtained from PIII$_2$ as follows. First we remove the Gaussian contribution of the contact term\footnote{This corresponds precisely to the contribution of the contact term coming from the states that are integrated out.}
\begin{equation}
\mathcal{B}_{NS}^{PIII_2}(s)=e^{\frac{\uns}{6}s^2}\mathcal{B}_{NS}^{PI}(m^{\frac{4}{5}}s) \ , 
\end{equation}
and then we change variables\footnote{As we will see later, in the $H_0$ theory we have $g_3^{PI}=-2u, g_2^{PI}=c$, where $c$ is the coupling of the $u$ observable. Therefore only $g_3$ is affected by the NS $\epsilon$ corrections.}
\begin{equation}\label{PIII_2,2.5}
s= m^{-\frac{4}{5}}\tilde s \ , \quad \uns = \frac{3}{\sqrt{2}}m^2 -\frac{1}{2\sqrt{2}}m^{\frac{6}{5}}g_2^{PI}-\frac{1}{4} m^{\frac{4}{5}}\gIII^{PI} \ ,\quad \Lambda^3 =\frac{3}{2}m^3 - \frac{3}{4} m^{\frac{11}{5}}g_2^{PI}\ ,
\end{equation}
which implies
\begin{equation}
g_2^{PIII_2}=m^{\frac{16}{5}}(g_2^{PI}+O(m^{-\frac{2}{5}})) \ , \quad g_3^{PIII_2}=m^{\frac{24}{5}}(g_3^{PI}+O(m^{-\frac{2}{5}})) \ .
\end{equation}
After the change of variables the equation \eqref{PIII_2,2} becomes (renaming $\tilde s \to s$ for simplicity) 
\begin{equation}
\HirD{\mathcal{B}_{NS}}{s}{4}+2\epsilon s \mathcal{B}_{NS}^2 +g_2^{PI}(2-3e^{\frac{\epsilon s}{m^{4/5}}})\mathcal{B}_{NS}^2+(6(e^{\frac{\epsilon s}{m^{4/5}}}-1)m^{4/5}-4\epsilon s)\mathcal{B}_{NS}^2+O(m^{-\frac{2}{5}})= 0 \ ,
\end{equation}
and in the limit $m \to \infty$ we obtain the PI Hirota equation 
\begin{equation}\label{PIII_2,3}
\HirD{\mathcal{B}_{NS}}{s}{4}+2\epsilon s \mathcal{B}_{NS}^2 -g_2^{PI}\mathcal{B}_{NS}^2= 0 \ .
\end{equation}
We observe that the parameters of the AD theory have fractional dimensions $[g_2^{PI}]=4/5, [g_3^{PI}]=6/5$ and that all the contact term contributions are subleading and decouple in the limit $m\to \infty$. This means that in the AD point we don't have contributions to the contact term coming from the extra massless degrees of freedom and the contact term of $H_0$ theory vanishes, $T^{PI}=0$.

\subsection{PI alias Argyres-Douglas $H_0$}
In the previous subsection we obtained the AD theory from holomorphic decoupling of the $N_f=1$ theory and we showed that the corresponding Painlevé equation is PI. In this section we will study in detail the Hurwitz expansion for this theory. This expansion was the one originally studied by HRZ in \cite{hone2013properties} and has more special properties with respect to the Lagrangian one because its SW curve is
\begin{equation}\label{PI,0}
y^2=4x^3-cx+2u \ ,
\end{equation}
therefore the parameters of the theory are exactly the elliptic invariants $g_2=c, g_3=-2u$. Furthermore, as we have shown before, the contact term vanishes, $T=0$. This implies that the structure of the Hurwitz expansion is particularly simple.

The parameter $u$ is the Coulomb branch parameter and has dimension $[u]=6/5$, and $c$ is the gauge coupling which corresponds to the source associated to the operator $u$ for the AD theory and has dimension $[c]=4/5$. The dimension of the source $s$ of the surface observable can be fixed from the fact that $[us/\epsilon]=0$, which implies $[s]=-1/5$.

We start now the analysis. The PI Hirota bilinear equation is \cite{hone2013properties}
\begin{equation}
\HirD{\Tau}{t}{4}+2t\Tau^2 = 0 \ ,
\end{equation}
and its sigma form for the hamiltonian $\zeta(t)=\partial_t \log \Tau$ is
\begin{equation}
\ddot\zeta(t)^2+4\dot\zeta(t)^3-2(\zeta-t\dot\zeta)=0\ .   
\end{equation}
We now consider a zero $t_0$ of $\Tau$ and changing variables\footnote{As already observed for PIV, for AD theories the effect of the surface observable is an \emph{additive} shift. From the analysis of PIII$_2$ this can be understood as a consequence of the decoupling limit \eqref{PIII_2,2.5} in which the shifted coupling $\Lambda \exp(\epsilon s)$ is truncated due to the scaling of $s$. }
\begin{equation}
t = \epsilon^{1/5} s+t_0\ , \quad c = -2\epsilon^{4/5}t_0 \ , \quad \Tau(s) =e^{\frac{\uns s}{\epsilon}}\mathcal{B}_{NS}\ , \quad  \zeta(s) =\epsilon^{-\frac{1}{5}}\eta(s)+ \epsilon^{-\frac{6}{5}}\uns \ ,   
\end{equation}
we get
\begin{equation}
\HirD{\mathcal{B}_{NS}}{s}{4}+2\epsilon s \mathcal{B}_{NS}^2 -c\mathcal{B}_{NS}^2= 0 \ ,
\end{equation}
which agrees with the result obtained from the coalescence limit \eqref{PIII_2,3}. The sigma-form equation becomes
\begin{equation}
\ddot\eta^2+4\dot\eta^3-2\epsilon (\eta-s\dot\eta)-c\dot\eta-2\uns=0\ .
\end{equation}
As usual, in the limit $\epsilon \to 0$, the equations become autonomous and reduces to the form \eqref{HRZ,0}, \eqref{HRZ,0'} with the following parameters
\begin{equation}
 g_2^{PI} = c \ , \quad g_3^{PI}=-2u\ , \quad T^{PI}=0\ ,   
\end{equation}
which agrees with the SW result \eqref{PI,0}. 
In this case, as already observed, the contact term vanishes, $T^{PI}=0$. 

We observe that the equation for $\mathcal{B}_{NS}$ is independent on $\uns$. This is due to the fact that the PI equation is invariant under the ``gauge transformations'' of $\Tau$
\begin{equation}
\Tau \to \Tau' = e^{a+bs}\Tau \ ,    
\end{equation}
as $D_s^{(k)}$ is covariant with respect to the gauge transformations
\begin{equation}
\HirD{\Tau'}{s}{k}=e^{2a+2bs}\HirD{\Tau}{s}{k} \ ,    
\end{equation}
which can be easily seen from the very definition of the Hirota derivative \eqref{HRZ,-1}. 

Substituting the general ansatz \eqref{HRZ,1} we obtain the following recursion relation for the coefficients $c_n$
\begin{equation}
\frac{2 n(n^2-1)(n-6)}{(n+1)!}c_n = -{F'}_n^4+g_2 F_{n-4}^0-2\epsilon F_{n-5}^0 \ .
\end{equation}
The coefficients are homogeneous polynomials in the modular parameters
\begin{equation}
\label{eq:cnPI}
c_n^{PI} = \sum_{4j+6k+5l=n}a_{jkl} \left(\frac{g_2}{2}\right)^j\left(2\gIII\right)^k\epsilon^l\ ,
\end{equation}
and numerically it can be verified that the coefficients $a_{mnl}$ are integers, see appendix~\ref{app:PI}. As usual, in the autonomous limit $\epsilon \to 0$ the coefficients $c_n^{PI}$ reduces to the Weierstrass $\sigma$-function coefficients.

We observe that this expansion is exactly centered around the AD superconformal point which corresponds to the cubic singularity $g_2=0,g_3=0$. We notice that the structure of the Hurwitz expansion for AD theory $H_0$ is particular simple due to the absence of the contact term.

\subsection{PIII$_3$ alias $N_f=0$}
As a last example we consider the $SU(2)$ pure gauge theory. The SW curve of this theory is 
\begin{equation}\label{PIII_3,-1}
y^2=(x^2-u)^2-4\Lambda^4 \ .
\end{equation}
The pure gauge theory corresponds to Painlevé PIII$_3$ and the Hirota bilinear equation for PIII$_3$ is the same as PIII$_2$
 \begin{equation}\label{PIII_3,0}
 \HirD{\Tau}{\log t}{4}-2t\frac{\partial }{\partial t}\HirD{\Tau}{\log t}{2}+\HirD{\Tau}{\log t}{2}+4t\Tau ^2=0 \ .
\end{equation}   
The hamiltonian is $\zeta =t\partial_t \log \Tau$, and the corresponding sigma-form of the equation is
\begin{equation}\label{PIII_3,1}
{(t\ddot\zeta)}^2 - 4{\dot\zeta}^2 (\zeta- t\dot\zeta) + 4\dot\zeta=0 \ ,
\end{equation}
obtained from the sigma form \eqref{PIII_2,1} of PIII$_2$ in the limit $\theta \to \infty$ which is the coalescence limit PIII$_2$ $\to$ PIII$_3$. From the gauge theory point of view, this limit corresponds exactly to the holomorphic decoupling $N_f=1 \to N_f=0$ obtained integrating out the hypermultiplet
\begin{equation}
\theta=\frac{m}{\epsilon}\ ,\quad m \to \infty\ , \quad m\Lambda_{N_f=1}^3 = \Lambda_{N_f=0}^4\ .
\end{equation}
To obtain the autonomous limit, similarly to PIII$_2$, we change variables as follows
\begin{equation}
    t= t_0e^{\epsilon s}\ ,  \quad \Lambda^4=\epsilon^4 t_0\ ,\quad  \Tau(s) =e^{\frac{\uns s}{\epsilon}}\mathcal{B}_{NS}(s)\ , \quad \zeta(s)=\epsilon^{-1}\eta(s)+\epsilon^{-2}\uns\ ,
\end{equation}
and the equation \eqref{PIII_3,0} becomes
\begin{equation}\label{PIII_3,2}
\HirD{\mathcal{B}_{NS}}{s}{4}-2\epsilon\frac{\partial }{\partial s}\HirD{\mathcal{B}_{NS}}{s}{2}+(\epsilon^2-4\uns)\HirD{\mathcal{B}_{NS}}{s}{2}+4\Lambda^4 e^{\epsilon s}{\mathcal{B}_{NS}}^2=0  
\end{equation}
with $\Lambda^4 = \epsilon^4 t_0$. The sigma-form equation reads
\begin{equation}\label{PIII_3,4}
{(\ddot\eta-\epsilon \dot\eta )}^2+ 4{\dot\eta}^3 - 4\epsilon\dot\eta^2\eta -4\uns\dot\eta^2+ 4\Lambda^4 e^{\epsilon s}\dot\eta=0 \ .    
\end{equation}
Again, in the limit $\epsilon \to 0$ the equations reduce to \eqref{HRZ,0},\eqref{HRZ,0'} with
\begin{equation}
\quad g_2^{PIII_3}=\frac{4}{3}u^2-4\Lambda^4 \ , \quad g_3^{PIII_3}=-\frac{8}{27}u^3+\frac{4}{3}u\Lambda^4 \ , \quad T^{PIII_3}=-\frac{u}{3}  \ .
\end{equation}
The parameters $g_2,g_3$ are exactly the elliptic invariants of the SW curve \eqref{PIII_3,-1}.

Substituting the ansatz \eqref{HRZ,1} we obtain the following recursion
\begin{equation}
\frac{2 n(n^2-1)(n-6)}{(n+1)!}c_n = -{F'}_n^4+2\epsilon(n-1)F_{n-1}^2-(\epsilon^2-4\uns)F_{n-2}^2-4\Lambda^4 e^{\epsilon \hat S}F_{n-4}^0 \ .    
\end{equation}
The coefficients $c_n$ are homogeneous polynomials of the parameters of the equation
\begin{equation}
\label{eq:cnPIII3}
c_n^{PIII_3} = \sum_{4j+6k+2p+l=n}a_{jkpl} \left(\frac{\gII}{2}\right)^j\left(2\gIII\right)^k \T^p \left(\frac{\epsilon}{2}\right)^l\ ,
\end{equation}
and $a_{jkpl}$ numerically are integer coefficients. Finally, in the autonomous limit the contact term $T$ contribution reduces to a Gaussian prefactor and can be removed so that the coefficients reduce to the Weierstrass $sigma$-function coefficients. We observe that we can easily study this expansion around the monopole or dyon point simply setting $u=\pm 2\Lambda^2$. At this point the expansion should simplify because the Weierstrass $\sigma$ reduces to a trigonometric function.

\newpage

\appendix
\section{Weierstrass elliptic functions} \label{appA}
In this section we review the theory of Weierstrass elliptic functions, we will follow the conventions of \cite{abramowitz+stegun}. Consider the lattice $\Lambda=\{2n \omega_1+2m\omega_2 \vert\ n,m \in \mathbb{Z}\}$ generated by the half-periods $\omega_1, \omega_2 \in \mathbb{C}$ which are linearly independent in $\mathbb{R}$. The Weierstrass elliptic function is 
\begin{equation}
\wp(z|\omega_1,\omega_2) \equiv \wp(z) \equiv \frac{1}{z^2} +\sum_{w \in \Lambda \setminus \{0\}} \frac{1}{(z-w)^2}-\frac{1}{w^2} \ .
\end{equation}
This function has a double pole in each point of the lattice $\Lambda$ and is doubly periodic with periods $2\omega_1, 2\omega_2$
\begin{equation}
\wp(z+2\omega_1) = \wp(z+2\omega_2) =\wp(z) \ .
\end{equation}
It is useful to define also the Eisenstein series
\begin{equation}
G_{2k}(\omega_1,\omega_2) =\sum_{w \in \Lambda \setminus \{0\}} \frac{1}{w^{2k}} \ .
\end{equation}
If we normalize the lattice using the generators $(1,\tau)$, $\tau = \omega_2/\omega_1$, then the Eisenstein series can be seen as function of the modular parameter $\tau$
\begin{equation}
G_{2k}(\tau) =\sum_{(n,m) \neq {0,0}} \frac{1}{(n+m\tau)^{2k}} \ , \quad G_{2k}(\omega_1,\omega_2) = (2\omega_1)^{-2k}G_{2k}(\tau) \ .
\end{equation}
For $2k>2$ $G_{2k}$ is a modular form of weight $2k$, which means that under a modular transformation of $\tau$ it transforms as
\begin{equation}
G_{2k}(A\tau) = (c\tau+d)^{2k}G_{2k}(\tau) \ , \quad A\tau = \frac{a\tau+b}{c\tau+d}\ , 
\quad A=
\begin{pmatrix}
a & b \\
c & d
\end{pmatrix}\in SL(2,\mathbb{Z})
\end{equation}
for $2k=2$ instead $G_2(\tau)$ is not a modular form, but a quasi-modular form, which means that under a modular transformation it get also a shift
\begin{equation}
G_2(A\tau) = (c\tau+d)^2\left(G_2(\tau)-\frac{2\pi ic}{c\tau+d}\right) \ . 
\end{equation}
It turns out that the Eisenstein series $G_{4},\ G_{6}$ are the generators for the space of the modular forms with respect to \(SL(2,\mathbb{Z})\). Therefore any $G_{2k}$ can be written as a polynomial in $G_4,G_6$. Finally, it is useful also to define the normalized\footnote{The normalization is chosen in such a way that the first coefficient of the Fourier series of $G_{2k}$ in terms of the nome $q=\exp(2\pi i \tau)$ is 1.} Eisenstein series
\begin{equation}
E_{2k}(\tau) = \frac{G_{2k}(\tau)}{2\zeta(2k)},
\end{equation}
where $\zeta(s)$ is the Riemann Zeta function. The Eisenstein series are strictly related to the Weierstrass elliptic function. Precisely, defining the modular invariants $g_2 = 60 G_4,\ g_3 = 140 G_6$, it turns out that the Weierstrass $\wp$-function satisfies the following differential equation
\begin{equation}\label{ODEWeier}
\dot\wp^2(z) = 4\wp^3(z)-g_2 \wp(z) -g_3 \ ,
\end{equation}
which gives a parametrization for an elliptic curve in the Weierstrass form
\begin{equation}
y^2 = 4x^3-g_2x-g_3 \ ,
\end{equation}
where $\dot\wp = \partial_z \wp$. Another form of the differential equation \eqref{ODEWeier}, which is useful in connection with the study of the $\Tau$-function, is obtained taking the derivative\footnote{Notice that in this way the parameter $g_3$ becomes an integration constant.} of \eqref{ODEWeier}
\begin{equation}\label{ODEWeier2}
 \ddot\wp(z) = 6\wp^2(z)-\frac{g_2}{2}\ . 
\end{equation}
We can express the $\wp$-function also as a function of the invariants $g_2, g_3$, in this case\footnote{It can be shown that the elliptic invariants $g_2,g_3$ and the half-periods $\omega_1,\omega_2$ are equivalent data so we can always perform the change of variables $\omega_i(g_2,g_3), i=1,2$.} we denote it as $\wp(z;g_2,g_3)$. 

From the $\wp$-function we can define the Weierstrass Zeta function $\zeta_W(z;g_2,g_3)$ as
\begin{equation}
-\dv{}{z}\zeta_W(z;g_2,g_3)\equiv \wp(z;g_2,g_3) \ .
\end{equation}
The function $\zeta_W(z;g_2,g_3)\equiv \zeta_W(z)$ is quasi-periodic in the following sense
\begin{equation}
\zeta_W(z+2\omega_1) = \zeta_W(z) +2\eta_1 \ , \quad  \zeta_W(z+2\omega_2) = \zeta_W(z) +2\eta_2 \ ,
\end{equation}
the constants $\eta_1, \eta_2$ are the half-quasi-periods and are given by
\begin{equation}
\eta_1 = \zeta_W(\omega_1), \quad \eta_2 = \zeta_W(\omega_2) \ ,
\end{equation}
which follows from the periodicity of $\wp$ and from the fact that $\zeta_W$ is a odd function. Finally, the quasi-period $\eta_1$ is directly related to the second Eisenstein series
\begin{equation}
\eta_1 = \frac{\pi^2}{6}\frac{E_2(\tau)}{2\omega_1} \ .
\end{equation}
From the Weierstrass Zeta function we can define the Weierstrass $\sigma$-function
\begin{equation}
\dv{}{z}\log\sigma(z;g_2,g_3)=\zeta_W(z;g_2,g_3) \ .
\end{equation}
The sigma function is an entire function with simple zeros at the lattice points and it is related to the Jacobi theta functions. Precisely, the following relation holds
\begin{equation}\label{2.1}
\sigma(z;\omega_1,\omega_2) = \frac{2\omega_1}{\pi} e^{\frac{\eta_1 z^2}{2\omega_1}}\frac{\theta_1(v)}{\theta_1'(0)} = \frac{2\omega_1}{\pi} e^{\frac{\pi^2}{6}\frac{E_2(\tau)}{4\omega_1^2}z^2}\frac{\theta_1(v)}{\theta_1'(0)} \ , \quad \theta_1'(0)=2\eta(\tau)^3 \ ,
\end{equation}
where $v= \pi z/2\omega_1$, $\eta(\tau)$ is the Dedekind eta function, and $\theta_1$ is the theta function defined as
\begin{equation}\label{2.2}
\theta_1(z,q) \equiv  \theta_1(z) \equiv -\sum_{n\in \mathbb{Z}+\frac{1}{2}} (-1)^{n} q^{n^2} e^{2niz} \ .
\end{equation}
In the above expression we have defined the nome $q=\exp(i\pi \tau)$, where $\tau =\omega_2/\omega_1$ is the modular parameter (with $\Im \tau>0$) and where the variable $z$ is the argument of the theta function. For convenience we define also the theta constants
\begin{equation}\label{tc}
\theta_2(\tau)=\sum_{n\in \mathbb{Z}+\frac{1}{2}}e^{i\pi \tau n^2}\ , \quad \theta_3(\tau)=\sum_{n\in \mathbb{Z}}e^{i\pi \tau n^2}, \quad 
\theta_4(\tau)=\sum_{n\in \mathbb{Z}}(-1)^ne^{i\pi \tau n^2}\ .
\end{equation}

The Weierstrass sigma satisfies the Hirota bilinear equation\footnote{As said previously $g_3$ appears in \eqref{ODEWeier2} as an integration constant. In the Hirota bilinear equation for $\sigma$ this corresponds exactly to the resonant coefficient $c_6^\sigma$.} 
\begin{equation}\label{HRZapp,0}
\HirD{\sigma}{z}{4}-g_2\sigma^2(z)= 0 \ ,
\end{equation}
which follows expressing $\wp$ in terms of $\sigma$ in the differential equation \eqref{ODEWeier2}.

Finally, the sigma function admits the following Hurwitz expansion around the origin
\begin{equation}\label{sigmaHRZ}
\sigma(z;g_2,g_3) = \sum_{m,n=0}^{\infty} a_{mn} \left(\frac{g_2}{2}\right)^m\left(2g_3\right)^n \frac{z^{4m+6n+1}}{(4m+6n+1)!} \ ,
\end{equation}
where $a_{00}=1$, $a_{mn}=0$ if some index is negative, the other coefficients are given by the following recurrence relation
\begin{equation}
a_{mn} = 3(m+1)a_{m+1,n+1}+\frac{16}{3}(n+1)a_{m-2,n+1}-\frac{1}{3}(2m+3n-1)(4m+6n-1)a_{m-1,n} \ ,
\end{equation}
and it can be shown that they are all integers\footnote{In general they can be negative. See \cite{onishi2010universalellipticfunctions}.}. We can rewrite the sigma function as
\begin{equation}\label{sigmaHRZ,1}
\sigma(z;g_2,g_3)= \sum_{n=0}^{+\infty}c_n^\sigma \frac{z^{n+1}}{(n+1)!} \ ,
\end{equation}
where we defined the coefficients
\begin{equation}
c_n^\sigma \equiv c_n^\sigma(g_2,g_3)= \sum_{4k+6l+1=n}a_{kl} \left(\frac{g_2}{2}\right)^k\left(2g_3\right)^l \ ,  
\end{equation}
and because $a_{nm}$ are integers we have that $c_n^\sigma$ are polynomials with integers coefficients, $c_n^\sigma \in \mathbb{Z}[g_2/2,2g_3]$.

\section{Computation of elliptic invariants and contact term for PVI}
\label{appB}
In this section we compute the elliptic invariants and the contact term of $N_f=4$ theory (PVI) in the massless case. In the massless case the prepotential for PVI is
\begin{equation}
\mathcal{F}_0=i\pi\tau a^2\ ,
\end{equation}
where $\tau$ is the IR coupling. The IR coupling $\tau$ is related to the UV coupling $\tau_0$ as follows
\begin{equation}
q=\exp(i\pi \tau_0)=\frac1{16}\frac{\theta_2^4(\tau)}{\theta_3^4(\tau)} = k^2(\tau) \ ,
\end{equation}
and we compute the contact term as
\begin{equation}
\tilde T=-q\pdv{q}u =-\left(q\pdv{q}\right)^2\mathcal{F}_0 \ .  
\end{equation}
To do the computation we use the following identities\footnote{See \cite{kiritsis1997introduction} appendix A.} from the theory of elliptic functions
\begin{align}
&i\pi q\pdv{q}\tau = \theta_4^{-4}(\tau)\ , \label{theta,0}  \\\
&\pdv{\tau}\log \theta_4(\tau) =\frac{i\pi}{12}(E_2(\tau) - \theta_2(\tau)^4-\theta_3(\tau)^4) \label{theta,1} \ , \\
& \frac{\theta_2^4(\tau)}{\theta_4^4(\tau)}=\frac{q}{1-q}\ , \quad \frac{\theta_3^4(\tau)}{\theta_4^4(\tau)}=\frac{1}{1-q} \label{theta,1.5} \ ,
\end{align}
where $E_2(\tau)$ is the second Eisenstein series. Using \eqref{theta,0} we have
\begin{equation}
u=q\pdv{q}\mathcal{F}_0=i\pi q\pdv{\tau}{q} a^2= \theta_4^{-4}(\tau) a^2 \ , \quad \frac{i\pi}{\omega_1}=\pdv{u}{a}= 2\theta_4^{-4}(\tau)a  \label{theta,2} \ ,
\end{equation}
and
\begin{equation}
\tilde T= -q\pdv{q}u = - a^2 q\pdv{\tau}{q}\pdv{}{\tau}\theta_4^{-4}(\tau) = \frac{4a^2}{i\pi}(\theta_4^{-4}(\tau))^2 \pdv{}{\tau}\log \theta_4(\tau) \ .
\end{equation}
By using \eqref{theta,1},\eqref{theta,1.5} and \eqref{theta,2} we get
\begin{equation}
\tilde T= \frac{a^2}{3}(\theta_4^{-4}(\tau))^2 (E_2(\tau) - \theta_2(\tau)^4-\theta_3(\tau)^4)=-\frac{\pi^2}{12}\frac{E_2(\tau)}{\omega_1^2}-\frac{u}{3}\frac{1+q}{1-q}\ ,
\end{equation}
i.e.
\begin{equation}
T=\tilde T + \frac{\pi^2}{12}\frac{E_2(\tau)}{\omega_1^2}=-\frac{u}{3}\frac{1+q}{1-q} \ ,
\end{equation}
which agrees with the PVI result \eqref{PVI,4} in the massless case. Consider the elliptic invariants
\begin{align}
&g_2(\omega_1,\omega_2) = 60 G_4(\omega_1,\omega_2)=\frac{4}{3}\left(\frac{\pi}{2\omega_1}\right)^{4}E_4(\tau)=\frac{4}{3}\left(\frac{1}{2i}\pdv{u}{a}\right)^{4}E_4(\tau)\ , \nonumber\\
&g_3(\omega_1,\omega_2) = 140 G_6(\omega_1,\omega_2)=\frac{8}{27}\left(\frac{\pi}{2\omega_1}\right)^{6}E_6(\tau)=\frac{8}{27}\left(\frac{1}{2i}\pdv{u}{a}\right)^{6}E_6(\tau)\ . \label{theta,3}
\end{align}
We consider now the following identities\footnote{See \cite{kiritsis1997introduction} appendix F.}
\begin{align}
&E_4(\tau)=\frac{1}{2}(\theta_2^8(\tau)+\theta_3^8(\tau)+\theta_4^8(\tau))\ , \\
&E_6(\tau)=\frac{1}{2}(\theta_2(\tau)^4+\theta_3(\tau)^4)(\theta_3(\tau)^4+\theta_4(\tau)^4)(\theta_4(\tau)^4-\theta_2(\tau)^4)\ ,
\end{align}
and using \eqref{theta,1.5} and \eqref{theta,3} we obtain 
\begin{align}
&g_2(\omega_1,\omega_2)=\frac{4}{3}\frac{(q^2-q+1)}{(q-1)^2}   \ ,\quad g_3(\omega_1,\omega_2)= \frac{4}{27} \frac{(2-3q-3q^2+2q^3)}{(q-1)^3}u^3\ ,
\end{align}
which again agree with \eqref{PVI,4}.

\section{Details on the derivation of the holomorphic anomaly equations} \label{appD}

\subsection{Equivalence of holomorphic anomaly equations with and without sources}\label{appD,1}
We want to find the relation between the holomorphic anomaly equations for the SD generating function $ \mathcal{F}_{\mathrm{st}}^{x,s}(a, \Lambda)$ and the SD free energy $\widehat{\mathcal{F}}(a,\Lambda)$. To do this we observe that
\begin{equation}\label{hol,7}
 \mathcal{F}_{\mathrm{st}}^{x,s}(a, \Lambda)=\widehat{\mathcal{F}}(a+\epsilon x,\Lambda e^{\epsilon s})+\mathcal{F}_{0,\mathrm{st}}^{x,s}(a, \Lambda) \Rightarrow e^{\mathcal{F}_{\mathrm{st}}^{x,s}(a, \Lambda)}=e^{\mathcal{F}_{0,\mathrm{st}}^{x,s}(a, \Lambda)}e^{\epsilon x\partial_a+\epsilon s\Lambda\partial\Lambda}e^{\widehat{\mathcal{F}}(a,\Lambda)} \ .
\end{equation}
We want to prove the following relation
\begin{equation}\label{hol,8app}
\left(D_{E_2} + \frac{\epsilon^2}{24}\partial_a^2\right)e^{\widehat{\mathcal{F}}(a,\Lambda)}=e^{-\epsilon x\partial_a-\epsilon s\Lambda\partial_\Lambda}e^{-\mathcal{F}_{0,\mathrm{st}}^{x,s}(a, \Lambda)}\left(D_{E_2}+\frac{\partial_x^2}{24} \right)e^{\mathcal{F}_{0,\mathrm{st}}^{x,s}(a, \Lambda)}e^{\epsilon x\partial_a+\epsilon s\Lambda\partial_\Lambda}e^{\widehat{\mathcal{F}}(a,\Lambda)}\ ,
\end{equation}
where\footnote{In this appendix we omit the explicit dependence of the free energies on $\epsilon$ for simplicity.}
\begin{align}
&\widehat{\mathcal{F}}(a,\Lambda)=\mathcal{F}(a,\Lambda,\epsilon)-\frac{1}{\epsilon^2}\mathcal{F}_0(a,\Lambda)\ , \\ 
&\mathcal{F}_{0,\mathrm{st}}^{x,s}(a, \Lambda)=   \frac{1}{\epsilon^2}\mathcal{F}_0(a+\epsilon x, \Lambda_{\epsilon s})-\frac1{\epsilon^2} \mathcal{F}_0(a, \Lambda)-\frac{x}{\epsilon} \partial_a\mathcal{F}_0(a, \Lambda)-\frac{x^2}{2} \partial_a^2\mathcal{F}_0(a, \Lambda)\ ,
\end{align}
and $\Lambda_{\epsilon s}=\Lambda \exp(\epsilon s)$. We recall that we have the following algebra of vector fields~\eqref{eq:OpAlgebra}\footnote{In this section we remove \(\partial_{\tau}'\) terms in the commutators, since this algebra will act only on the stable expressions that do not have an explicit \(\tau\) dependence.}
\begin{equation}\begin{gathered}
\label{eq:2,app}
[\partial_a, \Lambda \partial_{\Lambda}] = 0\ ,\\
[D_{E_2}, \partial_a] =  \frac1{12} \partial_a^3\mathcal{F}_0(a,\Lambda) \widehat{d}\ , \\
[D_{E_2}, \Lambda\partial_\Lambda] =
\frac1{12} \Lambda \partial_\Lambda\partial_a^2\mathcal{F}_0(a,\Lambda) \widehat{d}
- \frac1{12} \Lambda \partial_\Lambda\partial_a\mathcal{F}_0(a,\Lambda) \partial_a\ ,\\
[\widehat{d}, \Lambda \partial_{\Lambda}] = 0\ ,\qquad
[\widehat{d}, \partial_a] = -\partial_a\ ,\qquad
[\widehat{d}, D_{E_2}] = -2 D_{E_2}\ ,
\end{gathered}\end{equation}
with the following action on the amplitudes
\begin{equation}\begin{gathered}
\label{eq:3}
D_{E_2}\Lambda\partial_\Lambda\mathcal{F}_0(a,\Lambda) = 0\ ,\quad
D_{E_2}\partial_a^3\mathcal{F}_0(a,\Lambda)= 0\ ,\quad  \widehat{d} \mathcal{F}_{g,n}(a,\Lambda) = -\frac12\delta_{g,1}\delta_{n,0} \ ,\ g+n>0,\\
D_{E_2}\mathcal{F}_{1,0}(a,\Lambda)=0\ ,\quad
D_{E_2}\mathcal{F}_{0,1}(a,\Lambda)=0\ .
\end{gathered}\end{equation}
To prove the result we need to compute the conjugation which appears in \eqref{hol,8app}. We consider first the conjugation by $\exp(\mathcal{F}_{0,\mathrm{st}}^{x,s}(a, \Lambda))$
\begin{align}&e^{-\mathcal{F}_{0,\mathrm{st}}^{x,s}(a,\Lambda)}
\left(D_{E_2}+\frac{\partial_x^2}{24} \right)e^{\mathcal{F}_{0,\mathrm{st}}^{x,s}(a,\Lambda)}= \\
&=D_{E_2}+D_{E_2}\mathcal{F}_{0,\mathrm{st}}^{x,s}(a,\Lambda)+\frac{1}{24}\partial_x^2+\frac{1}{12}\partial_x\mathcal{F}_{0,\mathrm{st}}^{x,s}(a,\Lambda)\partial_x+\frac{1}{24}\partial_x^2\mathcal{F}_{0,\mathrm{st}}^{x,s}(a,\Lambda)+\frac{1}{24}\left(\partial_x\mathcal{F}_{0,\mathrm{st}}^{x,s}(a,\Lambda)\right)^2\ .
\end{align}
Using the algebra \eqref{eq:2,app} and the conditions \eqref{eq:3} we obtain (see next subsection \ref{appD,genusZero})
\begin{equation}\label{hae0}
D_{E_2}\mathcal{F}_{0,\mathrm{st}}^{x,s}(a,\Lambda)= -\frac{1}{24}\left(\partial_x\mathcal{F}_{0,\mathrm{st}}^{x,s}(a,\Lambda)\right)^2 \ ,
\end{equation}
which implies
\begin{equation}\label{holapp,4}
e^{-\mathcal{F}_{0,\mathrm{st}}^{x,s}(a,\Lambda)}
\left(D_{E_2}+\frac{\partial_x^2}{24} \right)e^{\mathcal{F}_{0,\mathrm{st}}^{x,s}(a,\Lambda)}=D_{E_2}+\frac{1}{24}\partial_x^2+\frac{1}{12}\partial_x\mathcal{F}_{0,\mathrm{st}}^{x,s}(a,\Lambda)\partial_x+\frac{1}{24}\partial_x^2\mathcal{F}_{0,\mathrm{st}}^{x,s}(a,\Lambda)\ .    
\end{equation}
We consider now the conjugation by the shift operator $\exp(\epsilon x\partial_a + \epsilon s\Lambda\partial_{\Lambda})$ of \eqref{holapp,4}. We have
\begin{equation}\label{holapp,5}
e^{-\epsilon x\partial_a - \epsilon s\Lambda\partial_{\Lambda}}\partial_x e^{\epsilon x\partial_a + \epsilon s\Lambda\partial_{\Lambda}}=\partial_x+\epsilon\partial_a\ , 
\end{equation}
and
\begin{align}
e^{-\epsilon x\partial_a - \epsilon s\Lambda\partial_{\Lambda}}&\partial_x\mathcal{F}_{0,\mathrm{st}}^{x,s}(a,\Lambda) e^{\epsilon x\partial_a + \epsilon s\Lambda\partial_{\Lambda}}= \nonumber\\ 
&=\frac{1}{\epsilon}\left(\partial_a\mathcal{F}_0(a,\Lambda)-
\partial_a\mathcal{F}_0(a-\epsilon x,\Lambda_{-\epsilon s})-
\epsilon x \partial_a^2\mathcal{F}_0(a-\epsilon x,\Lambda_{-\epsilon s})\right), \nonumber \\
e^{-\epsilon x\partial_a - \epsilon s\Lambda\partial_{\Lambda}}&\partial_x^2\mathcal{F}_{0,\mathrm{st}}^{x,s}(a,\Lambda) e^{\epsilon x\partial_a + \epsilon s\Lambda\partial_{\Lambda}}=\partial_a^2\mathcal{F}_0(a,\Lambda)-\partial_a^2\mathcal{F}_0(a-\epsilon x,\Lambda_{-\epsilon s})\ . \label{holapp,6}
\end{align}
For $D_{E_2}$ we define
\begin{equation}
D^{x,s}_{E_2} = e^{-\epsilon x\partial_a - \epsilon s\Lambda\partial_{\Lambda}}
D_{E_2} e^{\epsilon x\partial_a + \epsilon s\Lambda\partial_{\Lambda}}\ ,
\end{equation}
and taking the derivatives with respect to $x$ we obtain
\begin{equation}\label{holapp,1}
\partial_x(D_{E_2}^{x,s}) =
\frac{\epsilon}{12} e^{-\epsilon x\partial_a -\epsilon s\Lambda\partial_{\Lambda}}
\partial_a^3\mathcal{F}_0(a,\Lambda) \widehat{d} e^{\epsilon x\partial_a +\epsilon s\Lambda\partial_{\Lambda}}\ .
\end{equation}
From the algebra \eqref{eq:2,app} we have
\begin{equation}
e^{-\epsilon x\partial_a - \epsilon s\Lambda\partial_{\Lambda}}\widehat{d} e^{\epsilon x\partial_a +\epsilon s\Lambda\partial_{\Lambda}}=\hat d-\epsilon x \partial_a\ ,
\end{equation}
then \eqref{holapp,1} reads
\begin{align}
&\partial_x\left(D_{E_2}^{x,s}\right) = \frac{\epsilon}{12} \partial_a^3\mathcal{F}_0(a-\epsilon x,\Lambda_{-\epsilon s})\left(\widehat{d}-\epsilon x\partial_a \right)\ . \label{holapp,2}
\end{align}
Analogously for the $s$ derivative we have
\begin{equation}\label{holapp,3}
\partial_s \left(D_{E_2}^{x,s}\right) =
\frac{\epsilon}{12} \Lambda_{-\epsilon s}\partial_\Lambda\partial_a^2\mathcal{F}_0(a-\epsilon x,\Lambda_{-\epsilon s})
\left(\widehat{d} -\epsilon x \partial_a\right)
-\frac{\epsilon}{12} \Lambda_{-\epsilon s} \partial_\Lambda\partial_a\mathcal{F}_0(a-\epsilon x,\Lambda_{-\epsilon s}) \partial_a\ ,
\end{equation}
and integrating \eqref{holapp,2} and \eqref{holapp,3} in $x$ and $s$ we obtain
\begin{multline}
\label{eq:4}
D^{x,s}_{E_2} =
-\frac{1}{12}\partial_a^2\mathcal{F}_0(a-\epsilon x,\Lambda_{-\epsilon s})\left(\widehat{d}-\epsilon x\partial_a \right)+
\frac{1}{12}\partial_a^2\mathcal{F}_0(a,\Lambda)\widehat{d}\\ 
+\frac{1}{12}\left(\partial_a\mathcal{F}_0(a-\epsilon x,\Lambda_{-\epsilon s})-\partial_a\mathcal{F}_0(a,\Lambda)\right)\partial_a + D_{E_2}\ .
\end{multline}
Finally, combining \eqref{eq:4} with \eqref{holapp,5}, \eqref{holapp,6} we obtain
\begin{align}
&e^{-\epsilon x\partial_a - \epsilon s\Lambda\partial_{\Lambda}}\left(D_{E_2}+\frac{1}{24}\partial_x^2+\frac{1}{12}\partial_x\mathcal{F}_{0,\mathrm{st}}^{x,s}(a,\Lambda)\partial_x+\frac{1}{24}\partial_x^2\mathcal{F}_{0,\mathrm{st}}^{x,s}(a,\Lambda)\right)e^{\epsilon x\partial_a + \epsilon s\Lambda\partial_{\Lambda}}e^{\widehat{\mathcal{F}}(a,\Lambda)} = \nonumber \\ 
&=\left(D_{E_2}+\frac{\epsilon^2}{24}\partial_a^2+\frac{1}{24}\left(\partial_a^2\mathcal{F}_0(a,\Lambda)-\partial_a^2\mathcal{F}_0(a-\epsilon x,\Lambda_{-\epsilon s}) \right)(2\widehat{d}+1)\right)e^{\widehat{\mathcal{F}}(a,\Lambda)}\ , \label{holapp,7}
\end{align}
where we used $\partial_x\widehat{\mathcal{F}}(a,\Lambda)=0$. From \eqref{eq:3} the last term in \eqref{holapp,7} vanishes and we obtain
\begin{equation}\label{hol,8app2}
e^{-\epsilon x\partial_a-\epsilon s\Lambda\partial_\Lambda}e^{-\mathcal{F}_{0,\mathrm{st}}^{x,s}(a, \Lambda)}\left(D_{E_2}+\frac{\partial_x^2}{24} \right)e^{\mathcal{F}_{0,\mathrm{st}}^{x,s}(a, \Lambda)}e^{\epsilon x\partial_a+\epsilon s\Lambda\partial_\Lambda}e^{\widehat{\mathcal{F}}(a,\Lambda)}=\left(D_{E_2} + \frac{\epsilon^2}{24}\partial_a^2\right)e^{\widehat{\mathcal{F}}(a,\Lambda)}\ .
\end{equation}

\subsection{Holomorphic anomaly at genus zero} \label{appD,genusZero}
In this subsection we want to derive the holomorphic anomaly equation for genus zero \eqref{hae0}. 
To do this it is convenient to consider first the $E_2$ derivative of the derivatives of $\mathcal{F}_{0,\mathrm{st}}^{x,s}(a,\Lambda)$ in order to apply \eqref{eq:3}. We have
\begin{align}
\partial_x^3 D_{E_2}\mathcal{F}_{0,\mathrm{st}}^{x,s}(a,\Lambda) &= D_{E_2} \partial_x^3 \mathcal{F}_{0,\mathrm{st}}^{x,s}(a,\Lambda) = \epsilon D_{E_2} \partial_a^3 \mathcal{F}_0(a+\epsilon x,\Lambda_{\epsilon s})=\nonumber\\
&=\epsilon D_{E_2}e^{\epsilon x\partial_a+\epsilon s\Lambda\partial\Lambda}\partial_a^3 \mathcal{F}_0(a,\Lambda)= \epsilon e^{\epsilon x\partial_a+\epsilon s\Lambda\partial\Lambda}D_{E_2}^{x,s}\partial_a^3 \mathcal{F}_0(a,\Lambda)\ .
\end{align}
and using \eqref{eq:4} and \eqref{eq:3} we obtain
\begin{multline}
\label{eq:9}
\partial_x^3 D_{E_2}\mathcal{F}_{0,\mathrm{st}}^{x,s}(a,\Lambda) = \frac{\epsilon}{12}e^{\epsilon x\partial_a+\epsilon s \Lambda\partial_{\Lambda}}\Big(-\partial_a^2\mathcal{F}_0(a-\epsilon x,\Lambda_{-\epsilon s})\left(\widehat{d}-\epsilon x\partial_a \right)+
\partial_a^2\mathcal{F}_0(a,\Lambda)\widehat{d}+
\\+
\left(\partial_a\mathcal{F}_0(a- \epsilon x,\Lambda_{-\epsilon s})-\partial_a\mathcal{F}_0(a,\Lambda)\right)\partial_a + 12D_{E_2}\Big) \partial_a^3\mathcal{F}_0(a,\Lambda)=\\=
\frac{\epsilon}{12}e^{\epsilon x\partial_a+\epsilon s \Lambda\partial_{\Lambda}}\Big(-\partial_a^2\mathcal{F}_0(a-\epsilon x,\Lambda_{-\epsilon s})\left(-3-\epsilon x\partial_a \right)-3
\partial_a^2\mathcal{F}_0(a,\Lambda)
\\+
\left(\partial_a\mathcal{F}_0(a- \epsilon x,\Lambda_{-\epsilon s})-\partial_a\mathcal{F}_0(a,\Lambda)\right)\partial_a \Big) \partial_a^3\mathcal{F}_0(a,\Lambda)=\\
\\=\frac{\epsilon}{12}\Big(3\partial_a^2\mathcal{F}_0(a,\Lambda)\partial_a^3\mathcal{F}_0(a+\epsilon x,\Lambda_{\epsilon s})+\epsilon x\partial_a^2\mathcal{F}_0(a,\Lambda)\partial_a^4\mathcal{F}_0(a+\epsilon x,\Lambda_{\epsilon s})\\
-3 \partial_a^2\mathcal{F}_0(a+\epsilon x,\Lambda_{\epsilon s})\partial_a^3\mathcal{F}_0(a+\epsilon x,\Lambda_{\epsilon s})
+\partial_a\mathcal{F}_0(a,\Lambda)\partial_a^4\mathcal{F}_0(a+\epsilon x,\Lambda_{\epsilon s})
\\
- \partial_a\mathcal{F}_0(a+\epsilon x,\Lambda_{\epsilon s}) \partial_a^4\mathcal{F}_0(a+\epsilon x,\Lambda_{\epsilon s})\Big)=\\
=
\frac{1}{12\epsilon^2}\partial_x^3 \left(-\frac12 \partial_a\mathcal{F}_0(a+\epsilon x,\Lambda_{\epsilon s})^2+ \epsilon x \partial_a^2\mathcal{F}_0(a,\Lambda) \partial_a\mathcal{F}_0(a+\epsilon x,\Lambda_{\epsilon s}) + \partial_a\mathcal{F}_0(a,\Lambda) \partial_a\mathcal{F}_0(a+\epsilon x,\Lambda_{\epsilon s})\right)\ .
\end{multline}
Analogously, for the \(s\) derivative we have
\begin{multline}
\partial_s D_{E_2}\mathcal{F}_{0,\mathrm{st}}^{x,s}(a,\Lambda) =\frac{1}{\epsilon} e^{\epsilon x\partial_a+\epsilon s\Lambda\partial\Lambda}D_{E_2}^{x,s}\Lambda\partial_\Lambda \mathcal{F}_0(a,\Lambda)\ = \\
=\frac{1}{12\epsilon}e^{\epsilon x\partial_a+\epsilon s \Lambda\partial_{\Lambda}}\Big(-\partial_a^2\mathcal{F}_0(a-\epsilon x,\Lambda_{-\epsilon s})\left(\widehat{d}-\epsilon x\partial_a \right)+
\partial_a^2\mathcal{F}_0(a,\Lambda)\widehat{d}
\\+
\left(\partial_a\mathcal{F}_0(a- \epsilon x,\Lambda_{-\epsilon s})-\partial_a\mathcal{F}_0(a,\Lambda)\right)\partial_a + 12D_{E_2}\Big) \Lambda\partial_\Lambda\mathcal{F}_0(a,\Lambda)
=\\=
\frac{1}{12\epsilon}e^{\epsilon x\partial_a+\epsilon s \Lambda\partial_{\Lambda}}\Big(\epsilon x\partial_a^2\mathcal{F}_0(a-\epsilon x,\Lambda_{-\epsilon s})+
\partial_a\mathcal{F}_0(a- \epsilon x,\Lambda_{-\epsilon s})-\partial_a\mathcal{F}_0(a,\Lambda)\Big) \Lambda\partial_\Lambda\partial_a\mathcal{F}_0(a,\Lambda)
=\\=
\frac{1}{12\epsilon}\Big(\epsilon x\partial_a^2\mathcal{F}_0(a,\Lambda)+
\partial_a\mathcal{F}_0(a,\Lambda)-
\partial_a\mathcal{F}_0(a+\epsilon x,\Lambda_{\epsilon s})\Big) \Lambda\partial_\Lambda\partial_a\mathcal{F}_0(a+\epsilon x,\Lambda_{\epsilon s})
=\\=
\frac{1}{12\epsilon^2}\partial_s \left(-\frac12 \partial_a\mathcal{F}_0(a+\epsilon x,\Lambda_{\epsilon s})^2+ \epsilon x \partial_a^2\mathcal{F}_0(a,\Lambda) \partial_a\mathcal{F}_0(a+\epsilon x,\Lambda_{\epsilon s}) + \partial_a\mathcal{F}_0(a,\Lambda) \partial_a\mathcal{F}_0(a+\epsilon x,\Lambda_{\epsilon s})\right)\ .
\end{multline}
Integrating back we find
\begin{multline}
D_{E_2}\mathcal{F}_{0,\mathrm{st}}^{x,s}(a,\Lambda)=
-\frac1{24\epsilon^2} \partial_a\mathcal{F}_0(a+\epsilon x,\Lambda_{\epsilon s})^2 
\\
+
\frac1{12 \epsilon} x \partial_a^2\mathcal{F}_0(a,\Lambda) \partial_a\mathcal{F}_0(a+\epsilon x,\Lambda_{\epsilon s}) +
\frac1{12 \epsilon^2} \partial_a\mathcal{F}_0(a,\Lambda) \partial_a\mathcal{F}_0(a+\epsilon x,\Lambda_{\epsilon s})\\
- \frac1{24\epsilon^2} \partial_a\mathcal{F}_0(a,\Lambda)^2-
\frac1{12\epsilon} x \partial_a\mathcal{F}_0(a,\Lambda)\partial_a^2\mathcal{F}_0(a,\Lambda)
- \frac{x^2}{24} \partial_a^2\mathcal{F}_0(a,\Lambda)=\\
=-\frac1{24\epsilon^2}\left( \partial_a\mathcal{F}_0(a+\epsilon x,\Lambda_{\epsilon s})-\partial_a\mathcal{F}_0(a,\Lambda)-\epsilon x \partial_a^2\mathcal{F}_0(a,\Lambda) \right)^2=-\frac{1}{24}\left(\partial_x\mathcal{F}_{0,\mathrm{st}}^{x,s}(a,\Lambda)\right)^2\ ,
\end{multline}
therefore we obtain
\begin{equation}
D_{E_2}\mathcal{F}_{0,\mathrm{st}}^{x,s}(a,\Lambda)= -\frac{1}{24}\left(\partial_x\mathcal{F}_{0,\mathrm{st}}^{x,s}(a,\Lambda)\right)^2 \ .
\end{equation}

\subsection{Proof of decoupling of the NS and SD equations}\label{appD,2}
We have the equation~\eqref{holNS,7.2}
\begin{align}
&\left(D_{E_2}\mathcal{F}_{\mathrm{st}}^{x,s}(a, \Lambda)+\frac{1}{24}\partial_x^2\mathcal{F}_{\mathrm{st}}^{x,s}(a, \Lambda)+\frac{1}{24}\left(\partial_x\mathcal{F}_{\mathrm{st}}^{x,s}(a, \Lambda)\right)^2\right)\nonumber\\
&-\left(\frac{\rho_{\mathrm{st}}}{12}+\frac{D_{E_2}\uns }{\partial_a \uns }\right)\frac{1}{\epsilon}\partial_x\mathcal{F}_{\mathrm{st}}^{x,s}(a, \Lambda)\label{holNSapp,7.2}\\
&-\frac{x}{\epsilon}\left(D_{E_2}\rho_{\mathrm{st}}-\frac{D_{E_2}\uns }{\partial_a \uns }\partial_a\rho_{\mathrm{st}}\right)+\left(D_{E_2}\partial_\epsilon W_{\mathrm{st}}+\frac{(\rho_{\mathrm{st}})^2}{24 \epsilon^2}-\frac{D_{E_2}\uns }{\partial_a \uns }\partial_a\partial_\epsilon W_{\mathrm{st}}\right)=0 \nonumber \ ,
\end{align}
and we want to decouple the SD and NS contributions. To do this we first rewrite
\begin{align}
&\frac{\rho_{\mathrm{st}}}{12}+\frac{D_{E_2}\uns }{\partial_a \uns }= \frac{1}{\partial_a \uns }\left(\frac{1}{12}\rho_{\mathrm{st}}\partial_a \uns +D_{E_2}\uns \right)=\frac{\epsilon\Lambda\partial_\Lambda}{\partial_a \uns }\left(\frac{\epsilon}{24}(\partial_a W_{\mathrm{st}})^2+D_{E_2}W_{\mathrm{st}}\right)\nonumber\\
&\Rightarrow\frac{D_{E_2}\uns }{\partial_a \uns }=-\frac{\rho_{\mathrm{st}}}{12}+\frac{\epsilon\Lambda\partial_\Lambda}{\partial_a \uns }\left(D_{E_2}W_{\mathrm{st}}+\frac{\epsilon}{24}(\partial_a W_{\mathrm{st}})^2\right)\ ,
\end{align}
and substituting in \eqref{holNSapp,7.2} we obtain
\begin{align}
&\left(D_{E_2}\mathcal{F}_{\mathrm{st}}^{x,s}(a, \Lambda)+\frac{1}{24}\partial_x^2\mathcal{F}_{\mathrm{st}}^{x,s}(a, \Lambda)+\frac{1}{24}\left(\partial_x\mathcal{F}_{\mathrm{st}}^{x,s}(a, \Lambda)\right)^2\right)\nonumber\\
&-\left(x\partial_a-\partial_\epsilon+\frac{\partial_x\mathcal{F}_{\mathrm{st}}^{x,s}(a, \Lambda)-x\partial_a\rho_{\mathrm{st}}+\epsilon\partial_a\partial_\epsilon W_{\mathrm{st}}}{\partial_a \uns }\Lambda\partial_\Lambda\right)\times \nonumber \\
&\times\left(D_{E_2}W_{\mathrm{st}}+\frac{\epsilon}{24}(\partial_a W_{\mathrm{st}})^2\right)=0\ . \label{holNS,8}
\end{align}
We observe now that the SD contributions are always evaluated at the shifted scales ($y=\epsilon x+a, \Lambda'=\Lambda_{\epsilon s}=\Lambda\exp(\epsilon s)$), the NS contributions instead are evaluated at the unshifted scales $(a,\Lambda)$. To decouple the previous equation it is then useful to rewrite it as a function of the SD and NS scales. Using \eqref{hol,7}, \eqref{hol,8app} we have
\begin{align}
&e^{-\mathcal{F}_{\mathrm{st}}^{x,s}(a, \Lambda)}\left(D_{E_2}+\frac{\partial_x^2}{24} \right)e^{\mathcal{F}_{\mathrm{st}}^{x,s}(a, \Lambda)}=\nonumber\\
&=e^{-\mathcal{F}_{\mathrm{st}}^{x,s}(a, \Lambda)}\left(D_{E_2}+\frac{\partial_x^2}{24} \right)e^{\mathcal{F}_{0,\mathrm{st}}^{x,s}(a, \Lambda)}e^{\epsilon x\partial_a+s\Lambda\partial\Lambda}e^{\widehat{\mathcal{F}}(a,\Lambda)}=\nonumber\\
&=e^{-\mathcal{F}_{\mathrm{st}}^{x,s}(a, \Lambda)}e^{\mathcal{F}_{0,\mathrm{st}}^{x,s}(a, \Lambda)}e^{\epsilon x\partial_a+s\Lambda\partial\Lambda}\left(D_{E_2} + \frac{\epsilon^2}{24}\partial_a^2\right)e^{\widehat{\mathcal{F}}(a,\Lambda)}= \nonumber\\
&=e^{\epsilon x\partial_a+s\Lambda\partial\Lambda}e^{-\widehat{\mathcal{F}}(a,\Lambda)}\left(D_{E_2} + \frac{\epsilon^2}{24}\partial_a^2\right)e^{\widehat{\mathcal{F}}(a,\Lambda)}=\nonumber\\
&=D_{E_2}\widehat{\mathcal{F}}(y,\Lambda')+\frac{1}{24}\partial_y^2\widehat{\mathcal{F}}(y,\Lambda')+\frac{1}{24}\left(\partial_y\widehat{\mathcal{F}}(y,\Lambda')\right)^2 \ ,
\end{align}
and we can also rewrite
\begin{equation}
\partial_x\mathcal{F}_{\mathrm{st}}^{x,s}(a, \Lambda)-x\partial_a\rho_{\mathrm{st}}+\epsilon\partial_a\partial_\epsilon W_{\mathrm{st}}=\epsilon\partial_y\mathcal{F}(y,\Lambda')-x\partial_a\rho+\epsilon\partial_a\partial_\epsilon W\ .
\end{equation}
Then \eqref{holNS,8} becomes
\begin{align}
&\left(D_{E_2}\widehat{\mathcal{F}}(y,\Lambda')+\frac{\epsilon^2}{24}\partial_y^2\widehat{\mathcal{F}}(y,\Lambda')+\frac{\epsilon^2}{24}\left(\partial_y\widehat{\mathcal{F}}(y,\Lambda')\right)^2\right)\nonumber\\
&-\left(\left(\frac{y-a}{\epsilon}\right)\partial_a-\partial_\epsilon+\frac{\epsilon\partial_y\mathcal{F}(y,\Lambda')-(y-a)\partial_a\rho/\epsilon+\epsilon\partial_a\partial_\epsilon W}{\partial_a \uns }\Lambda\partial_\Lambda\right)\times \nonumber\\
&\times\left(D_{E_2}W_{\mathrm{st}}+\frac{\epsilon}{24}(\partial_a W_{\mathrm{st}})^2\right)=0\ . \label{holNS,9}
\end{align}
Taking the second derivative with respect to $y$ we get
\begin{align}
&\partial_y^2\left(D_{E_2}\widehat{\mathcal{F}}(y,\Lambda')+\frac{\epsilon^2}{24}\partial_y^2\widehat{\mathcal{F}}(y,\Lambda')+\frac{\epsilon^2}{24}\left(\partial_y\widehat{\mathcal{F}}(y,\Lambda')\right)^2\right)=\nonumber\\
&=\epsilon\frac{\partial_y^3\mathcal{F}(y,\Lambda')}{\partial_a \uns }\Lambda\partial_\Lambda\left(D_{E_2}W_{\mathrm{st}}+\frac{\epsilon}{24}(\partial_a W_{\mathrm{st}})^2\right)
\end{align}
and the first derivative in $\Lambda'$ gives
\begin{align}                                                                                                                                                                 
&\partial_{\Lambda'}\left(D_{E_2}\widehat{\mathcal{F}}(y,\Lambda')+\frac{\epsilon^2}{24}\partial_y^2\widehat{\mathcal{F}}(y,\Lambda')+\frac{\epsilon^2}{24}\left(\partial_y\widehat{\mathcal{F}}(y,\Lambda')\right)^2\right)=\nonumber\\
&=\epsilon\frac{\partial_{\Lambda'}\partial_y\mathcal{F}(y,\Lambda')}{\partial_a \uns }\Lambda\partial_\Lambda\left(D_{E_2}W_{\mathrm{st}}+\frac{\epsilon}{24}(\partial_a W_{\mathrm{st}})^2\right)\ .
\end{align}
From this equations it follows that
\begin{align}
&\Lambda\partial_\Lambda\left(D_{E_2}W_{\mathrm{st}}+\frac{\epsilon}{24}(\partial_a W_{\mathrm{st}})^2\right)=c_1 \partial_a \uns =\epsilon c_1\Lambda\partial_\Lambda\partial_a W\ , \label{holNS,10.1app} \\
&\left(D_{E_2}\widehat{\mathcal{F}}(y,\Lambda')+\frac{\epsilon^2}{24}\partial_y^2\widehat{\mathcal{F}}(y,\Lambda')+\frac{\epsilon^2}{24}\left(\partial_y\widehat{\mathcal{F}}(y,\Lambda')\right)^2\right)-\epsilon c_1\partial_y\mathcal{F}(y,\Lambda')=\frac{c_2}{\epsilon} y+c_3 \ , \label{holNS,10.2app} 
\end{align}
for some dimensionless\footnote{In the theories with flavour in principle these constants can actually be functions of dimensionless combinations of the mass parameters but we can rule out this dependence because it will produce extra terms in the holomorphic anomaly equations which are inconsistent with the limit $\epsilon \to 0$.} constants $c_1,c_2,c_3$. Integrating \eqref{holNS,10.1app} we get
\begin{equation}
\left(D_{E_2}W_{\mathrm{st}}+\frac{\epsilon}{24}(\partial_a W_{\mathrm{st}})^2\right)=c_1 \rho +f(a,\epsilon)\ ,    
\end{equation}
for some function $f(a,\epsilon)$. Substituting in \eqref{holNS,9} we obtain
\begin{align}
\frac{c_2}{\epsilon} y+c_3+c_1\partial_a W-\left(\left(\frac{y-a}{\epsilon}\right)\partial_a-\partial_\epsilon\right)f(a,\epsilon)=0\ ,
\end{align}
whose coefficients in $y$ give the two equations
\begin{align}
& \partial_a f(a,\epsilon)=c_2\ \Rightarrow f(a,\epsilon)=c_2 a+f_0 \epsilon\ , \\
& (c_3+f_0) \epsilon+c_1\rho+a c_2=0\ ,\label{holNS,11}
\end{align}
for some dimensionless constant $f_0$. We have now two cases: if $c_1=0$ then from \eqref{holNS,11} we get $c_2=0, c_3=-f_0$ and we obtain the following equations
\begin{align}
&D_{E_2}W_{\mathrm{st}}+\frac{\epsilon}{24}(\partial_a W_{\mathrm{st}})^2=\epsilon f_0\ , \label{holNS,17.1} \\
&D_{E_2}\widehat{\mathcal{F}}(y,\Lambda')+\frac{\epsilon^2}{24}\partial_y^2\widehat{\mathcal{F}}(y,\Lambda')+\frac{\epsilon^2}{24}\left(\partial_y\widehat{\mathcal{F}}(y,\Lambda')\right)^2=-f_0\ , \label{holNS,17.2} 
\end{align}
in the limit $\epsilon \to 0$ we have $\widehat{\mathcal{F}}(y,\Lambda')\to \mathcal{F}_{1,0}(y,\Lambda') $ and \eqref{holNS,17.2} becomes
\begin{equation}
D_{E_2}\mathcal{F}_{1,0}(y,\Lambda')=-f_0\ , \label{holNS,18} 
\end{equation}
from \eqref{eq:3} we have that $\mathcal{F}_{1,0}(y,\Lambda')$ is modular therefore we need $f_0=0$ and \eqref{holNS,17.1}, \eqref{holNS,17.2} become exactly the holomorphic anomaly equations for NS and SD respectively. 

The case $c_1\neq0$ in unphysical. Indeed, from \eqref{holNS,11} we get
\begin{equation}
\rho= -\frac{c_2}{c_1}a-\frac{c_3+f_0}{c_1}\epsilon \ .
\end{equation}
and integrating in $a$ we obtain
\begin{equation}\label{holNS,11bis}
W= -\frac{1}{2\epsilon}\frac{c_2}{c_1}a^2-\frac{c_3+f_0}{c_1} a +\Lambda w\left(\frac{\Lambda}{\epsilon}\right) \ .
\end{equation}
for some dimensionless function $w(\Lambda/\epsilon)$. In particular in the SW limit $\epsilon\to 0$ we have
\begin{equation}\label{holNS,11tris}
\lim_{\epsilon\to 0} \epsilon w\left(\frac{\Lambda}{\epsilon}\right)=w_0\Lambda\ ,\quad \mathcal{F}_0(a,\Lambda)= \lim_{\epsilon\to 0} \epsilon W=-\frac{1}{2}\frac{c_2}{c_1}a^2+\Lambda^2 w_0 \ ,
\end{equation} 
for some dimensionless constant $w_0$. The prepotential in \eqref{holNS,11tris} is unphysical because the IR coupling $\tau_{SW}$ is independent on the $UV$ coupling $\Lambda$. Indeed, in this case the structure constants will be trivial $\partial_a^3\mathcal{F}_0(a,\Lambda)$=0.

\newpage
\bibliographystyle{JHEP}
\bibliography{refPainlev}
\newpage

\section{Coefficients of the Hurwitz expansions} \label{appE}

In this extra appendix we collect some numerical evidence of the integrality of the polynomials in the Hurwitz expansions of 
the $\Tau$-functions in some trial cases.
Decomposing the numerical coefficients in prime factors we could not recognize any regularity to further reduce them in an obvious way. 
We computed many more coefficient polynomials evidencing their integrality, but we do not report them for brevity. For notational simplicity {\it in this appendix we denote all symbols without boldface}. We refer to section \ref{section5} for the corresponding assignments. We also refer to section \ref{section5} for the explicit expressions of the invariants $g_2,g_3,T$.
\subsection{Hurwitz expansion for PVI}
\label{app:PVI}
In this subsection we denote \(c_n=c_n^{PVI}\) from \eqref{eq:cnPVI}.

(Massive case: $\tilde e_2 =e_2/4\ , \ \tilde e_4 =e_4/2\ ,\ \tilde\epsilon=\epsilon/2$)\footnote{In the case of PVI the natural basis is not given by the elliptic invariants $g_2,g_3$ and the contact term $T$ due to the presence of the adimensional coupling $q$ which introduces a more complicated dependence. Nevertheless, in the SW limit the expansion can be reorganized in this basis because of the universality of the structure of the SW blowup factor.} 
{\footnotesize
  \quad \\ $c_{0}= 1
$   \quad \\ $c_{1}= 0
$   \quad \\ $c_{2}= \tilde e_2 q^{2}+u-q^{2} u-\tilde\epsilon^{2}+2 q \tilde\epsilon^{2}-q^{2} \tilde\epsilon^{2}
$   \quad \\ $c_{3}= -8 \tilde e_2 q^{2} \tilde\epsilon-8 q u \tilde\epsilon+8 q^{2} u \tilde\epsilon
$   \quad \\ $c_{4}= 2 p_4 q-\tilde e_4 q^{2}-3 p_4 q^{2}+\tilde e_4 q^{3}+p_4 q^{3}+\tilde e_2^2 q^{4}+6 \tilde e_2 q^{2} u-4 \tilde e_2 q^{3} u-2 \tilde e_2 q^{4} u+u^2+2 q u^2-6 q^{2} u^2+2 q^{3} u^2+q^{4} u^2+46 \tilde e_2 q^{2} \tilde\epsilon^{2}+28 \tilde e_2 q^{3} \tilde\epsilon^{2}-2 \tilde e_2 q^{4} \tilde\epsilon^{2}-2 u \tilde\epsilon^{2}+28 q u \tilde\epsilon^{2}-28 q^{3} u \tilde\epsilon^{2}+2 q^{4} u \tilde\epsilon^{2}+\tilde\epsilon^{4}-12 q \tilde\epsilon^{4}+22 q^{2} \tilde\epsilon^{4}-12 q^{3} \tilde\epsilon^{4}+q^{4} \tilde\epsilon^{4}
$   \quad \\ $c_{5}= -8 p_4 q \tilde\epsilon+16 \tilde e_4 q^{2} \tilde\epsilon-8 p_4 q^{2} \tilde\epsilon-12 \tilde e_4 q^{3} \tilde\epsilon+20 p_4 q^{3} \tilde\epsilon-24 \tilde e_2^2 q^{4} \tilde\epsilon-4 \tilde e_4 q^{4} \tilde\epsilon-4 p_4 q^{4} \tilde\epsilon-88 \tilde e_2 q^{2} u \tilde\epsilon+24 \tilde e_2 q^{3} u \tilde\epsilon+64 \tilde e_2 q^{4} u \tilde\epsilon-32 q u^2 \tilde\epsilon+32 q^{2} u^2 \tilde\epsilon+32 q^{3} u^2 \tilde\epsilon-32 q^{4} u^2 \tilde\epsilon-232 \tilde e_2 q^{2} \tilde\epsilon^{3}-496 \tilde e_2 q^{3} \tilde\epsilon^{3}-40 \tilde e_2 q^{4} \tilde\epsilon^{3}-40 q u \tilde\epsilon^{3}-264 q^{2} u \tilde\epsilon^{3}+264 q^{3} u \tilde\epsilon^{3}+40 q^{4} u \tilde\epsilon^{3}+32 q \tilde\epsilon^{5}-32 q^{2} \tilde\epsilon^{5}-32 q^{3} \tilde\epsilon^{5}+32 q^{4} \tilde\epsilon^{5}
$   \quad \\ $c_{6}= 6 e_6 q^{2}-12 e_6 q^{3}+24 \tilde e_2 p_4 q^{3}-24 \tilde e_2 \tilde e_4 q^{4}+6 e_6 q^{4}-36 \tilde e_2 p_4 q^{4}+24 \tilde e_2 \tilde e_4 q^{5}+12 \tilde e_2 p_4 q^{5}+64 \tilde e_2^3 q^{6}-6 p_4 q u+30 \tilde e_4 q^{2} u+3 p_4 q^{2} u-84 \tilde e_4 q^{3} u+30 p_4 q^{3} u-240 \tilde e_2^2 q^{4} u+72 \tilde e_4 q^{4} u-48 p_4 q^{4} u+432 \tilde e_2^2 q^{5} u-12 \tilde e_4 q^{5} u+24 p_4 q^{5} u-144 \tilde e_2^2 q^{6} u-6 \tilde e_4 q^{6} u-3 p_4 q^{6} u-48 \tilde e_2^2 q^{7} u+156 \tilde e_2 q^{2} u^2-528 \tilde e_2 q^{3} u^2+564 \tilde e_2 q^{4} u^2-96 \tilde e_2 q^{5} u^2-156 \tilde e_2 q^{6} u^2+48 \tilde e_2 q^{7} u^2+12 \tilde e_2 q^{8} u^2-u^3-3 q u^3+30 q^{2} u^3-62 q^{3} u^3+36 q^{4} u^3+36 q^{5} u^3-62 q^{6} u^3+30 q^{7} u^3-3 q^{8} u^3-q^{9} u^3+72 p_4 q \tilde\epsilon^{2}-1128 \tilde e_4 q^{2} \tilde\epsilon^{2}+916 p_4 q^{2} \tilde\epsilon^{2}-424 \tilde e_4 q^{3} \tilde\epsilon^{2}-636 p_4 q^{3} \tilde\epsilon^{2}+23360 \tilde e_2^2 q^{4} \tilde\epsilon^{2}+1480 \tilde e_4 q^{4} \tilde\epsilon^{2}-388 p_4 q^{4} \tilde\epsilon^{2}+7040 \tilde e_2^2 q^{5} \tilde\epsilon^{2}+72 \tilde e_4 q^{5} \tilde\epsilon^{2}+36 p_4 q^{5} \tilde\epsilon^{2}-192 \tilde e_2^2 q^{6} \tilde\epsilon^{2}-13792 \tilde e_2 q^{2} u \tilde\epsilon^{2}+4832 \tilde e_2 q^{3} u \tilde\epsilon^{2}+27712 \tilde e_2 q^{4} u \tilde\epsilon^{2}-14656 \tilde e_2 q^{5} u \tilde\epsilon^{2}-4192 \tilde e_2 q^{6} u \tilde\epsilon^{2}+96 \tilde e_2 q^{7} u \tilde\epsilon^{2}-12 u^2 \tilde\epsilon^{2}+536 q u^2 \tilde\epsilon^{2}-416 q^{2} u^2 \tilde\epsilon^{2}-2968 q^{3} u^2 \tilde\epsilon^{2}+5720 q^{4} u^2 \tilde\epsilon^{2}-2968 q^{5} u^2 \tilde\epsilon^{2}-416 q^{6} u^2 \tilde\epsilon^{2}+536 q^{7} u^2 \tilde\epsilon^{2}-12 q^{8} u^2 \tilde\epsilon^{2}+62656 \tilde e_2 q^{2} \tilde\epsilon^{4}+367872 \tilde e_2 q^{3} \tilde\epsilon^{4}+182400 \tilde e_2 q^{4} \tilde\epsilon^{4}+1280 \tilde e_2 q^{5} \tilde\epsilon^{4}+192 \tilde e_2 q^{6} \tilde\epsilon^{4}-48 u \tilde\epsilon^{4}-272 q u \tilde\epsilon^{4}-29616 q^{2} u \tilde\epsilon^{4}+29936 q^{3} u \tilde\epsilon^{4}+29936 q^{4} u \tilde\epsilon^{4}-29616 q^{5} u \tilde\epsilon^{4}-272 q^{6} u \tilde\epsilon^{4}-48 q^{7} u \tilde\epsilon^{4}-64 \tilde\epsilon^{6}-4224 q \tilde\epsilon^{6}-23488 q^{2} \tilde\epsilon^{6}+55552 q^{3} \tilde\epsilon^{6}-23488 q^{4} \tilde\epsilon^{6}-4224 q^{5} \tilde\epsilon^{6}-64 q^{6} \tilde\epsilon^{6}
$   \quad \\ $c_{7}= -96 e_6 q^{2} \tilde\epsilon+96 e_6 q^{3} \tilde\epsilon-288 \tilde e_2 p_4 q^{3} \tilde\epsilon+448 \tilde e_2 \tilde e_4 q^{4} \tilde\epsilon+96 e_6 q^{4} \tilde\epsilon-96 \tilde e_2 \tilde e_4 q^{5} \tilde\epsilon-96 e_6 q^{5} \tilde\epsilon+448 \tilde e_2 p_4 q^{5} \tilde\epsilon-1056 \tilde e_2^3 q^{6} \tilde\epsilon-352 \tilde e_2 \tilde e_4 q^{6} \tilde\epsilon-160 \tilde e_2 p_4 q^{6} \tilde\epsilon-1008 \tilde e_2^3 q^{7} \tilde\epsilon+160 p_4 q u \tilde\epsilon-400 \tilde e_4 q^{2} u \tilde\epsilon-64 p_4 q^{2} u \tilde\epsilon+736 \tilde e_4 q^{3} u \tilde\epsilon-624 p_4 q^{3} u \tilde\epsilon+3536 \tilde e_2^2 q^{4} u \tilde\epsilon+176 \tilde e_4 q^{4} u \tilde\epsilon+528 p_4 q^{4} u \tilde\epsilon-2688 \tilde e_2^2 q^{5} u \tilde\epsilon-864 \tilde e_4 q^{5} u \tilde\epsilon+320 p_4 q^{5} u \tilde\epsilon-4640 \tilde e_2^2 q^{6} u \tilde\epsilon+256 \tilde e_4 q^{6} u \tilde\epsilon-368 p_4 q^{6} u \tilde\epsilon+3024 \tilde e_2^2 q^{7} u \tilde\epsilon+96 \tilde e_4 q^{7} u \tilde\epsilon+48 p_4 q^{7} u \tilde\epsilon+768 \tilde e_2^2 q^{8} u \tilde\epsilon-2816 \tilde e_2 q^{2} u^2 \tilde\epsilon+6512 \tilde e_2 q^{3} u^2 \tilde\epsilon-544 \tilde e_2 q^{4} u^2 \tilde\epsilon-7904 \tilde e_2 q^{5} u^2 \tilde\epsilon+4128 \tilde e_2 q^{6} u^2 \tilde\epsilon+1776 \tilde e_2 q^{7} u^2 \tilde\epsilon-960 \tilde e_2 q^{8} u^2 \tilde\epsilon-192 \tilde e_2 q^{9} u^2 \tilde\epsilon+32 u^3 \tilde\epsilon+96 q u^3 \tilde\epsilon-464 q^{2} u^3 \tilde\epsilon+336 q^{3} u^3 \tilde\epsilon+592 q^{4} u^3 \tilde\epsilon-1120 q^{5} u^3 \tilde\epsilon+384 q^{6} u^3 \tilde\epsilon+496 q^{7} u^3 \tilde\epsilon-432 q^{8} u^3 \tilde\epsilon+64 q^{9} u^3 \tilde\epsilon+16 q^{10} u^3 \tilde\epsilon-896 p_4 q \tilde\epsilon^{3}+16688 \tilde e_4 q^{2} \tilde\epsilon^{3}-13616 p_4 q^{2} \tilde\epsilon^{3}+24624 \tilde e_4 q^{3} \tilde\epsilon^{3}-5360 p_4 q^{3} \tilde\epsilon^{3}-372432 \tilde e_2^2 q^{4} \tilde\epsilon^{3}-17232 \tilde e_4 q^{4} \tilde\epsilon^{3}+15632 p_4 q^{4} \tilde\epsilon^{3}-483216 \tilde e_2^2 q^{5} \tilde\epsilon^{3}-23056 \tilde e_4 q^{5} \tilde\epsilon^{3}+4688 p_4 q^{5} \tilde\epsilon^{3}-108144 \tilde e_2^2 q^{6} \tilde\epsilon^{3}-1024 \tilde e_4 q^{6} \tilde\epsilon^{3}-448 p_4 q^{6} \tilde\epsilon^{3}+3024 \tilde e_2^2 q^{7} \tilde\epsilon^{3}+227904 \tilde e_2 q^{2} u \tilde\epsilon^{3}+148288 \tilde e_2 q^{3} u \tilde\epsilon^{3}-517920 \tilde e_2 q^{4} u \tilde\epsilon^{3}-220576 \tilde e_2 q^{5} u \tilde\epsilon^{3}+298208 \tilde e_2 q^{6} u \tilde\epsilon^{3}+65632 \tilde e_2 q^{7} u \tilde\epsilon^{3}-1536 \tilde e_2 q^{8} u \tilde\epsilon^{3}+144 u^2 \tilde\epsilon^{3}-6704 q u^2 \tilde\epsilon^{3}+304 q^{2} u^2 \tilde\epsilon^{3}+50288 q^{3} u^2 \tilde\epsilon^{3}-47888 q^{4} u^2 \tilde\epsilon^{3}-41808 q^{5} u^2 \tilde\epsilon^{3}+55824 q^{6} u^2 \tilde\epsilon^{3}-1968 q^{7} u^2 \tilde\epsilon^{3}-8384 q^{8} u^2 \tilde\epsilon^{3}+192 q^{9} u^2 \tilde\epsilon^{3}-989952 \tilde e_2 q^{2} \tilde\epsilon^{5}-6832848 \tilde e_2 q^{3} \tilde\epsilon^{5}-8737600 \tilde e_2 q^{4} \tilde\epsilon^{5}-2905312 \tilde e_2 q^{5} \tilde\epsilon^{5}-23104 \tilde e_2 q^{6} \tilde\epsilon^{5}-3024 \tilde e_2 q^{7} \tilde\epsilon^{5}+816 u \tilde\epsilon^{5}+5568 q u \tilde\epsilon^{5}+499232 q^{2} u \tilde\epsilon^{5}+6032 q^{3} u \tilde\epsilon^{5}-969104 q^{4} u \tilde\epsilon^{5}-26144 q^{5} u \tilde\epsilon^{5}+477760 q^{6} u \tilde\epsilon^{5}+5072 q^{7} u \tilde\epsilon^{5}+768 q^{8} u \tilde\epsilon^{5}+1008 \tilde\epsilon^{7}+67664 q \tilde\epsilon^{7}+439920 q^{2} \tilde\epsilon^{7}-508592 q^{3} \tilde\epsilon^{7}-508592 q^{4} \tilde\epsilon^{7}+439920 q^{5} \tilde\epsilon^{7}+67664 q^{6} \tilde\epsilon^{7}+1008 q^{7} \tilde\epsilon^{7} $ \\
}

(Massless case: $\tilde\epsilon=\epsilon/2$)
{\footnotesize
   \quad \\ $c_{0}= 1
$   \quad \\ $c_{1}= 0
$   \quad \\ $c_{2}= u-q^{2} u-\tilde\epsilon^{2}+2 q \tilde\epsilon^{2}-q^{2} \tilde\epsilon^{2}
$   \quad \\ $c_{3}= -8 q u \tilde\epsilon+8 q^{2} u \tilde\epsilon
$   \quad \\ $c_{4}= u^2+2 q u^2-6 q^{2} u^2+2 q^{3} u^2+q^{4} u^2-2 u \tilde\epsilon^{2}+28 q u \tilde\epsilon^{2}-28 q^{3} u \tilde\epsilon^{2}+2 q^{4} u \tilde\epsilon^{2}+\tilde\epsilon^{4}-12 q \tilde\epsilon^{4}+22 q^{2} \tilde\epsilon^{4}-12 q^{3} \tilde\epsilon^{4}+q^{4} \tilde\epsilon^{4}
$   \quad \\ $c_{5}= -32 q u^2 \tilde\epsilon+32 q^{2} u^2 \tilde\epsilon+32 q^{3} u^2 \tilde\epsilon-32 q^{4} u^2 \tilde\epsilon-40 q u \tilde\epsilon^{3}-264 q^{2} u \tilde\epsilon^{3}+264 q^{3} u \tilde\epsilon^{3}+40 q^{4} u \tilde\epsilon^{3}+32 q \tilde\epsilon^{5}-32 q^{2} \tilde\epsilon^{5}-32 q^{3} \tilde\epsilon^{5}+32 q^{4} \tilde\epsilon^{5}
$   \quad \\ $c_{6}= -u^3-3 q u^3+30 q^{2} u^3-62 q^{3} u^3+36 q^{4} u^3+36 q^{5} u^3-62 q^{6} u^3+30 q^{7} u^3-3 q^{8} u^3-q^{9} u^3-12 u^2 \tilde\epsilon^{2}+536 q u^2 \tilde\epsilon^{2}-416 q^{2} u^2 \tilde\epsilon^{2}-2968 q^{3} u^2 \tilde\epsilon^{2}+5720 q^{4} u^2 \tilde\epsilon^{2}-2968 q^{5} u^2 \tilde\epsilon^{2}-416 q^{6} u^2 \tilde\epsilon^{2}+536 q^{7} u^2 \tilde\epsilon^{2}-12 q^{8} u^2 \tilde\epsilon^{2}-48 u \tilde\epsilon^{4}-272 q u \tilde\epsilon^{4}-29616 q^{2} u \tilde\epsilon^{4}+29936 q^{3} u \tilde\epsilon^{4}+29936 q^{4} u \tilde\epsilon^{4}-29616 q^{5} u \tilde\epsilon^{4}-272 q^{6} u \tilde\epsilon^{4}-48 q^{7} u \tilde\epsilon^{4}-64 \tilde\epsilon^{6}-4224 q \tilde\epsilon^{6}-23488 q^{2} \tilde\epsilon^{6}+55552 q^{3} \tilde\epsilon^{6}-23488 q^{4} \tilde\epsilon^{6}-4224 q^{5} \tilde\epsilon^{6}-64 q^{6} \tilde\epsilon^{6}
$   \quad \\ $c_{7}= 32 u^3 \tilde\epsilon+96 q u^3 \tilde\epsilon-464 q^{2} u^3 \tilde\epsilon+336 q^{3} u^3 \tilde\epsilon+592 q^{4} u^3 \tilde\epsilon-1120 q^{5} u^3 \tilde\epsilon+384 q^{6} u^3 \tilde\epsilon+496 q^{7} u^3 \tilde\epsilon-432 q^{8} u^3 \tilde\epsilon+64 q^{9} u^3 \tilde\epsilon+16 q^{10} u^3 \tilde\epsilon+144 u^2 \tilde\epsilon^{3}-6704 q u^2 \tilde\epsilon^{3}+304 q^{2} u^2 \tilde\epsilon^{3}+50288 q^{3} u^2 \tilde\epsilon^{3}-47888 q^{4} u^2 \tilde\epsilon^{3}-41808 q^{5} u^2 \tilde\epsilon^{3}+55824 q^{6} u^2 \tilde\epsilon^{3}-1968 q^{7} u^2 \tilde\epsilon^{3}-8384 q^{8} u^2 \tilde\epsilon^{3}+192 q^{9} u^2 \tilde\epsilon^{3}+816 u \tilde\epsilon^{5}+5568 q u \tilde\epsilon^{5}+499232 q^{2} u \tilde\epsilon^{5}+6032 q^{3} u \tilde\epsilon^{5}-969104 q^{4} u \tilde\epsilon^{5}-26144 q^{5} u \tilde\epsilon^{5}+477760 q^{6} u \tilde\epsilon^{5}+5072 q^{7} u \tilde\epsilon^{5}+768 q^{8} u \tilde\epsilon^{5}+1008 \tilde\epsilon^{7}+67664 q \tilde\epsilon^{7}+439920 q^{2} \tilde\epsilon^{7}-508592 q^{3} \tilde\epsilon^{7}-508592 q^{4} \tilde\epsilon^{7}+439920 q^{5} \tilde\epsilon^{7}+67664 q^{6} \tilde\epsilon^{7}+1008 q^{7} \tilde\epsilon^{7}
$   \quad \\ $c_{8}= -23 u^4-96 q u^4+500 q^{2} u^4-504 q^{3} u^4-450 q^{4} u^4+1380 q^{5} u^4-1048 q^{6} u^4-132 q^{7} u^4+697 q^{8} u^4-372 q^{9} u^4+36 q^{10} u^4+12 q^{11} u^4-408 u^3 \tilde\epsilon^{2}+2924 q u^3 \tilde\epsilon^{2}-6036 q^{2} u^3 \tilde\epsilon^{2}-20300 q^{3} u^3 \tilde\epsilon^{2}+51656 q^{4} u^3 \tilde\epsilon^{2}-1424 q^{5} u^3 \tilde\epsilon^{2}-55104 q^{6} u^3 \tilde\epsilon^{2}+22624 q^{7} u^3 \tilde\epsilon^{2}+10704 q^{8} u^3 \tilde\epsilon^{2}-3676 q^{9} u^3 \tilde\epsilon^{2}-812 q^{10} u^3 \tilde\epsilon^{2}-148 q^{11} u^3 \tilde\epsilon^{2}-1938 u^2 \tilde\epsilon^{4}+52980 q u^2 \tilde\epsilon^{4}-239440 q^{2} u^2 \tilde\epsilon^{4}-226516 q^{3} u^2 \tilde\epsilon^{4}+236644 q^{4} u^2 \tilde\epsilon^{4}+1319404 q^{5} u^2 \tilde\epsilon^{4}-1192832 q^{6} u^2 \tilde\epsilon^{4}-235052 q^{7} u^2 \tilde\epsilon^{4}+214766 q^{8} u^2 \tilde\epsilon^{4}+73760 q^{9} u^2 \tilde\epsilon^{4}-1776 q^{10} u^2 \tilde\epsilon^{4}-8308 u \tilde\epsilon^{6}-119728 q u \tilde\epsilon^{6}-5052088 q^{2} u \tilde\epsilon^{6}-10787312 q^{3} u \tilde\epsilon^{6}+15374592 q^{4} u \tilde\epsilon^{6}+15388784 q^{5} u \tilde\epsilon^{6}-10250568 q^{6} u \tilde\epsilon^{6}-4474640 q^{7} u \tilde\epsilon^{6}-63628 q^{8} u \tilde\epsilon^{6}-7104 q^{9} u \tilde\epsilon^{6}-9323 \tilde\epsilon^{8}-647824 q \tilde\epsilon^{8}-5581380 q^{2} \tilde\epsilon^{8}-4380976 q^{3} \tilde\epsilon^{8}+21239006 q^{4} \tilde\epsilon^{8}-4380976 q^{5} \tilde\epsilon^{8}-5581380 q^{6} \tilde\epsilon^{8}-647824 q^{7} \tilde\epsilon^{8}-9323 q^{8} \tilde\epsilon^{8}
$   \quad \\ $c_{9}= 544 u^4 \tilde\epsilon+3120 q u^4 \tilde\epsilon-7248 q^{2} u^4 \tilde\epsilon-11424 q^{3} u^4 \tilde\epsilon+39712 q^{4} u^4 \tilde\epsilon-30352 q^{5} u^4 \tilde\epsilon-14288 q^{6} u^4 \tilde\epsilon+42480 q^{7} u^4 \tilde\epsilon-25776 q^{8} u^4 \tilde\epsilon-2592 q^{9} u^4 \tilde\epsilon+7328 q^{10} u^4 \tilde\epsilon-1232 q^{11} u^4 \tilde\epsilon-272 q^{12} u^4 \tilde\epsilon+4464 u^3 \tilde\epsilon^{3}-81856 q u^3 \tilde\epsilon^{3}+53728 q^{2} u^3 \tilde\epsilon^{3}+820416 q^{3} u^3 \tilde\epsilon^{3}-633264 q^{4} u^3 \tilde\epsilon^{3}-2223680 q^{5} u^3 \tilde\epsilon^{3}+2615424 q^{6} u^3 \tilde\epsilon^{3}+347584 q^{7} u^3 \tilde\epsilon^{3}-998320 q^{8} u^3 \tilde\epsilon^{3}-59904 q^{9} u^3 \tilde\epsilon^{3}+145824 q^{10} u^3 \tilde\epsilon^{3}+8576 q^{11} u^3 \tilde\epsilon^{3}+1008 q^{12} u^3 \tilde\epsilon^{3}+22944 u^2 \tilde\epsilon^{5}-232624 q u^2 \tilde\epsilon^{5}+5638608 q^{2} u^2 \tilde\epsilon^{5}+8714736 q^{3} u^2 \tilde\epsilon^{5}-13976112 q^{4} u^2 \tilde\epsilon^{5}-25363760 q^{5} u^2 \tilde\epsilon^{5}+14857008 q^{6} u^2 \tilde\epsilon^{5}+21165968 q^{7} u^2 \tilde\epsilon^{5}-6094896 q^{8} u^2 \tilde\epsilon^{5}-4296416 q^{9} u^2 \tilde\epsilon^{5}-447552 q^{10} u^2 \tilde\epsilon^{5}+12096 q^{11} u^2 \tilde\epsilon^{5}+68544 u \tilde\epsilon^{7}+1903008 q u \tilde\epsilon^{7}+42567648 q^{2} u \tilde\epsilon^{7}+240765024 q^{3} u \tilde\epsilon^{7}-47883680 q^{4} u \tilde\epsilon^{7}-452964448 q^{5} u \tilde\epsilon^{7}-40121952 q^{6} u \tilde\epsilon^{7}+223471904 q^{7} u \tilde\epsilon^{7}+31468384 q^{8} u \tilde\epsilon^{7}+677184 q^{9} u \tilde\epsilon^{7}+48384 q^{10} u \tilde\epsilon^{7}+63504 \tilde\epsilon^{9}+4746736 q \tilde\epsilon^{9}+60633440 q^{2} \tilde\epsilon^{9}+185948064 q^{3} \tilde\epsilon^{9}-251391744 q^{4} \tilde\epsilon^{9}-251391744 q^{5} \tilde\epsilon^{9}+185948064 q^{6} \tilde\epsilon^{9}+60633440 q^{7} \tilde\epsilon^{9}+4746736 q^{8} \tilde\epsilon^{9}+63504 q^{9} \tilde\epsilon^{9}
$   \quad \\ $c_{10}= -227 u^5-982 q u^5+5037 q^{2} u^5-3532 q^{3} u^5-8852 q^{4} u^5+15618 q^{5} u^5-3178 q^{6} u^5-13952 q^{7} u^5+14241 q^{8} u^5-698 q^{9} u^5-6691 q^{10} u^5+3660 q^{11} u^5-330 q^{12} u^5-114 q^{13} u^5-7935 u^4 \tilde\epsilon^{2}-15410 q u^4 \tilde\epsilon^{2}-104571 q^{2} u^4 \tilde\epsilon^{2}+243112 q^{3} u^4 \tilde\epsilon^{2}+193506 q^{4} u^4 \tilde\epsilon^{2}-474240 q^{5} u^4 \tilde\epsilon^{2}+294894 q^{6} u^4 \tilde\epsilon^{2}-573888 q^{7} u^4 \tilde\epsilon^{2}+425601 q^{8} u^4 \tilde\epsilon^{2}+319574 q^{9} u^4 \tilde\epsilon^{2}-334515 q^{10} u^4 \tilde\epsilon^{2}+5496 q^{11} u^4 \tilde\epsilon^{2}+24924 q^{12} u^4 \tilde\epsilon^{2}+3452 q^{13} u^4 \tilde\epsilon^{2}-47660 u^3 \tilde\epsilon^{4}+969358 q u^3 \tilde\epsilon^{4}-563718 q^{2} u^3 \tilde\epsilon^{4}-12739128 q^{3} u^3 \tilde\epsilon^{4}-8482126 q^{4} u^3 \tilde\epsilon^{4}+67525642 q^{5} u^3 \tilde\epsilon^{4}-20856020 q^{6} u^3 \tilde\epsilon^{4}-68358536 q^{7} u^3 \tilde\epsilon^{4}+35114656 q^{8} u^3 \tilde\epsilon^{4}+14697338 q^{9} u^3 \tilde\epsilon^{4}-5081062 q^{10} u^3 \tilde\epsilon^{4}-2089280 q^{11} u^3 \tilde\epsilon^{4}-84070 q^{12} u^3 \tilde\epsilon^{4}-5394 q^{13} u^3 \tilde\epsilon^{4}-231790 u^2 \tilde\epsilon^{6}-1075792 q u^2 \tilde\epsilon^{6}-76154206 q^{2} u^2 \tilde\epsilon^{6}-362284976 q^{3} u^2 \tilde\epsilon^{6}+253282108 q^{4} u^2 \tilde\epsilon^{6}+812743968 q^{5} u^2 \tilde\epsilon^{6}-10559868 q^{6} u^2 \tilde\epsilon^{6}-856176896 q^{7} u^2 \tilde\epsilon^{6}+2370826 q^{8} u^2 \tilde\epsilon^{6}+181460208 q^{9} u^2 \tilde\epsilon^{6}+54946010 q^{10} u^2 \tilde\epsilon^{6}+1745136 q^{11} u^2 \tilde\epsilon^{6}-64728 q^{12} u^2 \tilde\epsilon^{6}-492565 u \tilde\epsilon^{8}-22512472 q u \tilde\epsilon^{8}-360209783 q^{2} u \tilde\epsilon^{8}-3520510256 q^{3} u \tilde\epsilon^{8}-3823274802 q^{4} u \tilde\epsilon^{8}+7616770240 q^{5} u \tilde\epsilon^{8}+7183654450 q^{6} u \tilde\epsilon^{8}-3887390064 q^{7} u \tilde\epsilon^{8}-2993196809 q^{8} u \tilde\epsilon^{8}-186098536 q^{9} u \tilde\epsilon^{8}-6480491 q^{10} u \tilde\epsilon^{8}-258912 q^{11} u \tilde\epsilon^{8}-339823 \tilde\epsilon^{10}-29314242 q \tilde\epsilon^{10}-596120771 q^{2} \tilde\epsilon^{10}-3320927384 q^{3} \tilde\epsilon^{10}-583386318 q^{4} \tilde\epsilon^{10}+9060177076 q^{5} \tilde\epsilon^{10}-583386318 q^{6} \tilde\epsilon^{10}-3320927384 q^{7} \tilde\epsilon^{10}-596120771 q^{8} \tilde\epsilon^{10}-29314242 q^{9} \tilde\epsilon^{10}-339823 q^{10} \tilde\epsilon^{10}
$   \quad \\ $c_{11}= 6880 u^5 \tilde\epsilon+43512 q u^5 \tilde\epsilon-87016 q^{2} u^5 \tilde\epsilon-259920 q^{3} u^5 \tilde\epsilon+665392 q^{4} u^5 \tilde\epsilon-124144 q^{5} u^5 \tilde\epsilon-979488 q^{6} u^5 \tilde\epsilon+1179904 q^{7} u^5 \tilde\epsilon-214816 q^{8} u^5 \tilde\epsilon-654312 q^{9} u^5 \tilde\epsilon+519704 q^{10} u^5 \tilde\epsilon-22064 q^{11} u^5 \tilde\epsilon-93808 q^{12} u^5 \tilde\epsilon+16736 q^{13} u^5 \tilde\epsilon+3440 q^{14} u^5 \tilde\epsilon+94000 u^4 \tilde\epsilon^{3}-398160 q u^4 \tilde\epsilon^{3}+2391472 q^{2} u^4 \tilde\epsilon^{3}+6769808 q^{3} u^4 \tilde\epsilon^{3}-14925760 q^{4} u^4 \tilde\epsilon^{3}-21450816 q^{5} u^4 \tilde\epsilon^{3}+32532416 q^{6} u^4 \tilde\epsilon^{3}+26092576 q^{7} u^4 \tilde\epsilon^{3}-30064976 q^{8} u^4 \tilde\epsilon^{3}-15830480 q^{9} u^4 \tilde\epsilon^{3}+12204976 q^{10} u^4 \tilde\epsilon^{3}+5204304 q^{11} u^4 \tilde\epsilon^{3}-2200608 q^{12} u^4 \tilde\epsilon^{3}-387232 q^{13} u^4 \tilde\epsilon^{3}-31520 q^{14} u^4 \tilde\epsilon^{3}+503888 u^3 \tilde\epsilon^{5}-6155632 q u^3 \tilde\epsilon^{5}+16279584 q^{2} u^3 \tilde\epsilon^{5}+247497088 q^{3} u^3 \tilde\epsilon^{5}+239168656 q^{4} u^3 \tilde\epsilon^{5}-1082651072 q^{5} u^3 \tilde\epsilon^{5}-906934448 q^{6} u^3 \tilde\epsilon^{5}+2334744544 q^{7} u^3 \tilde\epsilon^{5}+2252240 q^{8} u^3 \tilde\epsilon^{5}-1018685200 q^{9} u^3 \tilde\epsilon^{5}+45344 q^{10} u^3 \tilde\epsilon^{5}+152266400 q^{11} u^3 \tilde\epsilon^{5}+20868432 q^{12} u^3 \tilde\epsilon^{5}+777792 q^{13} u^3 \tilde\epsilon^{5}+22384 q^{14} u^3 \tilde\epsilon^{5}+2025696 u^2 \tilde\epsilon^{7}+41614400 q u^2 \tilde\epsilon^{7}+825804672 q^{2} u^2 \tilde\epsilon^{7}+8183829792 q^{3} u^2 \tilde\epsilon^{7}+3733549440 q^{4} u^2 \tilde\epsilon^{7}-23517612864 q^{5} u^2 \tilde\epsilon^{7}-13488461056 q^{6} u^2 \tilde\epsilon^{7}+22073888064 q^{7} u^2 \tilde\epsilon^{7}+12170374112 q^{8} u^2 \tilde\epsilon^{7}-6245951808 q^{9} u^2 \tilde\epsilon^{7}-3242848640 q^{10} u^2 \tilde\epsilon^{7}-536036192 q^{11} u^2 \tilde\epsilon^{7}-444224 q^{12} u^2 \tilde\epsilon^{7}+268608 q^{13} u^2 \tilde\epsilon^{7}+3127344 u \tilde\epsilon^{9}+214179160 q u \tilde\epsilon^{9}+3427104616 q^{2} u \tilde\epsilon^{9}+42785299568 q^{3} u \tilde\epsilon^{9}+121506039168 q^{4} u \tilde\epsilon^{9}-49056388784 q^{5} u \tilde\epsilon^{9}-232327603536 q^{6} u \tilde\epsilon^{9}-31985656448 q^{7} u \tilde\epsilon^{9}+113291176592 q^{8} u \tilde\epsilon^{9}+31031467672 q^{9} u \tilde\epsilon^{9}+1053878440 q^{10} u \tilde\epsilon^{9}+56301776 q^{11} u \tilde\epsilon^{9}+1074432 q^{12} u \tilde\epsilon^{9}+1410192 \tilde\epsilon^{11}+161669168 q \tilde\epsilon^{11}+5376925424 q^{2} \tilde\epsilon^{11}+46049526480 q^{3} \tilde\epsilon^{11}+83574372256 q^{4} \tilde\epsilon^{11}-135163903520 q^{5} \tilde\epsilon^{11}-135163903520 q^{6} \tilde\epsilon^{11}+83574372256 q^{7} \tilde\epsilon^{11}+46049526480 q^{8} \tilde\epsilon^{11}+5376925424 q^{9} \tilde\epsilon^{11}+161669168 q^{10} \tilde\epsilon^{11}+1410192 q^{11} \tilde\epsilon^{11}
$   \quad \\ $c_{12}= -2095 u^6-9078 q u^6+48528 q^{2} u^6-23474 q^{3} u^6-129063 q^{4} u^6+185244 q^{5} u^6+30572 q^{6} u^6-259812 q^{7} u^6+194445 q^{8} u^6+50998 q^{9} u^6-145038 q^{10} u^6+47790 q^{11} u^6+24215 q^{12} u^6-10680 q^{13} u^6-4200 q^{14} u^6+1384 q^{15} u^6+306 q^{16} u^6-36 q^{17} u^6-6 q^{18} u^6-122562 u^5 \tilde\epsilon^{2}-696484 q u^5 \tilde\epsilon^{2}-1585664 q^{2} u^5 \tilde\epsilon^{2}+8649980 q^{3} u^5 \tilde\epsilon^{2}+1092198 q^{4} u^5 \tilde\epsilon^{2}-25276456 q^{5} u^5 \tilde\epsilon^{2}+23798528 q^{6} u^5 \tilde\epsilon^{2}+3069072 q^{7} u^5 \tilde\epsilon^{2}-27264438 q^{8} u^5 \tilde\epsilon^{2}+29064076 q^{9} u^5 \tilde\epsilon^{2}-3105504 q^{10} u^5 \tilde\epsilon^{2}-15515444 q^{11} u^5 \tilde\epsilon^{2}+7903906 q^{12} u^5 \tilde\epsilon^{2}+743248 q^{13} u^5 \tilde\epsilon^{2}-722464 q^{14} u^5 \tilde\epsilon^{2}-37848 q^{15} u^5 \tilde\epsilon^{2}+6000 q^{16} u^5 \tilde\epsilon^{2}-144 q^{17} u^5 \tilde\epsilon^{2}-1011815 u^4 \tilde\epsilon^{4}+7911736 q u^4 \tilde\epsilon^{4}-12963746 q^{2} u^4 \tilde\epsilon^{4}-268314856 q^{3} u^4 \tilde\epsilon^{4}-17564777 q^{4} u^4 \tilde\epsilon^{4}+1314213496 q^{5} u^4 \tilde\epsilon^{4}-149488788 q^{6} u^4 \tilde\epsilon^{4}-2233056456 q^{7} u^4 \tilde\epsilon^{4}+411414663 q^{8} u^4 \tilde\epsilon^{4}+1737762696 q^{9} u^4 \tilde\epsilon^{4}-402803314 q^{10} u^4 \tilde\epsilon^{4}-522708952 q^{11} u^4 \tilde\epsilon^{4}+82292393 q^{12} u^4 \tilde\epsilon^{4}+51120584 q^{13} u^4 \tilde\epsilon^{4}+2901992 q^{14} u^4 \tilde\epsilon^{4}+296584 q^{15} u^4 \tilde\epsilon^{4}-1440 q^{16} u^4 \tilde\epsilon^{4}-5097748 u^3 \tilde\epsilon^{6}-9261520 q u^3 \tilde\epsilon^{6}-356062544 q^{2} u^3 \tilde\epsilon^{6}-6490402408 q^{3} u^3 \tilde\epsilon^{6}-6107918756 q^{4} u^3 \tilde\epsilon^{6}+17941186688 q^{5} u^3 \tilde\epsilon^{6}+41531871264 q^{6} u^3 \tilde\epsilon^{6}-43926625784 q^{7} u^3 \tilde\epsilon^{6}-44271863580 q^{8} u^3 \tilde\epsilon^{6}+38422840048 q^{9} u^3 \tilde\epsilon^{6}+11624715600 q^{10} u^3 \tilde\epsilon^{6}-5763518424 q^{11} u^3 \tilde\epsilon^{6}-2409191980 q^{12} u^3 \tilde\epsilon^{6}-174148768 q^{13} u^3 \tilde\epsilon^{6}-6452256 q^{14} u^3 \tilde\epsilon^{6}-69832 q^{15} u^3 \tilde\epsilon^{6}-15584031 u^2 \tilde\epsilon^{8}-634599106 q u^2 \tilde\epsilon^{8}-8881939426 q^{2} u^2 \tilde\epsilon^{8}-133082515706 q^{3} u^2 \tilde\epsilon^{8}-293238795809 q^{4} u^2 \tilde\epsilon^{8}+402189035004 q^{5} u^2 \tilde\epsilon^{8}+752791517668 q^{6} u^2 \tilde\epsilon^{8}-293875984420 q^{7} u^2 \tilde\epsilon^{8}-679780164641 q^{8} u^2 \tilde\epsilon^{8}+42160617574 q^{9} u^2 \tilde\epsilon^{8}+161286154718 q^{10} u^2 \tilde\epsilon^{8}+46946160414 q^{11} u^2 \tilde\epsilon^{8}+4203236577 q^{12} u^2 \tilde\epsilon^{8}-66369952 q^{13} u^2 \tilde\epsilon^{8}-768864 q^{14} u^2 \tilde\epsilon^{8}-17343354 u \tilde\epsilon^{10}-1739159756 q u \tilde\epsilon^{10}-35956817784 q^{2} u \tilde\epsilon^{10}-484521910596 q^{3} u \tilde\epsilon^{10}-2444019540258 q^{4} u \tilde\epsilon^{10}-1579662303032 q^{5} u \tilde\epsilon^{10}+4589163693824 q^{6} u \tilde\epsilon^{10}+4120456235704 q^{7} u \tilde\epsilon^{10}-1834619270494 q^{8} u \tilde\epsilon^{10}-2047448543292 q^{9} u \tilde\epsilon^{10}-274105094024 q^{10} u \tilde\epsilon^{10}-7081298868 q^{11} u \tilde\epsilon^{10}-445627910 q^{12} u \tilde\epsilon^{10}-3020160 q^{13} u \tilde\epsilon^{10}-3940145 \tilde\epsilon^{12}-839692852 q \tilde\epsilon^{12}-44941372642 q^{2} \tilde\epsilon^{12}-564907036260 q^{3} \tilde\epsilon^{12}-2096766693055 q^{4} \tilde\epsilon^{12}+232862848152 q^{5} \tilde\epsilon^{12}+4949191773604 q^{6} \tilde\epsilon^{12}+232862848152 q^{7} \tilde\epsilon^{12}-2096766693055 q^{8} \tilde\epsilon^{12}-564907036260 q^{9} \tilde\epsilon^{12}-44941372642 q^{10} \tilde\epsilon^{12}-839692852 q^{11} \tilde\epsilon^{12}-3940145 q^{12} \tilde\epsilon^{12}
$   \quad \\ $c_{13}= 80160 u^6 \tilde\epsilon+543664 q u^6 \tilde\epsilon-1008288 q^{2} u^6 \tilde\epsilon-4405328 q^{3} u^6 \tilde\epsilon+10358032 q^{4} u^6 \tilde\epsilon+1182640 q^{5} u^6 \tilde\epsilon-21282752 q^{6} u^6 \tilde\epsilon+18173808 q^{7} u^6 \tilde\epsilon+7904768 q^{8} u^6 \tilde\epsilon-23259648 q^{9} u^6 \tilde\epsilon+11467280 q^{10} u^6 \tilde\epsilon+5888480 q^{11} u^6 \tilde\epsilon-7605040 q^{12} u^6 \tilde\epsilon+1719728 q^{13} u^6 \tilde\epsilon+159520 q^{14} u^6 \tilde\epsilon+171440 q^{15} u^6 \tilde\epsilon-76032 q^{16} u^6 \tilde\epsilon-15120 q^{17} u^6 \tilde\epsilon+2352 q^{18} u^6 \tilde\epsilon+336 q^{19} u^6 \tilde\epsilon+1679120 u^5 \tilde\epsilon^{3}+6063128 q u^5 \tilde\epsilon^{3}+59581432 q^{2} u^5 \tilde\epsilon^{3}+15539752 q^{3} u^5 \tilde\epsilon^{3}-453033016 q^{4} u^5 \tilde\epsilon^{3}+129458032 q^{5} u^5 \tilde\epsilon^{3}+828875920 q^{6} u^5 \tilde\epsilon^{3}-577250480 q^{7} u^5 \tilde\epsilon^{3}+158350416 q^{8} u^5 \tilde\epsilon^{3}-362734312 q^{9} u^5 \tilde\epsilon^{3}-205142088 q^{10} u^5 \tilde\epsilon^{3}+556489256 q^{11} u^5 \tilde\epsilon^{3}+21849864 q^{12} u^5 \tilde\epsilon^{3}-222090912 q^{13} u^5 \tilde\epsilon^{3}+20779328 q^{14} u^5 \tilde\epsilon^{3}+22629216 q^{15} u^5 \tilde\epsilon^{3}-724784 q^{16} u^5 \tilde\epsilon^{3}-327936 q^{17} u^5 \tilde\epsilon^{3}+8064 q^{18} u^5 \tilde\epsilon^{3}+10540336 u^4 \tilde\epsilon^{5}-48758512 q u^4 \tilde\epsilon^{5}-146430464 q^{2} u^4 \tilde\epsilon^{5}+5927211920 q^{3} u^4 \tilde\epsilon^{5}+8171570528 q^{4} u^4 \tilde\epsilon^{5}-27324761648 q^{5} u^4 \tilde\epsilon^{5}-41111180864 q^{6} u^4 \tilde\epsilon^{5}+71526240336 q^{7} u^4 \tilde\epsilon^{5}+49660106624 q^{8} u^4 \tilde\epsilon^{5}-75283786576 q^{9} u^4 \tilde\epsilon^{5}-23253844736 q^{10} u^4 \tilde\epsilon^{5}+30521782576 q^{11} u^4 \tilde\epsilon^{5}+7395140768 q^{12} u^4 \tilde\epsilon^{5}-5376568464 q^{13} u^4 \tilde\epsilon^{5}-720673472 q^{14} u^4 \tilde\epsilon^{5}+58559728 q^{15} u^4 \tilde\epsilon^{5}-5228720 q^{16} u^4 \tilde\epsilon^{5}+80640 q^{17} u^4 \tilde\epsilon^{5}+47733936 u^3 \tilde\epsilon^{7}+926836880 q u^3 \tilde\epsilon^{7}+6859073968 q^{2} u^3 \tilde\epsilon^{7}+142796345744 q^{3} u^3 \tilde\epsilon^{7}+293758004144 q^{4} u^3 \tilde\epsilon^{7}-356712778208 q^{5} u^3 \tilde\epsilon^{7}-1296174868064 q^{6} u^3 \tilde\epsilon^{7}+240599275744 q^{7} u^3 \tilde\epsilon^{7}+2209854110656 q^{8} u^3 \tilde\epsilon^{7}-467844491056 q^{9} u^3 \tilde\epsilon^{7}-1084210989072 q^{10} u^3 \tilde\epsilon^{7}+114958749136 q^{11} u^3 \tilde\epsilon^{7}+170707905904 q^{12} u^3 \tilde\epsilon^{7}+22943797888 q^{13} u^3 \tilde\epsilon^{7}+1450749632 q^{14} u^3 \tilde\epsilon^{7}+40107456 q^{15} u^3 \tilde\epsilon^{7}+435312 q^{16} u^3 \tilde\epsilon^{7}+107141376 u^2 \tilde\epsilon^{9}+7320589328 q u^2 \tilde\epsilon^{9}+106823752176 q^{2} u^2 \tilde\epsilon^{9}+1804988806192 q^{3} u^2 \tilde\epsilon^{9}+8594883308624 q^{4} u^2 \tilde\epsilon^{9}-342312893024 q^{5} u^2 \tilde\epsilon^{9}-24749640963296 q^{6} u^2 \tilde\epsilon^{9}-7542804263776 q^{7} u^2 \tilde\epsilon^{9}+22582860496288 q^{8} u^2 \tilde\epsilon^{9}+8946197122128 q^{9} u^2 \tilde\epsilon^{9}-5946722572496 q^{10} u^2 \tilde\epsilon^{9}-2847205125392 q^{11} u^2 \tilde\epsilon^{9}-589072505840 q^{12} u^2 \tilde\epsilon^{9}-26185588480 q^{13} u^2 \tilde\epsilon^{9}+761343168 q^{14} u^2 \tilde\epsilon^{9}+1353024 q^{15} u^2 \tilde\epsilon^{9}+82129536 u \tilde\epsilon^{11}+12666136632 q u \tilde\epsilon^{11}+383227385048 q^{2} u \tilde\epsilon^{11}+5518178943944 q^{3} u \tilde\epsilon^{11}+40539198198504 q^{4} u \tilde\epsilon^{11}+73603296578992 q^{5} u \tilde\epsilon^{11}-46053336387728 q^{6} u \tilde\epsilon^{11}-148264380364912 q^{7} u \tilde\epsilon^{11}-27691173531312 q^{8} u \tilde\epsilon^{11}+70214551368216 q^{9} u \tilde\epsilon^{11}+29421863741752 q^{10} u \tilde\epsilon^{11}+2247502639656 q^{11} u \tilde\epsilon^{11}+65033742024 q^{12} u \tilde\epsilon^{11}+3287104128 q^{13} u \tilde\epsilon^{11}+2315520 q^{14} u \tilde\epsilon^{11}+1705536 \tilde\epsilon^{13}+4425054528 q \tilde\epsilon^{13}+355470697856 q^{2} \tilde\epsilon^{13}+6457627511168 q^{3} \tilde\epsilon^{13}+38190119195840 q^{4} \tilde\epsilon^{13}+45140353971136 q^{5} \tilde\epsilon^{13}-90147998136064 q^{6} \tilde\epsilon^{13}-90147998136064 q^{7} \tilde\epsilon^{13}+45140353971136 q^{8} \tilde\epsilon^{13}+38190119195840 q^{9} \tilde\epsilon^{13}+6457627511168 q^{10} \tilde\epsilon^{13}+355470697856 q^{11} \tilde\epsilon^{13}+4425054528 q^{12} \tilde\epsilon^{13}+1705536 q^{13} \tilde\epsilon^{13}$
}

\subsection{Hurwitz expansion for PIV}
\label{app:PIV}
In this subsection \(c_n=c_n^{PIV}\) from \eqref{eq:cnPIV}.

($\alpha=g_2^{PIV}/2\ ,\ \beta=2g_3^{PIV}$)
{\footnotesize
  \quad \\ $c_{0}= 1
$   \quad \\ $c_{1}= 0
$   \quad \\ $c_{2}= -3 T
$   \quad \\ $c_{3}= 4 \epsilon \Lambda
$   \quad \\ $c_{4}= 15 T^{2}-\alpha+8 \epsilon^{2}
$   \quad \\ $c_{5}= 24 u \epsilon-36 T \epsilon \Lambda
$   \quad \\ $c_{6}= -105 T^{3}+21 T \alpha-3 \beta-168 T \epsilon^{2}+106 \epsilon^{2} \Lambda^2
$   \quad \\ $c_{7}= -672 T u \epsilon+168 T^{2} \epsilon \Lambda+8 \alpha \epsilon \Lambda+176 \epsilon^{3} \Lambda
$   \quad \\ $c_{8}= 945 T^{4}-378 T^{2} \alpha-9 \alpha^2+108 T \beta+3024 T^{2} \epsilon^{2}+48 \alpha \epsilon^{2}+192 \epsilon^{4}+816 u \epsilon^{2} \Lambda-3000 T \epsilon^{2} \Lambda^2
$   \quad \\ $c_{9}= 15120 T^{2} u \epsilon+336 u \alpha \epsilon+672 u \epsilon^{3}+2520 T^{3} \epsilon \Lambda-24 T \alpha \epsilon \Lambda-216 \beta \epsilon \Lambda-7248 T \epsilon^{3} \Lambda+5392 \epsilon^{3} \Lambda^3
$   \quad \\ $c_{10}= -10395 T^{5}+6930 T^{3} \alpha+495 T \alpha^2-2970 T^{2} \beta-18 \alpha \beta-55440 T^{3} \epsilon^{2}-4704 u^2 \epsilon^{2}-2640 T \alpha \epsilon^{2}-432 \beta \epsilon^{2}-10560 T \epsilon^{4}-54288 T u \epsilon^{2} \Lambda+55356 T^{2} \epsilon^{2} \Lambda^2+3020 \alpha \epsilon^{2} \Lambda^2+10592 \epsilon^{4} \Lambda^2
$   \quad \\ $c_{11}= -332640 T^{3} u \epsilon-22176 T u \alpha \epsilon+864 u \beta \epsilon-44352 T u \epsilon^{3}-124740 T^{4} \epsilon \Lambda-10296 T^{2} \alpha \epsilon \Lambda-1404 \alpha^2 \epsilon \Lambda+15120 T \beta \epsilon \Lambda+217008 T^{2} \epsilon^{3} \Lambda+20400 \alpha \epsilon^{3} \Lambda-12864 \epsilon^{5} \Lambda+8832 u \epsilon^{3} \Lambda^2-347040 T \epsilon^{3} \Lambda^3
$   \quad \\ $c_{12}= 135135 T^{6}-135135 T^{4} \alpha-19305 T^{2} \alpha^2+69 \alpha^3+77220 T^{3} \beta+1404 T \alpha \beta-54 \beta^2+1081080 T^{4} \epsilon^{2}+366912 T u^2 \epsilon^{2}+102960 T^{2} \alpha \epsilon^{2}-5592 \alpha^2 \epsilon^{2}+33696 T \beta \epsilon^{2}+411840 T^{2} \epsilon^{4}+45504 \alpha \epsilon^{4}-41472 \epsilon^{6}+2484144 T^{2} u \epsilon^{2} \Lambda+67632 u \alpha \epsilon^{2} \Lambda-119616 u \epsilon^{4} \Lambda-611208 T^{3} \epsilon^{2} \Lambda^2-167928 T \alpha \epsilon^{2} \Lambda^2-16776 \beta \epsilon^{2} \Lambda^2-945792 T \epsilon^{4} \Lambda^2+409288 \epsilon^{4} \Lambda^4
$   \quad \\ $c_{13}= 7567560 T^{4} u \epsilon+1009008 T^{2} u \alpha \epsilon-6600 u \alpha^2 \epsilon-78624 T u \beta \epsilon+2018016 T^{2} u \epsilon^{3}+328416 u \alpha \epsilon^{3}-874368 u \epsilon^{5}+3783780 T^{5} \epsilon \Lambda+648648 T^{3} \alpha \epsilon \Lambda+121164 T \alpha^2 \epsilon \Lambda-727272 T^{2} \beta \epsilon \Lambda-4392 \alpha \beta \epsilon \Lambda-5909904 T^{3} \epsilon^{3} \Lambda-1337856 u^2 \epsilon^{3} \Lambda-1527984 T \alpha \epsilon^{3} \Lambda-87984 \beta \epsilon^{3} \Lambda+296256 T \epsilon^{5} \Lambda-3479424 T u \epsilon^{3} \Lambda^2+14050608 T^{2} \epsilon^{3} \Lambda^3+451760 \alpha \epsilon^{3} \Lambda^3+1461920 \epsilon^{5} \Lambda^3
$   \quad \\ $c_{14}= -2027025 T^{7}+2837835 T^{5} \alpha+675675 T^{3} \alpha^2-7245 T \alpha^3-2027025 T^{4} \beta-73710 T^{2} \alpha \beta+513 \alpha^2 \beta+5670 T \beta^2-22702680 T^{5} \epsilon^{2}-19262880 T^{2} u^2 \epsilon^{2}-3603600 T^{3} \alpha \epsilon^{2}+300384 u^2 \alpha \epsilon^{2}+587160 T \alpha^2 \epsilon^{2}-1769040 T^{2} \beta \epsilon^{2}-32112 \alpha \beta \epsilon^{2}-14414400 T^{3} \epsilon^{4}-6878592 u^2 \epsilon^{4}-4777920 T \alpha \epsilon^{4}-52416 \beta \epsilon^{4}+4354560 T \epsilon^{6}-99786960 T^{3} u \epsilon^{2} \Lambda-6500592 T u \alpha \epsilon^{2} \Lambda+325008 u \beta \epsilon^{2} \Lambda-1197504 T u \epsilon^{4} \Lambda-8902530 T^{4} \epsilon^{2} \Lambda^2+5565924 T^{2} \alpha \epsilon^{2} \Lambda^2-213918 \alpha^2 \epsilon^{2} \Lambda^2+2086488 T \beta \epsilon^{2} \Lambda^2+49055328 T^{2} \epsilon^{4} \Lambda^2+5165088 \alpha \epsilon^{4} \Lambda^2-5384832 \epsilon^{6} \Lambda^2-6088416 u \epsilon^{4} \Lambda^3-49063656 T \epsilon^{4} \Lambda^4
$   \quad \\ $c_{15}= -181621440 T^{5} u \epsilon-40360320 T^{3} u \alpha \epsilon+792000 T u \alpha^2 \epsilon+4717440 T^{2} u \beta \epsilon-74880 u \alpha \beta \epsilon-80720640 T^{3} u \epsilon^{3}-5031936 u^3 \epsilon^{3}-39409920 T u \alpha \epsilon^{3}+2006784 u \beta \epsilon^{3}+104924160 T u \epsilon^{5}-105945840 T^{6} \epsilon \Lambda-29549520 T^{4} \alpha \epsilon \Lambda-6873840 T^{2} \alpha^2 \epsilon \Lambda+16752 \alpha^3 \epsilon \Lambda+30663360 T^{3} \beta \epsilon \Lambda+452160 T \alpha \beta \epsilon \Lambda-26784 \beta^2 \epsilon \Lambda+157116960 T^{4} \epsilon^{3} \Lambda+145446912 T u^2 \epsilon^{3} \Lambda+71974080 T^{2} \alpha \epsilon^{3} \Lambda-1699104 \alpha^2 \epsilon^{3} \Lambda+12564864 T \beta \epsilon^{3} \Lambda+34686720 T^{2} \epsilon^{5} \Lambda+20543232 \alpha \epsilon^{5} \Lambda-29585664 \epsilon^{7} \Lambda+354212352 T^{2} u \epsilon^{3} \Lambda^2+12129792 u \alpha \epsilon^{3} \Lambda^2-84409344 u \epsilon^{5} \Lambda^2-443953536 T^{3} \epsilon^{3} \Lambda^3-42081408 T \alpha \epsilon^{3} \Lambda^3-1292160 \beta \epsilon^{3} \Lambda^3-259839744 T \epsilon^{5} \Lambda^3+41785216 \epsilon^{5} \Lambda^5
$   \quad \\ $c_{16}= 34459425 T^{8}-64324260 T^{6} \alpha-22972950 T^{4} \alpha^2+492660 T^{2} \alpha^3+321 \alpha^4+55135080 T^{5} \beta+3341520 T^{3} \alpha \beta-69768 T \alpha^2 \beta-385560 T^{2} \beta^2+4968 \alpha \beta^2+514594080 T^{6} \epsilon^{2}+873250560 T^{3} u^2 \epsilon^{2}+122522400 T^{4} \alpha \epsilon^{2}-40852224 T u^2 \alpha \epsilon^{2}-39926880 T^{2} \alpha^2 \epsilon^{2}+22560 \alpha^3 \epsilon^{2}+80196480 T^{3} \beta \epsilon^{2}+2306304 u^2 \beta \epsilon^{2}+4367232 T \alpha \beta \epsilon^{2}-188352 \beta^2 \epsilon^{2}+490089600 T^{4} \epsilon^{4}+935488512 T u^2 \epsilon^{4}+324898560 T^{2} \alpha \epsilon^{4}-3758208 \alpha^2 \epsilon^{4}+7128576 T \beta \epsilon^{4}-296110080 T^{2} \epsilon^{6}+34498560 \alpha \epsilon^{6}-48328704 \epsilon^{8}+3829381920 T^{4} u \epsilon^{2} \Lambda+401188032 T^{2} u \alpha \epsilon^{2} \Lambda-1199136 u \alpha^2 \epsilon^{2} \Lambda-39588480 T u \beta \epsilon^{2} \Lambda+1016918784 T^{2} u \epsilon^{4} \Lambda+108258048 u \alpha \epsilon^{4} \Lambda-422165760 u \epsilon^{6} \Lambda+1008025200 T^{5} \epsilon^{2} \Lambda^2-118592544 T^{3} \alpha \epsilon^{2} \Lambda^2+27893712 T \alpha^2 \epsilon^{2} \Lambda^2-161675424 T^{2} \beta \epsilon^{2} \Lambda^2-661920 \alpha \beta \epsilon^{2} \Lambda^2-1884868608 T^{3} \epsilon^{4} \Lambda^2-342421248 u^2 \epsilon^{4} \Lambda^2-594193920 T \alpha \epsilon^{4} \Lambda^2-13396992 \beta \epsilon^{4} \Lambda^2+310171392 T \epsilon^{6} \Lambda^2+143182080 T u \epsilon^{4} \Lambda^3+3407919648 T^{2} \epsilon^{4} \Lambda^4+69198368 \alpha \epsilon^{4} \Lambda^4+281868032 \epsilon^{6} \Lambda^4\\
$   \quad  $c_{17}= 4631346720 T^{6} u \epsilon+1543782240 T^{4} u \alpha \epsilon-60588000 T^{2} u \alpha^2 \epsilon-209952 u \alpha^3 \epsilon-240589440 T^{3} u \beta \epsilon+11456640 T u \alpha \beta \epsilon-357696 u \beta^2 \epsilon+3087564480 T^{4} u \epsilon^{3}+769886208 T u^3 \epsilon^{3}+3014858880 T^{2} u \alpha \epsilon^{3}-245952 u \alpha^2 \epsilon^{3}-307037952 T u \beta \epsilon^{3}-8026698240 T^{2} u \epsilon^{5}+274853376 u \alpha \epsilon^{5}-1037753856 u \epsilon^{7}+2977294320 T^{7} \epsilon \Lambda+1212971760 T^{5} \alpha \epsilon \Lambda+330369840 T^{3} \alpha^2 \epsilon \Lambda-2773008 T \alpha^3 \epsilon \Lambda-1233020880 T^{4} \beta \epsilon \Lambda-28861920 T^{2} \alpha \beta \epsilon \Lambda-14256 \alpha^2 \beta \epsilon \Lambda+3740256 T \beta^2 \epsilon \Lambda-4190266080 T^{5} \epsilon^{3} \Lambda-9971859456 T^{2} u^2 \epsilon^{3} \Lambda-2665725120 T^{3} \alpha \epsilon^{3} \Lambda+82096128 u^2 \alpha \epsilon^{3} \Lambda+259716960 T \alpha^2 \epsilon^{3} \Lambda-1114731072 T^{2} \beta \epsilon^{3} \Lambda-7960896 \alpha \beta \epsilon^{3} \Lambda-4444588800 T^{3} \epsilon^{5} \Lambda-3243423744 u^2 \epsilon^{5} \Lambda-2868261120 T \alpha \epsilon^{5} \Lambda-46994688 \beta \epsilon^{5} \Lambda+3488852736 T \epsilon^{7} \Lambda-24712736256 T^{3} u \epsilon^{3} \Lambda^2-1691665920 T u \alpha \epsilon^{3} \Lambda^2+100500480 u \beta \epsilon^{3} \Lambda^2+6427782144 T u \epsilon^{5} \Lambda^2+10803038688 T^{4} \epsilon^{3} \Lambda^3+2373394752 T^{2} \alpha \epsilon^{3} \Lambda^3-41310432 \alpha^2 \epsilon^{3} \Lambda^3+298200960 T \beta \epsilon^{3} \Lambda^3+23091631488 T^{2} \epsilon^{5} \Lambda^3+1259829120 \alpha \epsilon^{5} \Lambda^3-862029312 \epsilon^{7} \Lambda^3-1992082176 u \epsilon^{5} \Lambda^4-8385220224 T \epsilon^{5} \Lambda^5\\
$   \quad $c_{18}= -654729075 T^{9}+1571349780 T^{7} \alpha+785674890 T^{5} \alpha^2-28081620 T^{3} \alpha^3-54891 T \alpha^4-1571349780 T^{6} \beta-142849980 T^{4} \alpha \beta+5965164 T^{2} \alpha^2 \beta+33588 \alpha^3 \beta+21976920 T^{3} \beta^2-849528 T \alpha \beta^2+14904 \beta^3-12570798240 T^{7} \epsilon^{2}-37331461440 T^{4} u^2 \epsilon^{2}-4190266080 T^{5} \alpha \epsilon^{2}+3492865152 T^{2} u^2 \alpha \epsilon^{2}+2275832160 T^{3} \alpha^2 \epsilon^{2}+17798976 u^2 \alpha^2 \epsilon^{2}-3857760 T \alpha^3 \epsilon^{2}-3428399520 T^{4} \beta \epsilon^{2}-394377984 T u^2 \beta \epsilon^{2}-373398336 T^{2} \alpha \beta \epsilon^{2}-1124064 \alpha^2 \beta \epsilon^{2}+32208192 T \beta^2 \epsilon^{2}-16761064320 T^{5} \epsilon^{4}-79984267776 T^{2} u^2 \epsilon^{4}-18519217920 T^{3} \alpha \epsilon^{4}+176721408 u^2 \alpha \epsilon^{4}+642653568 T \alpha^2 \epsilon^{4}-609493248 T^{2} \beta \epsilon^{4}-25733376 \alpha \beta \epsilon^{4}+16878274560 T^{3} \epsilon^{6}-8399144448 u^2 \epsilon^{6}-5899253760 T \alpha \epsilon^{6}+12883968 \beta \epsilon^{6}+8264208384 T \epsilon^{8}-145897446240 T^{5} u \epsilon^{2} \Lambda-20539141056 T^{3} u \alpha \epsilon^{2} \Lambda+240650208 T u \alpha^2 \epsilon^{2} \Lambda+2990437056 T^{2} u \beta \epsilon^{2} \Lambda-22496832 u \alpha \beta \epsilon^{2} \Lambda-111287215872 T^{3} u \epsilon^{4} \Lambda-2279107584 u^3 \epsilon^{4} \Lambda-18158683392 T u \alpha \epsilon^{4} \Lambda+1114801920 u \beta \epsilon^{4} \Lambda+55392056064 T u \epsilon^{6} \Lambda-53044959240 T^{6} \epsilon^{2} \Lambda^2-64954008 T^{4} \alpha \epsilon^{2} \Lambda^2-2264587272 T^{2} \alpha^2 \epsilon^{2} \Lambda^2+8415432 \alpha^3 \epsilon^{2} \Lambda^2+10212311520 T^{3} \beta \epsilon^{2} \Lambda^2+90691488 T \alpha \beta \epsilon^{2} \Lambda^2-9951984 \beta^2 \epsilon^{2} \Lambda^2+52756329024 T^{4} \epsilon^{4} \Lambda^2+51716710656 T u^2 \epsilon^{4} \Lambda^2+41724238464 T^{2} \alpha \epsilon^{4} \Lambda^2-586131264 \alpha^2 \epsilon^{4} \Lambda^2+3405687552 T \beta \epsilon^{4} \Lambda^2+1176374016 T^{2} \epsilon^{6} \Lambda^2+8935803648 \alpha \epsilon^{6} \Lambda^2-14061376512 \epsilon^{8} \Lambda^2+39474642816 T^{2} u \epsilon^{4} \Lambda^3+2405404032 u \alpha \epsilon^{4} \Lambda^3-34279371264 u \epsilon^{6} \Lambda^3-181093205664 T^{3} \epsilon^{4} \Lambda^4-9427516896 T \alpha \epsilon^{4} \Lambda^4-35708256 \beta \epsilon^{4} \Lambda^4-82478804736 T \epsilon^{6} \Lambda^4+5457048256 \epsilon^{6} \Lambda^6
$   \quad \\ $c_{19}= -125707982400 T^{7} u \epsilon-58663725120 T^{5} u \alpha \epsilon+3837240000 T^{3} u \alpha^2 \epsilon+39890880 T u \alpha^3 \epsilon+11427998400 T^{4} u \beta \epsilon-1088380800 T^{2} u \alpha \beta \epsilon-5434560 u \alpha^2 \beta \epsilon+67962240 T u \beta^2 \epsilon-117327450240 T^{5} u \epsilon^{3}-73139189760 T^{2} u^3 \epsilon^{3}-190941062400 T^{3} u \alpha \epsilon^{3}-554932224 u^3 \alpha \epsilon^{3}+46730880 T u \alpha^2 \epsilon^{3}+29168605440 T^{2} u \beta \epsilon^{3}-37718784 u \alpha \beta \epsilon^{3}+508357555200 T^{3} u \epsilon^{5}-6944329728 u^3 \epsilon^{5}-52222141440 T u \alpha \epsilon^{5}+3166497792 u \beta \epsilon^{5}+197173232640 T u \epsilon^{7}-86424237900 T^{8} \epsilon \Lambda-48188059920 T^{6} \alpha \epsilon \Lambda-14733257400 T^{4} \alpha^2 \epsilon \Lambda+283381200 T^{2} \alpha^3 \epsilon \Lambda-1425420 \alpha^4 \epsilon \Lambda+49140393120 T^{5} \beta \epsilon \Lambda+1465128000 T^{3} \alpha \beta \epsilon \Lambda-2725920 T \alpha^2 \beta \epsilon \Lambda-321343200 T^{2} \beta^2 \epsilon \Lambda+2293920 \alpha \beta^2 \epsilon \Lambda+113137184160 T^{6} \epsilon^{3} \Lambda+558411909120 T^{3} u^2 \epsilon^{3} \Lambda+78886677600 T^{4} \alpha \epsilon^{3} \Lambda-17263060992 T u^2 \alpha \epsilon^{3} \Lambda-24649745760 T^{2} \alpha^2 \epsilon^{3} \Lambda+96553440 \alpha^3 \epsilon^{3} \Lambda+80322503040 T^{3} \beta \epsilon^{3} \Lambda+1313998848 u^2 \beta \epsilon^{3} \Lambda+1474851456 T \alpha \beta \epsilon^{3} \Lambda-134830656 \beta^2 \epsilon^{3} \Lambda+338207356800 T^{4} \epsilon^{5} \Lambda+595417522176 T u^2 \epsilon^{5} \Lambda+246373735680 T^{2} \alpha \epsilon^{5} \Lambda-3282111360 \alpha^2 \epsilon^{5} \Lambda+12095488512 T \beta \epsilon^{5} \Lambda-232854393600 T^{2} \epsilon^{7} \Lambda+31279299840 \alpha \epsilon^{7} \Lambda-48489661440 \epsilon^{9} \Lambda+1453060926720 T^{4} u \epsilon^{3} \Lambda^2+143445201408 T^{2} u \alpha \epsilon^{3} \Lambda^2-382435584 u \alpha^2 \epsilon^{3} \Lambda^2-16467093504 T u \beta \epsilon^{3} \Lambda^2-15221781504 T^{2} u \epsilon^{5} \Lambda^2+34225959936 u \alpha \epsilon^{5} \Lambda^2-228048786432 u \epsilon^{7} \Lambda^2-119903284800 T^{5} \epsilon^{3} \Lambda^3-102499933824 T^{3} \alpha \epsilon^{3} \Lambda^3+7466546496 T \alpha^2 \epsilon^{3} \Lambda^3-36562637952 T^{2} \beta \epsilon^{3} \Lambda^3-49680000 \alpha \beta \epsilon^{3} \Lambda^3-1467543921408 T^{3} \epsilon^{5} \Lambda^3-96641872896 u^2 \epsilon^{5} \Lambda^3-205141572864 T \alpha \epsilon^{5} \Lambda^3-973419264 \beta \epsilon^{5} \Lambda^3-64263217152 T \epsilon^{7} \Lambda^3+185211867648 T u \epsilon^{5} \Lambda^4+889201855104 T^{2} \epsilon^{5} \Lambda^5+11814500480 \alpha \epsilon^{5} \Lambda^5+63492215552 \epsilon^{7} \Lambda^5
$   \quad \\ $c_{20}= 13749310575 T^{10}-41247931725 T^{8} \alpha-27498621150 T^{6} \alpha^2+1474285050 T^{4} \alpha^3+5763555 T^{2} \alpha^4+160839 \alpha^5+47140493400 T^{7} \beta+5999699160 T^{5} \alpha \beta-417561480 T^{3} \alpha^2 \beta-7053480 T \alpha^3 \beta-1153788300 T^{4} \beta^2+89200440 T^{2} \alpha \beta^2+257580 \alpha^2 \beta^2-3129840 T \beta^3+329983453800 T^{8} \epsilon^{2}+1567921380480 T^{5} u^2 \epsilon^{2}+146659312800 T^{6} \alpha \epsilon^{2}-244500560640 T^{3} u^2 \alpha \epsilon^{2}-119481188400 T^{4} \alpha^2 \epsilon^{2}-3737784960 T u^2 \alpha^2 \epsilon^{2}+405064800 T^{2} \alpha^3 \epsilon^{2}-14713560 \alpha^4 \epsilon^{2}+143992779840 T^{5} \beta \epsilon^{2}+41409688320 T^{2} u^2 \beta \epsilon^{2}+26137883520 T^{3} \alpha \beta \epsilon^{2}+240848640 u^2 \alpha \beta \epsilon^{2}+236053440 T \alpha^2 \beta \epsilon^{2}-3381860160 T^{2} \beta^2 \epsilon^{2}+7801920 \alpha \beta^2 \epsilon^{2}+586637251200 T^{6} \epsilon^{4}+5598898744320 T^{3} u^2 \epsilon^{4}+7100061696 u^4 \epsilon^{4}+972258940800 T^{4} \alpha \epsilon^{4}-37111495680 T u^2 \alpha \epsilon^{4}-67478624640 T^{2} \alpha^2 \epsilon^{4}+306453888 \alpha^3 \epsilon^{4}+42664527360 T^{3} \beta \epsilon^{4}+4434269184 u^2 \beta \epsilon^{4}+5404008960 T \alpha \beta \epsilon^{4}-419302656 \beta^2 \epsilon^{4}-886109414400 T^{4} \epsilon^{6}+1763820334080 T u^2 \epsilon^{6}+619421644800 T^{2} \alpha \epsilon^{6}-6472535040 \alpha^2 \epsilon^{6}-2705633280 T \beta \epsilon^{6}-867741880320 T^{2} \epsilon^{8}+47779098624 \alpha \epsilon^{8}-69898567680 \epsilon^{10}+5629051078560 T^{6} u \epsilon^{2} \Lambda+956054625120 T^{4} u \alpha \epsilon^{2} \Lambda-29006056800 T^{2} u \alpha^2 \epsilon^{2} \Lambda+90390240 u \alpha^3 \epsilon^{2} \Lambda-181724135040 T^{3} u \beta \epsilon^{2} \Lambda+5206032000 T u \alpha \beta \epsilon^{2} \Lambda-269075520 u \beta^2 \epsilon^{2} \Lambda+8642028205440 T^{4} u \epsilon^{4} \Lambda+507012839424 T u^3 \epsilon^{4} \Lambda+1869550260480 T^{2} u \alpha \epsilon^{4} \Lambda-5092460928 u \alpha^2 \epsilon^{4} \Lambda-225239864832 T u \beta \epsilon^{4} \Lambda-4052345552640 T^{2} u \epsilon^{6} \Lambda+208936915200 u \alpha \epsilon^{6} \Lambda-887114769408 u \epsilon^{8} \Lambda+2395498931280 T^{7} \epsilon^{2} \Lambda^2+193938993360 T^{5} \alpha \epsilon^{2} \Lambda^2+148852423440 T^{3} \alpha^2 \epsilon^{2} \Lambda^2-1676850480 T \alpha^3 \epsilon^{2} \Lambda^2-581577388560 T^{4} \beta \epsilon^{2} \Lambda^2-6919590240 T^{2} \alpha \beta \epsilon^{2} \Lambda^2-61207920 \alpha^2 \beta \epsilon^{2} \Lambda^2+1820841120 T \beta^2 \epsilon^{2} \Lambda^2-487360177920 T^{5} \epsilon^{4} \Lambda^2-4669735359744 T^{2} u^2 \epsilon^{4} \Lambda^2-2297513272320 T^{3} \alpha \epsilon^{4} \Lambda^2+20166181632 u^2 \alpha \epsilon^{4} \Lambda^2+117995104512 T \alpha^2 \epsilon^{4} \Lambda^2-470217125376 T^{2} \beta \epsilon^{4} \Lambda^2-116370432 \alpha \beta \epsilon^{4} \Lambda^2-1433128032000 T^{3} \epsilon^{6} \Lambda^2-1374321567744 u^2 \epsilon^{6} \Lambda^2-1667581850880 T \alpha \epsilon^{6} \Lambda^2-18139292928 \beta \epsilon^{6} \Lambda^2+2065774298112 T \epsilon^{8} \Lambda^2-5876381903616 T^{3} u \epsilon^{4} \Lambda^3-464802483456 T u \alpha \epsilon^{4} \Lambda^3+33048255744 u \beta \epsilon^{4} \Lambda^3+4450024829952 T u \epsilon^{6} \Lambda^3+8038297821456 T^{4} \epsilon^{4} \Lambda^4+757488032352 T^{2} \alpha \epsilon^{4} \Lambda^4-10803767952 \alpha^2 \epsilon^{4} \Lambda^4+40546989504 T \beta \epsilon^{4} \Lambda^4+10885286912256 T^{2} \epsilon^{6} \Lambda^4+324650481408 \alpha \epsilon^{6} \Lambda^4+5879732736 \epsilon^{8} \Lambda^4-594605991168 u \epsilon^{6} \Lambda^5-1740586124928 T \epsilon^{6} \Lambda^6$
}

\subsection{Hurwitz expansion for PIII$_2$ and PIII$_3$}
\label{app:PIII}
In this subsection, for PIII$_2$ we denote \(c_n=c_n^{PIII_2}\) from~\eqref{eq:cnPIII} and for PIII$_3$ we denote \(c_n=c_n^{PIII_3}\) from~\eqref{eq:cnPIII3}.

\noindent(PIII$_2$: $\alpha=g_2^{PIII_2}/2\ ,\ \beta=2g_3^{PIII_2},\ \tilde\epsilon=\epsilon/2$;\ PIII$_3$: $\alpha=g_2^{PIII_3}/2\ ,\ \beta=2g_3^{PIII_3},\ \tilde\epsilon=\epsilon/2$)\footnote{The coefficients of the Hurwitz expansion for PIII$_2$ and PIII$_3$ in the basis of $g_2,g_3,T,\epsilon$ are the same. What changes is the parametrization of this basis in terms of the gauge theory parameters $u,\Lambda,m$ for PIII$_2\ (N_f=1)$  and in terms of $u,\Lambda$ for PIII$_3\ (N_f=0)$.}
{\footnotesize
  \quad \\ $c_{0}= 1
$   \quad \\ $c_{1}= 0
$   \quad \\ $c_{2}= -3 T-\tilde\epsilon^{2}
$   \quad \\ $c_{3}= 0
$   \quad \\ $c_{4}= 15 T^2-\alpha+6 T \tilde\epsilon^{2}+\tilde\epsilon^{4}
$   \quad \\ $c_{5}= 24 T^2 \tilde\epsilon-4 \alpha \tilde\epsilon
$   \quad \\ $c_{6}= -105 T^3+21 T \alpha-3 \beta+27 T^2 \tilde\epsilon^{2}-9 \alpha \tilde\epsilon^{2}-9 T \tilde\epsilon^{4}-\tilde\epsilon^{6}
$   \quad \\ $c_{7}= -672 T^3 \tilde\epsilon+240 T \alpha \tilde\epsilon-48 \beta \tilde\epsilon+96 T^2 \tilde\epsilon^{3}-16 \alpha \tilde\epsilon^{3}
$   \quad \\ $c_{8}= 945 T^4-378 T^2 \alpha-9 \alpha^2+108 T \beta-4380 T^3 \tilde\epsilon^{2}+1932 T \alpha \tilde\epsilon^{2}-444 \beta \tilde\epsilon^{2}+186 T^2 \tilde\epsilon^{4}-22 \alpha \tilde\epsilon^{4}+12 T \tilde\epsilon^{6}+\tilde\epsilon^{8}
$   \quad \\ $c_{9}= 15120 T^4 \tilde\epsilon-7944 T^2 \alpha \tilde\epsilon-184 \alpha^2 \tilde\epsilon+2448 T \beta \tilde\epsilon-26784 T^3 \tilde\epsilon^{3}+12528 T \alpha \tilde\epsilon^{3}-3024 \beta \tilde\epsilon^{3}+48 T^2 \tilde\epsilon^{5}-8 \alpha \tilde\epsilon^{5}
$   \quad \\ $c_{10}= -10395 T^5+6930 T^3 \alpha+495 T \alpha^2-2970 T^2 \beta-18 \alpha \beta+185211 T^4 \tilde\epsilon^{2}-101766 T^2 \alpha \tilde\epsilon^{2}-2003 \alpha^2 \tilde\epsilon^{2}+31068 T \beta \tilde\epsilon^{2}-136026 T^3 \tilde\epsilon^{4}+65778 T \alpha \tilde\epsilon^{4}-16182 \beta \tilde\epsilon^{4}-582 T^2 \tilde\epsilon^{6}+82 \alpha \tilde\epsilon^{6}-15 T \tilde\epsilon^{8}-\tilde\epsilon^{10}
$   \quad \\ $c_{11}= -332640 T^5 \tilde\epsilon+223344 T^3 \alpha \tilde\epsilon+13296 T \alpha^2 \tilde\epsilon-89424 T^2 \beta \tilde\epsilon-576 \alpha \beta \tilde\epsilon+1749600 T^4 \tilde\epsilon^{3}-955824 T^2 \alpha \tilde\epsilon^{3}-15376 \alpha^2 \tilde\epsilon^{3}+283680 T \beta \tilde\epsilon^{3}-549600 T^3 \tilde\epsilon^{5}+270672 T \alpha \tilde\epsilon^{5}-67152 \beta \tilde\epsilon^{5}-2208 T^2 \tilde\epsilon^{7}+368 \alpha \tilde\epsilon^{7}
$   \quad \\ $c_{12}= 135135 T^6-135135 T^4 \alpha-19305 T^2 \alpha^2+69 \alpha^3+77220 T^3 \beta+1404 T \alpha \beta-54 \beta^2-6286878 T^5 \tilde\epsilon^{2}+3968340 T^3 \alpha \tilde\epsilon^{2}+191382 T \alpha^2 \tilde\epsilon^{2}-1469196 T^2 \beta \tilde\epsilon^{2}-9420 \alpha \beta \tilde\epsilon^{2}+12987711 T^4 \tilde\epsilon^{4}-7002870 T^2 \alpha \tilde\epsilon^{4}-91015 \alpha^2 \tilde\epsilon^{4}+2019204 T \beta \tilde\epsilon^{4}-1507596 T^3 \tilde\epsilon^{6}+748668 T \alpha \tilde\epsilon^{6}-186492 \beta \tilde\epsilon^{6}-6207 T^2 \tilde\epsilon^{8}+1057 \alpha \tilde\epsilon^{8}+18 T \tilde\epsilon^{10}+\tilde\epsilon^{12}
$   \quad \\ $c_{13}= 7567560 T^6 \tilde\epsilon-6018012 T^4 \alpha \tilde\epsilon-663984 T^2 \alpha^2 \tilde\epsilon+2764 \alpha^3 \tilde\epsilon+2948400 T^3 \beta \tilde\epsilon+56736 T \alpha \beta \tilde\epsilon-3024 \beta^2 \tilde\epsilon-82356768 T^5 \tilde\epsilon^{3}+49342800 T^3 \alpha \tilde\epsilon^{3}+1938000 T \alpha^2 \tilde\epsilon^{3}-17086608 T^2 \beta \tilde\epsilon^{3}-105024 \alpha \beta \tilde\epsilon^{3}+76266000 T^4 \tilde\epsilon^{5}-40546056 T^2 \alpha \tilde\epsilon^{5}-419192 \alpha^2 \tilde\epsilon^{5}+11381328 T \beta \tilde\epsilon^{5}-137952 T^3 \tilde\epsilon^{7}+70224 T \alpha \tilde\epsilon^{7}-17712 \beta \tilde\epsilon^{7}-14328 T^2 \tilde\epsilon^{9}+2388 \alpha \tilde\epsilon^{9}
$   \quad \\ $c_{14}= -2027025 T^7+2837835 T^5 \alpha+675675 T^3 \alpha^2-7245 T \alpha^3-2027025 T^4 \beta-73710 T^2 \alpha \beta+513 \alpha^2 \beta+5670 T \beta^2+199971135 T^6 \tilde\epsilon^{2}-138607371 T^4 \alpha \tilde\epsilon^{2}-12259593 T^2 \alpha^2 \tilde\epsilon^{2}+62233 \alpha^3 \tilde\epsilon^{2}+60239196 T^3 \beta \tilde\epsilon^{2}+1237284 T \alpha \beta \tilde\epsilon^{2}-89982 \beta^2 \tilde\epsilon^{2}-815543415 T^5 \tilde\epsilon^{4}+469961274 T^3 \alpha \tilde\epsilon^{4}+15178059 T \alpha^2 \tilde\epsilon^{4}-154077642 T^2 \beta \tilde\epsilon^{4}-889554 \alpha \beta \tilde\epsilon^{4}+337815549 T^4 \tilde\epsilon^{6}-176882202 T^2 \alpha \tilde\epsilon^{6}-1369029 \alpha^2 \tilde\epsilon^{6}+48297492 T \beta \tilde\epsilon^{6}+30505701 T^3 \tilde\epsilon^{8}-15215169 T \alpha \tilde\epsilon^{8}+3799023 \beta \tilde\epsilon^{8}-25827 T^2 \tilde\epsilon^{10}+4273 \alpha \tilde\epsilon^{10}-21 T \tilde\epsilon^{12}-\tilde\epsilon^{14}
$   \quad \\ $c_{15}= -181621440 T^7 \tilde\epsilon+162882720 T^5 \alpha \tilde\epsilon+29184000 T^3 \alpha^2 \tilde\epsilon-375456 T \alpha^3 \tilde\epsilon-94741920 T^4 \beta \tilde\epsilon-3738240 T^2 \alpha \beta \tilde\epsilon+25824 \alpha^2 \beta \tilde\epsilon+369792 T \beta^2 \tilde\epsilon+3412528704 T^6 \tilde\epsilon^{3}-2170399968 T^4 \alpha \tilde\epsilon^{3}-160843392 T^2 \alpha^2 \tilde\epsilon^{3}+1016416 \alpha^3 \tilde\epsilon^{3}+860360832 T^3 \beta \tilde\epsilon^{3}+19816704 T \alpha \beta \tilde\epsilon^{3}-1904256 \beta^2 \tilde\epsilon^{3}-6350387328 T^5 \tilde\epsilon^{5}+3550870848 T^3 \alpha \tilde\epsilon^{5}+95124288 T \alpha^2 \tilde\epsilon^{5}-1112400576 T^2 \beta \tilde\epsilon^{5}-6051072 \alpha \beta \tilde\epsilon^{5}+874094976 T^4 \tilde\epsilon^{7}-445619904 T^2 \alpha \tilde\epsilon^{7}-1503808 \alpha^2 \tilde\epsilon^{7}+115860096 T \beta \tilde\epsilon^{7}+228803136 T^3 \tilde\epsilon^{9}-114240864 T \alpha \tilde\epsilon^{9}+28540128 \beta \tilde\epsilon^{9}-30912 T^2 \tilde\epsilon^{11}+5152 \alpha \tilde\epsilon^{11}
$   \quad \\ $c_{16}= 34459425 T^8-64324260 T^6 \alpha-22972950 T^4 \alpha^2+492660 T^2 \alpha^3+321 \alpha^4+55135080 T^5 \beta+3341520 T^3 \alpha \beta-69768 T \alpha^2 \beta-385560 T^2 \beta^2+4968 \alpha \beta^2-6300435960 T^7 \tilde\epsilon^{2}+4642373736 T^5 \alpha \tilde\epsilon^{2}+679251816 T^3 \alpha^2 \tilde\epsilon^{2}-10488216 T \alpha^3 \tilde\epsilon^{2}-2334296232 T^4 \beta \tilde\epsilon^{2}-104974704 T^2 \alpha \beta \tilde\epsilon^{2}+696744 \alpha^2 \beta \tilde\epsilon^{2}+12826512 T \beta^2 \tilde\epsilon^{2}+42861778596 T^6 \tilde\epsilon^{4}-25659822276 T^4 \alpha \tilde\epsilon^{4}-1686752316 T^2 \alpha^2 \tilde\epsilon^{4}+13223724 \alpha^3 \tilde\epsilon^{4}+9466218288 T^3 \beta \tilde\epsilon^{4}+268275024 T \alpha \beta \tilde\epsilon^{4}-32226408 \beta^2 \tilde\epsilon^{4}-38745843480 T^5 \tilde\epsilon^{6}+21108908304 T^3 \alpha \tilde\epsilon^{6}+476468472 T \alpha^2 \tilde\epsilon^{6}-6356572848 T^2 \beta \tilde\epsilon^{6}-34950960 \alpha \beta \tilde\epsilon^{6}-1865304090 T^4 \tilde\epsilon^{8}+1046601828 T^2 \alpha \tilde\epsilon^{8}+18910602 \alpha^2 \tilde\epsilon^{8}-318442680 T \beta \tilde\epsilon^{8}+1019463288 T^3 \tilde\epsilon^{10}-509252952 T \alpha \tilde\epsilon^{10}+127253496 \beta \tilde\epsilon^{10}+4836 T^2 \tilde\epsilon^{12}-764 \alpha \tilde\epsilon^{12}+24 T \tilde\epsilon^{14}+\tilde\epsilon^{16}
$   \quad \\ $c_{17}= 4631346720 T^8 \tilde\epsilon-4521076560 T^6 \alpha \tilde\epsilon-1226778480 T^4 \alpha^2 \tilde\epsilon+31861296 T^2 \alpha^3 \tilde\epsilon-4176 \alpha^4 \tilde\epsilon+3067515360 T^5 \beta \tilde\epsilon+211507200 T^3 \alpha \beta \tilde\epsilon-4816800 T \alpha^2 \beta \tilde\epsilon-29175552 T^2 \beta^2 \tilde\epsilon+417312 \alpha \beta^2 \tilde\epsilon-134183604672 T^7 \tilde\epsilon^{3}+88037605152 T^5 \alpha \tilde\epsilon^{3}+11393315328 T^3 \alpha^2 \tilde\epsilon^{3}-206444064 T \alpha^3 \tilde\epsilon^{3}-39765148128 T^4 \beta \tilde\epsilon^{3}-2206304640 T^2 \alpha \beta \tilde\epsilon^{3}+13223328 \alpha^2 \beta \tilde\epsilon^{3}+315709056 T \beta^2 \tilde\epsilon^{3}+420217345440 T^6 \tilde\epsilon^{5}-239168412720 T^4 \alpha \tilde\epsilon^{5}-15315721920 T^2 \alpha^2 \tilde\epsilon^{5}+143574256 \alpha^3 \tilde\epsilon^{5}+83363343552 T^3 \beta \tilde\epsilon^{5}+3334566528 T \alpha \beta \tilde\epsilon^{5}-464003136 \beta^2 \tilde\epsilon^{5}-170517748608 T^5 \tilde\epsilon^{7}+90090243264 T^3 \alpha \tilde\epsilon^{7}+1851307200 T \alpha^2 \tilde\epsilon^{7}-26117293248 T^2 \beta \tilde\epsilon^{7}-195771648 \alpha \beta \tilde\epsilon^{7}-40879485216 T^4 \tilde\epsilon^{9}+21519881616 T^2 \alpha \tilde\epsilon^{9}+180156656 \alpha^2 \tilde\epsilon^{9}-5920340256 T \beta \tilde\epsilon^{9}+2361830976 T^3 \tilde\epsilon^{11}-1179776352 T \alpha \tilde\epsilon^{11}+294801696 \beta \tilde\epsilon^{11}+172896 T^2 \tilde\epsilon^{13}-28816 \alpha \tilde\epsilon^{13}
$   \quad \\ $c_{18}= -654729075 T^9+1571349780 T^7 \alpha+785674890 T^5 \alpha^2-28081620 T^3 \alpha^3-54891 T \alpha^4-1571349780 T^6 \beta-142849980 T^4 \alpha \beta+5965164 T^2 \alpha^2 \beta+33588 \alpha^3 \beta+21976920 T^3 \beta^2-849528 T \alpha \beta^2+14904 \beta^3+201644650935 T^8 \tilde\epsilon^{2}-153472402188 T^6 \alpha \tilde\epsilon^{2}-35572332186 T^4 \alpha^2 \tilde\epsilon^{2}+1100992764 T^2 \alpha^3 \tilde\epsilon^{2}-970985 \alpha^4 \tilde\epsilon^{2}+88837568424 T^5 \beta \tilde\epsilon^{2}+7497411408 T^3 \alpha \beta \tilde\epsilon^{2}-184446216 T \alpha^2 \beta \tilde\epsilon^{2}-1174529160 T^2 \beta^2 \tilde\epsilon^{2}+18883800 \alpha \beta^2 \tilde\epsilon^{2}-2060347538916 T^7 \tilde\epsilon^{4}+1239855532236 T^5 \alpha \tilde\epsilon^{4}+157039543692 T^3 \alpha^2 \tilde\epsilon^{4}-3177138132 T \alpha^3 \tilde\epsilon^{4}-517001103540 T^4 \beta \tilde\epsilon^{4}-39298810392 T^2 \alpha \beta \tilde\epsilon^{4}+195686388 \alpha^2 \beta \tilde\epsilon^{4}+6195471192 T \beta^2 \tilde\epsilon^{4}+3262065628212 T^6 \tilde\epsilon^{6}-1754323769700 T^4 \alpha \tilde\epsilon^{6}-130475616300 T^2 \alpha^2 \tilde\epsilon^{6}+1334176332 \alpha^3 \tilde\epsilon^{6}+585720087696 T^3 \beta \tilde\epsilon^{6}+39701917296 T \alpha \beta \tilde\epsilon^{6}-5898782952 \beta^2 \tilde\epsilon^{6}-311489753418 T^5 \tilde\epsilon^{8}+145322283996 T^3 \alpha \tilde\epsilon^{8}+5586917346 T \alpha^2 \tilde\epsilon^{8}-39363783084 T^2 \beta \tilde\epsilon^{8}-1372421628 \alpha \beta \tilde\epsilon^{8}-298175944446 T^4 \tilde\epsilon^{10}+154807935708 T^2 \alpha \tilde\epsilon^{10}+954504302 \alpha^2 \tilde\epsilon^{10}-41564614680 T \beta \tilde\epsilon^{10}-6370678260 T^3 \tilde\epsilon^{12}+3187499940 T \alpha \tilde\epsilon^{12}-797145228 \beta \tilde\epsilon^{12}+668292 T^2 \tilde\epsilon^{14}-111436 \alpha \tilde\epsilon^{14}-27 T \tilde\epsilon^{16}-\tilde\epsilon^{18}
$   \quad \\ $c_{19}= -125707982400 T^9 \tilde\epsilon+129898248480 T^7 \alpha \tilde\epsilon+51199944480 T^5 \alpha^2 \tilde\epsilon-2203382880 T^3 \alpha^3 \tilde\epsilon-2789280 T \alpha^4 \tilde\epsilon-101709185760 T^6 \beta \tilde\epsilon-11218694400 T^4 \alpha \beta \tilde\epsilon+534405600 T^2 \alpha^2 \beta \tilde\epsilon+3124800 \alpha^3 \beta \tilde\epsilon+1926581760 T^3 \beta^2 \tilde\epsilon-86028480 T \alpha \beta^2 \tilde\epsilon+1788480 \beta^3 \tilde\epsilon+5184937293120 T^8 \tilde\epsilon^{3}-3392667185184 T^6 \alpha \tilde\epsilon^{3}-755365406688 T^4 \alpha^2 \tilde\epsilon^{3}+27019448544 T^2 \alpha^3 \tilde\epsilon^{3}-42949280 \alpha^4 \tilde\epsilon^{3}+1757337685056 T^5 \beta \tilde\epsilon^{3}+199084939776 T^3 \alpha \beta \tilde\epsilon^{3}-5206935744 T \alpha^2 \beta \tilde\epsilon^{3}-33530029056 T^2 \beta^2 \tilde\epsilon^{3}+610030656 \alpha \beta^2 \tilde\epsilon^{3}-24434174308032 T^7 \tilde\epsilon^{5}+13490703299232 T^5 \alpha \tilde\epsilon^{5}+1943858549760 T^3 \alpha^2 \tilde\epsilon^{5}-40343194272 T \alpha^3 \tilde\epsilon^{5}-5319113995296 T^4 \beta \tilde\epsilon^{5}-629630011008 T^2 \alpha \beta \tilde\epsilon^{5}+2375785056 \alpha^2 \beta \tilde\epsilon^{5}+103152458880 T \beta^2 \tilde\epsilon^{5}+19185746159808 T^6 \tilde\epsilon^{7}-9298051961760 T^4 \alpha \tilde\epsilon^{7}-1134597817728 T^2 \alpha^2 \tilde\epsilon^{7}+10732608544 \alpha^3 \tilde\epsilon^{7}+3038929881984 T^3 \beta \tilde\epsilon^{7}+456898496256 T \alpha \beta \tilde\epsilon^{7}-67794102144 \beta^2 \tilde\epsilon^{7}+3190569641280 T^5 \tilde\epsilon^{9}-1871874604320 T^3 \alpha \tilde\epsilon^{9}+23216772576 T \alpha^2 \tilde\epsilon^{9}+528275642976 T^2 \beta \tilde\epsilon^{9}-12995167872 \alpha \beta \tilde\epsilon^{9}-1384094869056 T^4 \tilde\epsilon^{11}+711029387808 T^2 \alpha \tilde\epsilon^{11}+3168265824 \alpha^2 \tilde\epsilon^{11}-187258689216 T \beta \tilde\epsilon^{11}-106997127744 T^3 \tilde\epsilon^{13}+53501343072 T \alpha \tilde\epsilon^{13}-13375683168 \beta \tilde\epsilon^{13}+1809984 T^2 \tilde\epsilon^{15}-301664 \alpha \tilde\epsilon^{15}
$   \quad \\ $c_{20}= 13749310575 T^{10}-41247931725 T^8 \alpha-27498621150 T^6 \alpha^2+1474285050 T^4 \alpha^3+5763555 T^2 \alpha^4+160839 \alpha^5+47140493400 T^7 \beta+5999699160 T^5 \alpha \beta-417561480 T^3 \alpha^2 \beta-7053480 T \alpha^3 \beta-1153788300 T^4 \beta^2+89200440 T^2 \alpha \beta^2+257580 \alpha^2 \beta^2-3129840 T \beta^3-6638452845570 T^9 \tilde\epsilon^{2}+5065584826680 T^7 \alpha \tilde\epsilon^{2}+1831520977020 T^5 \alpha^2 \tilde\epsilon^{2}-92808482040 T^3 \alpha^3 \tilde\epsilon^{2}-100320690 T \alpha^4 \tilde\epsilon^{2}-3383275732200 T^6 \beta \tilde\epsilon^{2}-495222462360 T^4 \alpha \beta \tilde\epsilon^{2}+26590227480 T^2 \alpha^2 \beta \tilde\epsilon^{2}+157321800 \alpha^3 \beta \tilde\epsilon^{2}+90006169680 T^3 \beta^2 \tilde\epsilon^{2}-4693522320 T \alpha \beta^2 \tilde\epsilon^{2}+114404400 \beta^3 \tilde\epsilon^{2}+94286884183821 T^8 \tilde\epsilon^{4}-54151412884020 T^6 \alpha \tilde\epsilon^{4}-13457360481822 T^4 \alpha^2 \tilde\epsilon^{4}+531612223812 T^2 \alpha^3 \tilde\epsilon^{4}-1182882771 \alpha^4 \tilde\epsilon^{4}+26102723565096 T^5 \beta \tilde\epsilon^{4}+4429354286160 T^3 \alpha \beta \tilde\epsilon^{4}-121555350216 T \alpha^2 \beta \tilde\epsilon^{4}-762687063864 T^2 \beta^2 \tilde\epsilon^{4}+15750994824 \alpha \beta^2 \tilde\epsilon^{4}-227746405182600 T^7 \tilde\epsilon^{6}+110666587239576 T^5 \alpha \tilde\epsilon^{6}+23106793925784 T^3 \alpha^2 \tilde\epsilon^{6}-436128912360 T \alpha^3 \tilde\epsilon^{6}-42619899934584 T^4 \beta \tilde\epsilon^{6}-9306495406800 T^2 \alpha \beta \tilde\epsilon^{6}+24363330936 \alpha^2 \beta \tilde\epsilon^{6}+1513339160688 T \beta^2 \tilde\epsilon^{6}+65674071269550 T^6 \tilde\epsilon^{8}-18825895417230 T^4 \alpha \tilde\epsilon^{8}-10589311873890 T^2 \alpha^2 \tilde\epsilon^{8}+74566082042 \alpha^3 \tilde\epsilon^{8}+7022789376840 T^3 \beta \tilde\epsilon^{8}+5032466431992 T \alpha \beta \tilde\epsilon^{8}-715199813964 \beta^2 \tilde\epsilon^{8}+44185477656852 T^5 \tilde\epsilon^{10}-24650424710712 T^3 \alpha \tilde\epsilon^{10}+278062980636 T \alpha^2 \tilde\epsilon^{10}+6649057289160 T^2 \beta \tilde\epsilon^{10}-132063679800 \alpha \beta \tilde\epsilon^{10}-3267687087630 T^4 \tilde\epsilon^{12}+1645818368460 T^2 \alpha \tilde\epsilon^{12}+2009429662 \alpha^2 \tilde\epsilon^{12}-417472660872 T \beta \tilde\epsilon^{12}-676344299592 T^3 \tilde\epsilon^{14}+338171790312 T \alpha \tilde\epsilon^{14}-84542902440 \beta \tilde\epsilon^{14}+3897315 T^2 \tilde\epsilon^{16}-649485 \alpha \tilde\epsilon^{16}+30 T \tilde\epsilon^{18}+\tilde\epsilon^{20}$
}

\subsection{Hurwitz expansion for PI}
\label{app:PI}
In this section \(c_n=c_n^{PI}\) from~\eqref{eq:cnPI}.

($\alpha=g_2^{PI}/2\ ,\ \beta=2g_3^{PI}$)
{\footnotesize
    \quad \\ $c_{0}= 1
$   \quad \\ $c_{1}= 0
$   \quad \\ $c_{2}= 0
$   \quad \\ $c_{3}= 0
$   \quad \\ $c_{4}= -\alpha
$   \quad \\ $c_{5}= 6 \epsilon
$   \quad \\ $c_{6}= -3 \beta
$   \quad \\ $c_{7}= 0
$   \quad \\ $c_{8}= -9 \alpha^{2}
$   \quad \\ $c_{9}= 84 \alpha \epsilon
$   \quad \\ $c_{10}= -18 \alpha \beta-294 \epsilon^{2}
$   \quad \\ $c_{11}= 216 \beta \epsilon
$   \quad \\ $c_{12}= 69 \alpha^{3}-54 \beta^{2}
$   \quad \\ $c_{13}= -1650 \alpha^{2} \epsilon
$   \quad \\ $c_{14}= 513 \alpha^{2} \beta+18774 \alpha \epsilon^{2}
$   \quad \\ $c_{15}= -18720 \alpha \beta \epsilon-78624 \epsilon^{3}
$   \quad \\ $c_{16}= 321 \alpha^{4}+4968 \alpha \beta^{2}+144144 \beta \epsilon^{2}
$   \quad \\ $c_{17}= -52488 \alpha^{3} \epsilon-89424 \beta^{2} \epsilon
$   \quad \\ $c_{18}= 33588 \alpha^{3} \beta+14904 \beta^{3}+1112436 \alpha^{2} \epsilon^{2}
$   \quad \\ $c_{19}= -1358640 \alpha^{2} \beta \epsilon-8670816 \alpha \epsilon^{3}
$   \quad \\ $c_{20}= 160839 \alpha^{5}+257580 \alpha^{2} \beta^{2}+15053040 \alpha \beta \epsilon^{2}+27734616 \epsilon^{4}
$   \quad \\ $c_{21}= -9642690 \alpha^{4} \epsilon-5786640 \alpha \beta^{2} \epsilon-67223520 \beta \epsilon^{3}
$   \quad \\ $c_{22}= 2808945 \alpha^{4} \beta+502200 \alpha \beta^{3}+221121900 \alpha^{3} \epsilon^{2}+47585880 \beta^{2} \epsilon^{2}
$   \quad \\ $c_{23}= -143272800 \alpha^{3} \beta \epsilon-12052800 \beta^{3} \epsilon-2725500960 \alpha^{2} \epsilon^{3}
$   \quad \\ $c_{24}= 1416951 \alpha^{6}+20019960 \alpha^{3} \beta^{2}+1506600 \beta^{4}+3160803600 \alpha^{2} \beta \epsilon^{2}+18370803024 \alpha \epsilon^{4}
$   \quad \\ $c_{25}= -17890596 \alpha^{5} \epsilon-1168920720 \alpha^{2} \beta^{2} \epsilon-33552639936 \alpha \beta \epsilon^{3}-55673733024 \epsilon^{5}
$   \quad \\ $c_{26}= -41843142 \alpha^{5} \beta+162100440 \alpha^{2} \beta^{3}-781903530 \alpha^{4} \epsilon^{2}+22914336240 \alpha \beta^{2} \epsilon^{2}+150406965072 \beta \epsilon^{4}
$   \quad \\ $c_{27}= 2956347720 \alpha^{4} \beta \epsilon-7272970560 \alpha \beta^{3} \epsilon+46552246560 \alpha^{3} \epsilon^{3}-165826388160 \beta^{2} \epsilon^{3}
$   \quad \\ $c_{28}= -388946691 \alpha^{7}-376375410 \alpha^{4} \beta^{2}+796330440 \alpha \beta^{4}-123190094160 \alpha^{3} \beta \epsilon^{2}+92545558560 \beta^{3} \epsilon^{2}-1059992932824 \alpha^{2} \epsilon^{4}
$   \quad    $c_{29}= 35919307926 \alpha^{6} \epsilon+52917876720 \alpha^{3} \beta^{2} \epsilon-23889913200 \beta^{4} \epsilon+3205542167136 \alpha^{2} \beta \epsilon^{3}+11682746528544 \alpha \epsilon^{5}
$   \quad \\ $c_{30}= -6519779667 \alpha^{6} \beta-9465715080 \alpha^{3} \beta^{3}+2388991320 \beta^{5}-1720632974886 \alpha^{5} \epsilon^{2}-2734614623160 \alpha^{2} \beta^{2} \epsilon^{2}-45309600930384 \alpha \beta \epsilon^{4}-48569368118544 \epsilon^{6}
$   \quad \\ $c_{31}= 978517329984 \alpha^{5} \beta \epsilon+1075789382400 \alpha^{2} \beta^{3} \epsilon+52109440015680 \alpha^{4} \epsilon^{3}+61723536910848 \alpha \beta^{2} \epsilon^{3}+239259812553216 \beta \epsilon^{5}
$   \quad    $c_{32}= 25514578881 \alpha^{8}-210469286736 \alpha^{5} \beta^{2}-144916218720 \alpha^{2} \beta^{4}-62015233698720 \alpha^{4} \beta \epsilon^{2}-40204399921920 \alpha \beta^{3} \epsilon^{2}-1021069206582336 \alpha^{3} \epsilon^{4}-453518093142144 \beta^{2} \epsilon^{4}
$   \quad \\ $c_{33}= -1666422337680 \alpha^{7} \epsilon+30257070700320 \alpha^{4} \beta^{2} \epsilon+11884499521920 \alpha \beta^{4} \epsilon+2042944257556224 \alpha^{3} \beta \epsilon^{3}+428809408012800 \beta^{3} \epsilon^{3}+12260416976841600 \alpha^{2} \epsilon^{5}
$   \quad \\ $c_{34}= -485174610648 \alpha^{7} \beta-4582619446320 \alpha^{4} \beta^{3}-1289959784640 \alpha \beta^{5}-16630361205624 \alpha^{6} \epsilon^{2}-1654258781960640 \alpha^{3} \beta^{2} \epsilon^{2}-203961254126400 \beta^{4} \epsilon^{2}-35333389387302336 \alpha^{2} \beta \epsilon^{4}-82954144406480256 \alpha \epsilon^{6}
$   \quad \\ $c_{35}= 152134674660000 \alpha^{6} \beta \epsilon+543425169335040 \alpha^{3} \beta^{3} \epsilon+46438552247040 \beta^{5} \epsilon+4829402276650944 \alpha^{5} \epsilon^{3}+40951207980949248 \alpha^{2} \beta^{2} \epsilon^{3}+315176201202046464 \alpha \beta \epsilon^{5}+252377177931680256 \epsilon^{7}
$   \quad \\ $c_{36}= -7647989401521 \alpha^{9}-41767502762088 \alpha^{6} \beta^{2}-60648644233440 \alpha^{3} \beta^{4}-3869879353920 \beta^{6}-13963139340227424 \alpha^{5} \beta \epsilon^{2}-21252172506998400 \alpha^{2} \beta^{3} \epsilon^{2}-188196130875388464 \alpha^{4} \epsilon^{4}-483274189528564608 \alpha \beta^{2} \epsilon^{4}-1221327388362122496 \beta \epsilon^{6}
$   \quad    $c_{37}= 1597613431426038 \alpha^{8} \epsilon+7287575315043744 \alpha^{5} \beta^{2} \epsilon+4860000915779520 \alpha^{2} \beta^{4} \epsilon+580850811563253696 \alpha^{4} \beta \epsilon^{3}+352379520821205504 \alpha \beta^{3} \epsilon^{3}+3629347149856321152 \alpha^{3} \epsilon^{5}+2426497407436144896 \beta^{2} \epsilon^{5}
$   \quad \\ $c_{38}= -544306979739483 \alpha^{8} \beta-1028311276281264 \alpha^{5} \beta^{3}-383302865236320 \alpha^{2} \beta^{5}-145957471490813256 \alpha^{7} \epsilon^{2}-450223680165654960 \alpha^{4} \beta^{2} \epsilon^{2}-126206480913710400 \alpha \beta^{4} \epsilon^{2}-13083603812270471232 \alpha^{3} \beta \epsilon^{4}-2419559327517745536 \beta^{3} \epsilon^{4}-39970888866183167424 \alpha^{2} \epsilon^{6}
$   \quad \\ $c_{39}= 109258810510910400 \alpha^{7} \beta \epsilon+125438691795788160 \alpha^{4} \beta^{3} \epsilon+20973065667632640 \alpha \beta^{5} \epsilon+7465233450031396416 \alpha^{6} \epsilon^{3}+13786503282072709632 \alpha^{3} \beta^{2} \epsilon^{3}+1276425643405248000 \beta^{4} \epsilon^{3}+170610428557450446336 \alpha^{2} \beta \epsilon^{5}+244708656405420340224 \alpha \epsilon^{7}
$   \quad    $c_{40}= -1013917176434889 \alpha^{10}-19199774752012080 \alpha^{7} \beta^{2}-10665863758194480 \alpha^{4} \beta^{4}-1343956990999680 \alpha \beta^{6}-8993959815439913760 \alpha^{6} \beta \epsilon^{2}-5880776768276140800 \alpha^{3} \beta^{3} \epsilon^{2}-364153149094199040 \beta^{5} \epsilon^{2}-239802121803604192992 \alpha^{5} \epsilon^{4}-234230113062827683200 \alpha^{2} \beta^{2} \epsilon^{4}-1230173253340180886016 \alpha \beta \epsilon^{6}-676706834479537540224 \epsilon^{8}
$   \quad \\ $c_{41}= 257198900059618020 \alpha^{9} \epsilon+3280225716831278880 \alpha^{6} \beta^{2} \epsilon+1038805296985975680 \alpha^{3} \beta^{4} \epsilon+56446193621986560 \beta^{6} \epsilon+414215430149742286464 \alpha^{5} \beta \epsilon^{3}+140048930941054152192 \alpha^{2} \beta^{3} \epsilon^{3}+5141394022880077283520 \alpha^{4} \epsilon^{5}+\\2123543703270210356736 \alpha \beta^{2} \epsilon^{5}+4007543942148905733120 \beta \epsilon^{7}
$   \quad \\ $c_{42}= -86437871519050170 \alpha^{9} \beta-349661947783523760 \alpha^{6} \beta^{3}-66227538979062720 \alpha^{3} \beta^{5}-4031870972999040 \beta^{7}-26244455690738520270 \alpha^{8} \epsilon^{2}-241771461681284791200 \alpha^{5} \beta^{2} \epsilon^{2}-39868882997418498240 \alpha^{2} \beta^{4} \epsilon^{2}-\\12027054880512357793632 \alpha^{4} \beta \epsilon^{4}-1677410179866363591936 \alpha \beta^{3} \epsilon^{4}-74935202706816131514240 \alpha^{3} \epsilon^{6}-\\8614891199805709029120 \beta^{2} \epsilon^{6}
$   \quad \\ $c_{43}= 17141216361034885560 \alpha^{8} \beta \epsilon+56634559538257975680 \alpha^{5} \beta^{3} \epsilon+5650331551179759360 \alpha^{2} \beta^{5} \epsilon+\\1528293063185125594560 \alpha^{7} \epsilon^{3}+10214299002936131574912 \alpha^{4} \beta^{2} \epsilon^{3}+672980146760631338496 \alpha \beta^{4} \epsilon^{3}+\\229035416163274962860544 \alpha^{3} \beta \epsilon^{5}+8883793454187394206720 \beta^{3} \epsilon^{5}+727102801248129087828480 \alpha^{2} \epsilon^{7}
$   \quad \\ $c_{44}= -155812911328032651 \alpha^{11}-2149281495195098670 \alpha^{8} \beta^{2}-4849557774995228400 \alpha^{5} \beta^{4}-342039979512624960 \alpha^{2} \beta^{6}-1531812482581505090400 \alpha^{7} \beta \epsilon^{2}-4022774166702416529600 \alpha^{4} \beta^{3} \epsilon^{2}-148507847339176846080 \alpha \beta^{5} \epsilon^{2}-\\58065576264612339606288 \alpha^{6} \epsilon^{4}-268718195338965949207680 \alpha^{3} \beta^{2} \epsilon^{4}-4878046619186157889920 \beta^{4} \epsilon^{4}-\\2837052581447767856673024 \alpha^{2} \beta \epsilon^{6}-4310436803047532713535616 \alpha \epsilon^{8}
$   \quad \\ $c_{45}= 25671108050154272286 \alpha^{10} \epsilon+422251379341336369440 \alpha^{7} \beta^{2} \epsilon+784023150124579024800 \alpha^{4} \beta^{4} \epsilon+\\17225302294257120000 \alpha \beta^{6} \epsilon+82274221325423641453632 \alpha^{6} \beta \epsilon^{3}+158923721697922624826880 \alpha^{3} \beta^{3} \epsilon^{3}+\\1583865250498050670080 \beta^{5} \epsilon^{3}+1516158412144555288845888 \alpha^{5} \epsilon^{5}+4452130384760643114481920 \alpha^{2} \beta^{2} \epsilon^{5}+21104914153748833029178368 \alpha \beta \epsilon^{7}+12020271555723070195535616 \epsilon^{9}
$   \quad \\ $c_{46}= 457002245380426137 \alpha^{10} \beta-39651545649351066480 \alpha^{7} \beta^{3}-61833558106241727120 \alpha^{4} \beta^{5}-721722455986066560 \alpha \beta^{7}-1805147237470952543490 \alpha^{9} \epsilon^{2}-38367530463572033104080 \alpha^{6} \beta^{2} \epsilon^{2}-52327000215657788644800 \alpha^{3} \beta^{4} \epsilon^{2}-\\311016531583424177280 \beta^{6} \epsilon^{2}-2900240815540915434105504 \alpha^{5} \beta \epsilon^{4}-3721509112622871754961280 \alpha^{2} \beta^{3} \epsilon^{4}-27607115343501169107968160 \alpha^{4} \epsilon^{6}-43174042490952140854636800 \alpha \beta^{2} \epsilon^{6}-72893706065960944569245568 \beta \epsilon^{8}
$   \quad    $c_{47}= -483101965679791580640 \alpha^{9} \beta \epsilon+8356279903195965592320 \alpha^{6} \beta^{3} \epsilon+9122936489578121333760 \alpha^{3} \beta^{5} \epsilon+34642677887331194880 \beta^{7} \epsilon+66910821915558650914080 \alpha^{8} \epsilon^{3}+2028231018986670690393600 \alpha^{5} \beta^{2} \epsilon^{3}+\\1841387437469129282933760 \alpha^{2} \beta^{4} \epsilon^{3}+68795016642651589252554240 \alpha^{4} \beta \epsilon^{5}+48876224832570019585904640 \alpha \beta^{3} \epsilon^{5}+344513520088978516236011520 \alpha^{3} \epsilon^{7}+190264013634279261361950720 \beta^{2} \epsilon^{7}
$   \quad \\ $c_{48}= 58581931023896704641 \alpha^{12}+153743102407096274520 \alpha^{9} \beta^{2}-734120023212871152480 \alpha^{6} \beta^{4}-\\645698596347331349760 \alpha^{3} \beta^{6}-2165167367958199680 \beta^{8}+79849101718087266766320 \alpha^{8} \beta \epsilon^{2}-\\739717563888387747774720 \alpha^{5} \beta^{3} \epsilon^{2}-537112321930721091985920 \alpha^{2} \beta^{5} \epsilon^{2}-840265898152433500383552 \alpha^{7} \epsilon^{4}-66770871086413117846212480 \alpha^{4} \beta^{2} \epsilon^{4}-33900934471184505688588800 \alpha \beta^{4} \epsilon^{4}-1075554734572492499863065600 \alpha^{3} \beta \epsilon^{6}-282037154509044583098808320 \beta^{3} \epsilon^{6}-2791244626252018773872926464 \alpha^{2} \epsilon^{8}
$   \quad \\ $c_{49}= -16476027621241804816536 \alpha^{11} \epsilon-39194435351714972721840 \alpha^{8} \beta^{2} \epsilon+137663740476617294620800 \alpha^{5} \beta^{4} \epsilon+84745287778992732864000 \alpha^{2} \beta^{6} \epsilon-7207137489190369762467072 \alpha^{7} \beta \epsilon^{3}+35311919958400774248967680 \alpha^{4} \beta^{3} \epsilon^{3}+14616501092920672278681600 \alpha \beta^{5} \epsilon^{3}-47383566530308128990453888 \alpha^{6} \epsilon^{5}+1348488812982089169033661440 \alpha^{3} \beta^{2} \epsilon^{5}+261411483394198530073728000 \beta^{4} \epsilon^{5}+10385029165695067572253673472 \alpha^{2} \beta \epsilon^{7}+13092369581815270041904487424 \alpha \epsilon^{9}
$   \quad    $c_{50}= 3569731062346847916252 \alpha^{11} \beta+4033242497662632885960 \alpha^{8} \beta^{3}-9637760723299526905920 \alpha^{5} \beta^{5}-5620752638745660576000 \alpha^{2} \beta^{7}+2137215465815076126656316 \alpha^{10} \epsilon^{2}+5208488693311600246374720 \alpha^{7} \beta^{2} \epsilon^{2}-10300858021676147714308800 \alpha^{4} \beta^{4} \epsilon^{2}-3798800159018699241792000 \alpha \beta^{6} \epsilon^{2}+439458744707798918632100928 \alpha^{6} \beta \epsilon^{4}-923015921693960711625361920 \alpha^{3} \beta^{3} \epsilon^{4}-155160789279963572108505600 \beta^{5} \epsilon^{4}+3259176861802252188541131648 \alpha^{5} \epsilon^{6}-15363008628250975060494887424 \alpha^{2} \beta^{2} \epsilon^{6}-54039305524932333061285455360 \alpha \beta \epsilon^{8}-27265762054599700941736487424 \epsilon^{10}$
}


\end{document}